\documentclass[a4paper,12pt]{article}

\usepackage[numbers]{natbib}
\usepackage[a4paper,margin=2.5cm]{geometry}
\usepackage{tocloft}
\usepackage{caption}
\usepackage{authblk}
\usepackage{subcaption}
\usepackage{amssymb}
\usepackage{amsmath,bm}
\usepackage{layout}
\usepackage{soul}
\usepackage{color}
\usepackage{braket}
\usepackage{amsmath} 
\usepackage{hyperref}
\usepackage{dirtytalk}
\usepackage{mathrsfs}
\usepackage{xcolor}
\usepackage{qcircuit}
\usepackage{graphicx}
\usepackage{listings}

\title{Notes on Quantum Computing for Thermal Science}

\author[1,2]{Pietro Asinari
\thanks{Corresponding author: pietro.asinari@polito.it}}

\author[1]{Nada Alghamdi}
\author[1]{Paolo De Angelis}
\author[1]{Giulio Barletta}
\author[1,3]{Giovanni Trezza}
\author[1]{Marina Provenzano}
\author[1]{Matteo Maria Piredda}
\author[1]{Matteo Fasano}
\author[1,2]{Eliodoro Chiavazzo}

\affil[1]{{\footnotesize Dipartimento Energia, Politecnico di Torino, Corso Duca degli Abruzzi 24, 10129, Torino (TO), Italy}}

\affil[2]{{\footnotesize Istituto Nazionale di Ricerca Metrologica, Strada delle Cacce 91, 10135, Torino (TO), Italy}}

\affil[3]{{\footnotesize Université Grenoble Alpes, 1130 Rue de la Piscine, St Martin D’Heres, 38402, France}}

\date{Version v01.22}

\begin{document}
\maketitle

\hrule
\vspace{0.5em}
\begin{abstract}
This document explores the potential of quantum computing for solving linear systems of interest in engineering. In particular, we focus on heat conduction as a paradigmatic example in thermal science. Conceived as a living document, it will be continuously updated with experimental findings and insights for the research community in Thermal Science. 
By experiments, we refer both to the search for the most effective algorithms and to the performance of real quantum hardware. Those fields are currently evolving rapidly, driving a technological race to define the best architectures. The development of novel algorithms for engineering problems aims at harnessing the unique strengths of quantum computing. Expectations are high, as users seek concrete evidence of quantum supremacy -- a true game changer for engineering applications. Among all heat transfer mechanisms (conduction, convection, radiation), we start with conduction as a paradigmatic test case in the field being characterized by a rich mathematical foundation for our investigations.
\end{abstract}
\vspace{0.5em}
\hrule

\newpage

\tableofcontents

\newpage

\section{Introduction}
\label{intro}

Quantum computing could transform fields like computational science and engineering with possibly strong impact on material science, renewable energy and even finance by revolutionizing data processing. 
The ambitious goal is quantum advantage possibly outperforming classical computers in specific tasks.
Here, in order to overcome this challenge, we focus on solving the heat conduction equation numerically as a paradigmatic application of quantum computing within the heat transfer and thermal science community. 
In most of the engineering applications, the computational domain - where heat conduction occurs - is discretized by a spatial mesh with $N$ nodes.
State-of-the-art CFD simulations can utilize up to 780 billion mesh cells on advanced supercomputers. For example, a study documented the use of a grid with 780 billion cells ($N\sim 7.8 \times 10^{11}$) on \href{https://it.wikipedia.org/wiki/Tianhe-2}{Tianhe-2}, leveraging over 1.376 million heterogeneous cores \citep{wang_performance_2018}. This is where -- potentially -- a quantum advantage could become interesting.

To better understand the potential quantum advantage, let us first recall how a classical computer processes real numbers. 
A common approach is to represent them in scientific notation as $\mu \times 10^{\mathbf{\mathrm{exponent}}}$, where $1 \leq \mu < 10$ is the significand (or mantissa). 
The precision by which a classical computer can store a real number depends on the number of bits available for encoding the mantissa. 
To illustrate this, consider a simplified scenario where the computer has only three bits to store the mantissa. 
With $N = 2^3 = 8$ possible binary configurations, the mantissa must be approximated to the closest available value within the range $\mu\in\left[1,10\right[$. 
A reasonable discretization scheme is $\mu \in \{\mu_0, \mu_1,\dots, \mu_{N-1}\}$, where the generic significand can be expressed as $\mu_b = 1 + (10-1) \,b/ 2^3$ and $b$ is an integer from $0$ to $2^3-1$ included. The difference between two consecutive significands, namely the precision, is $\Delta\mu = \mu_{b+1} - \mu_b = (10-1) / 2^3 = 1.125$. Naturally, real-world classical computers operate with far greater precision. 
A widely used standard for representing real numbers is the IEEE 754 double-precision floating-point format, which allocates 64 bits per number. 
Within this format, 53 bits are dedicated to encoding the mantissa, leading to a much finer discretization $(10-1) / 2^{53} \approx 1 \times 10^{-15}$. 
This implies that, in a classical system, 53 bits are used solely to encode the significand of a single real number. 
More generally, given $n_{\text{classic}}$ available classical bits, the number of real numbers that can be stored (with double-precision) is given by 
\begin{equation}
\label{classical}
    N_{\text{classic}} = \left\lfloor n_{\text{classic}} / 64 \right\rfloor,
\end{equation}
where $\left\lfloor \cdot \right\rfloor$ denotes the floor function, which returns the largest integer less than or equal to its argument. 
This constraint on classical storage is a key limitation that quantum computing aims to overcome.

The key distinction lies in the following fundamental property: a quantum system with $n$ qubits has 
\begin{equation}\label{key-point}
    N = 2^n\;\;(\gg n\text{  typically})
\end{equation}
computational basis states, similar to a classical system. However -- and this is the crucial difference -- a quantum system can exist in a superposition of all these basis states, with each state weighted by a complex probability amplitude. 
These amplitudes, which may be continuous real (or complex) numbers, determine the probability of measuring each basis state upon observation. 
As a result, a quantum computer can encode and manipulate a number of real values that is at least proportional to the number of computational basis states. 
This \underline{exponential scaling} in the number of quantum states provides a potential advantage over classical systems. 
However, the actual precision of stored information is constrained by interactions with the environment, which cause decoherence, i.e. a process that disrupts quantum superpositions and limits computational performance. 
Moreover, there is a limit in the measurement resolution due to the actual hardware.
To fix ideas, let us consider the following example: In November 2024, IBM released \href{https://newsroom.ibm.com/2024-11-13-ibm-launches-its-most-advanced-quantum-computers,-fueling-new-scientific-value-and-progress-towards-quantum-advantage}{the IBM Heron R2 processor} with $n = 156$ qubits, which could -- in principle -- accommodate $N = 2^{156} \sim 10^{46}$ field values on a mesh. On the other hand, a classical computer with the same number of classical bits $n_{\text{classic}} = 156$ can store at most $N_{\text{classic}} = 2$ real field values in double-precision. 
Hence it is clear that quantum computing may pave the way to significantly larger meshes than those currently used.

However, there is a problem. Currently, we are in the \href{https://en.wikipedia.org/wiki/Noisy_intermediate-scale_quantum_era}{noisy intermediate-scale quantum (NISQ) era}, with processors of up to 1,000 non-fault-tolerant qubits. 
Overcoming noise and decoherence remains a significant challenge, making it crucial to align quantum hardware advancements with specific application needs. 
In reality, the current NISQ computers are still rarely advantageous over classical computers for most of applications, which therefore must be investigated individually. Let us focus here on solving the heat conduction equation.

\subsection{Philosophical remark}
\label{philosophical}

Before proceeding further, it is worth first discussing how an ideally reversible quantum computer can model an irreversible phenomenon. To clarify this point, let us consider an ideal quantum computer as a specific example of a generic quantum system. Among all possible quantum systems, a particularly illustrative one is the Schr\"{o}dinger equation for a free particle in one dimension: $i\,\hbar\,\partial_t \Psi = \hat{p}/(2\,m)\,\Psi = -\hbar^2/(2\,m)\,\partial_z^2\Psi$ or equivalently $\partial_t \Psi = i\,\hbar/(2\,m)\,\partial_z^2\Psi$, where $\Psi$ represents the wavefunction of the quantum system. This equation is classified as dispersive, meaning it supports wave-like solutions with frequency-dependent phase velocities. Its purely imaginary time evolution results in phase oscillations without any decay. In other words, the Schr\"{o}dinger equation is not dissipative -- it describes reversible, wave-like behavior rather than an irreversible process. By contrast, the heat conduction equation, which we aim to model here, namely $\partial_t T=D\,\partial_z^2 T$, is 
inherently dissipative. 

\vspace{0.4cm}  
\fbox{\begin{minipage}{0.9\textwidth}  
In an ideal quantum computer, where there is no interaction with the environment, state evolution is unitary and therefore reversible. The only source of irreversibility in such a system is measurement.  
\end{minipage}}  
\vspace{0.4cm}  

The key idea, therefore, is to construct a reversible quantum evolution that, upon measurement, collapses onto the target dissipative dynamics. The measurement in quantum mechanics consists of three ingredients: (i) the state, (ii) the observable and (iii) its expectation value. The state of a system is represented by its wavefunction $\Psi$, which contains all the information about the system. The observable refers to any physical quantity that can be measured, such as position, momentum, or energy, and it is represented by an operator $\hat{O}$. The expectation value of an observable is the average result one would obtain from many measurements of that observable, and it is calculated by taking the inner product $\braket{\Psi|\hat{O}|\Psi}$, where $\hat{O}$ is the operator corresponding to the observable. When an observable is measured, the system’s wavefunction collapses to one of the eigenstates of the corresponding operator, and the measurement result will be one of the associated eigenvalues. The expectation value is the weighted average of these eigenvalues, with the weights being the probabilities of the system being found in each eigenstate. Hence the measurement process in quantum mechanics is often used as a way to model irreversible phenomena.

\section{Heat conduction equation}
\label{hce}

Let us consider the one-dimensional heat conduction equation, as a paradigmatic application to the heat transfer community, namely
\begin{equation}\label{heat-conduction}
    \frac{\partial T}{\partial t} = 
    D\,\frac{\partial^2 T}{\partial z^2},
\end{equation}
with the function $T = T(z,t)$ being the local temperature, and the positive coefficient $D$ the thermal diffusivity of the medium.
Let us consider a constant diffusivity, a given initial profile $T(z,0)$ and the periodic spatial boundary condition. 
This problem can be solved analytically using the Fourier transform and it is usually trivial for most of the current classical numerical techniques. 

Let us solve the previous equation by the classical finite-difference (FD) method, which consists in solving differential equations by approximating derivatives with finite differences. 
Both the spatial domain and time domain are discretized by a regular mesh: the unknown function $T$ is evaluated at the generic $l$-th mesh node and at the $\tau$-th time step, namely $T_l^\tau = T(z_l,t_\tau)$ where $z_l = l\,\Delta z$ with $0\leq l\leq (N-1)$ and $t_\tau = \tau\,\Delta t$ with $0\leq \tau\leq N_t$ ($\tau=0$ identifies the given initial profile). 
The quantities $\Delta z$ and $\Delta t$ are the spatial and temporal partitions of the grid, while $N$ is the number of space mesh nodes and $N_t$ is the number of time steps. 
Let us use a fully implicit FD scheme that yields the stability of the solution for arbitrary diffusivity of the equation and the grid size:
\begin{equation}\label{fd-heat-conduction}
    \frac{T_l^{\tau+1}-T_l^{\tau}}{\Delta t} = 
    D\,\frac{T_{l-1}^{\tau+1}-2\,T_l^{\tau+1}+T_{l+1}^{\tau+1}}{\Delta z^2}.
\end{equation}
The previous formula can be reformulated as
\begin{equation}\label{line-heat-conduction}
    -r\,T_{l-1}^{\tau+1}+\left(1+2\,r\right)\,T_l^{\tau+1}
    -r\,T_{l+1}^{\tau+1} = T_l^{\tau},
\end{equation}
where $r=D\,\Delta t/\Delta z^2$ is the (dimensionless) numerical Fourier number. Let us define a new operator $\hat{C}$ as
\begin{equation}\label{conduction-matrix}
\hat{C} := \begin{bmatrix}
\left(1+2\,r\right) & -r & 0 & 0 & \dots & -r \\
-r & \left(1+2\,r\right) & -r & 0 & \dots & 0 \\
 0 & -r & \left(1+2\,r\right) & -r & \dots & 0 \\
\dots & \dots & \dots & \dots & \dots & \dots \\
0 & \dots & -r & \left(1+2\,r\right) & -r & 0 \\
0 & \dots & 0 & -r & \left(1+2\,r\right) & -r \\
-r & \dots & 0 & 0 & -r & \left(1+2\,r\right) \\
\end{bmatrix},
\end{equation}
which can be used to formulate a linear system of equations which is consistent with Eq.~(\ref{fd-heat-conduction}). In particular, using the operator $\hat{C}$ defined by Eq.~(\ref{conduction-matrix}), Eq.~(\ref{fd-heat-conduction}) becomes in compact form
\begin{equation}\label{matrix-heat-conduction}
    \hat{C}\,\vec{T}^+ = \vec{T},
\end{equation}
where $\vec{T}^+$ stands for $\vec{T}^{\tau+1} = \left(T_0^{\tau+1}, T_1^{\tau+1}, T_2^{\tau+1}, \dots,T_{N-1}^{\tau+1}\right)^T$ and $\vec{T}$ stands for $\vec{T}^{\tau} = \left(T_0^\tau, T_1^\tau,\right.$ $\left. T_2^\tau, \dots,T_{N-1}^\tau\right)^T$.
Clearly the inverse of operator $\hat{C}$ can be used as a time-progress operator for the temperature profile subject to heat conduction, namely
\begin{equation}\label{one-step-matrix-heat-conduction}
    \vec{T}^+ = \vec{T}(t+\Delta t) = \hat{C}^{-1}\,\vec{T},
\end{equation}
which can be also generalized by the following formula
\begin{equation}\label{multi-matrix-heat-conduction}
    \vec{T}(t+\tau\,\Delta t) = 
    (\hat{C}^{-1})^\tau\,\vec{T}.
\end{equation}
In the following section, for the sake of simplicity and without loss of generality, we will focus on Eq.~(\ref{one-step-matrix-heat-conduction}) only.

\subsection{Discrete Fourier Transform (FT)}
The one-dimensional heat conduction problem defined by Eq.~(\ref{matrix-heat-conduction}) can be solved numerically by means of the direct method given by Eq.~(\ref{one-step-matrix-heat-conduction}), which requires to invert the operator $\hat{C}$ of Eq.~(\ref{conduction-matrix}).
There are also other alternative methods for special cases, which involve some transformations. 
When discretizing the heat equation using finite differences, the resulting matrix $\hat{C}$ is diagonalizable by the Fourier transform only if the problem involves a periodic domain, leading to a circulant matrix structure, as in the present one-dimensional case. In this case, the discrete Fourier Transform (FT) efficiently diagonalizes $\hat{C}$, which seems quite natural also in the context of quantum computing \citep{nielsen_quantum_2010}. For non-periodic boundary conditions, such as Dirichlet or Neumann, $\hat{C}$ becomes a standard tridiagonal (non-circulant) matrix; here, the discrete Sine Transform (ST) and discrete Cosine Transform (CT) serve as the appropriate diagonalizing tools, matching the boundary constraints (Dirichlet for ST, Neumann for CT). In more complex cases -- such as variable coefficients, irregular domains, or non-uniform grids -- no standard transform diagonalizes $\hat{C}$, and numerical methods like eigen-decomposition or iterative solvers are typically used instead.

In the present one-dimensional example, because the mesh is regular and the domain is periodic, the discrete FT efficiently diagonalizes $\hat{C}$. In the usual mathematical notation, the discrete FT takes as input the column vector $\vec{T}$, which can be defined by a proper orthonormal basis $\vec{e}_0, \vec{e}_1, \vec{e}_2, \dots \vec{e}_{N-1}$, namely
\begin{equation}\label{T_field}
    \vec{T} = \sum_{l = 0}^{N-1} T_l\,\vec{e}_l,
\end{equation}
where $T_l$ is the nodal value for the $l$-th mesh node and $\|\vec{e}_l\|=1$. The discrete FT outputs the transformed data, a column vector of complex numbers $\vec{\tilde{T}}$ defined in the same orthonormal basis, namely
\begin{equation}\label{tildeT_field}
    \vec{\tilde{T}} = \sum_{{{m}} = 0}^{N-1} \tilde{T}_{{m}}\,\vec{e}_{{m}}.
\end{equation}
Please note that using the subscript ${{m}}$ instead of $l$ in the previous expression is unessential because the nodes are the same: it is just a matter of convention for making more evident the meaning of this sum. Each component of the transformed data is defined as\footnote{In some numerical routines, e.g. the \say{\textit{scipy.fft}} function of the SciPy platform \citep{virtanen_scipy_2020}, the standard Fourier transform is defined with regards to $e^{-i\,2\pi/N}$ and without the prefactor $1/\sqrt{N}$. It is possible to convert the results based on the standard form to those in the present document, by (i) multiplying them by $1/\sqrt{N}$ and (ii) taking the complex conjugate.} 
\begin{equation}\label{FT}
    \tilde{T}_{{m}} = \frac{1}{\sqrt{N}}\,\sum_{l = 0}^{N-1} \omega_N^{{{m}}l}\,T_l,
\end{equation}
where
\begin{equation}\label{omega}
    \omega_N = e^{i\,2\pi/N},
\end{equation}
and the parameter $i$ is the usual imaginary unit ($i=\sqrt{-1}$). Please note that the factor $1/\sqrt{N}$ in front of Eq.~(\ref{FT}) is chosen to realize a unitary transformation by construction, which allows to implement this transformation by a unitary quantum circuit \citep{nielsen_quantum_2010}. Moreover, the positive sign of the argument of the exponent of Eq.~(\ref{omega}) is quite common in the quantum community and it implies an anti-clockwise rotation in the complex plane (Argand plane). See Appendix \ref{appedix-discrete-FT-works} for details about the physical meaning of the discrete FT. It is also useful to compute the wavenumber spectrum by the transformed field $\vec{\tilde{T}}$, which describes how the variance of the temperature field is distributed over different harmonic components. In case of a classical field, the wavenumber spectrum $\vec{p}^{\;c}$ is defined as 
\begin{equation}\label{spectrum}
    \vec{p}^{\;c} = \frac{1}{N}\,\vec{\tilde{T}} \odot \vec{\tilde{T}}^*,
\end{equation}
where $\odot$ represents the Hadamard (element-wise) product and the superscript $^*$ means the complex conjugate. Another useful concept is the inverse transform, which is given by:
\begin{equation}\label{antiFT}
    T_l = \frac{1}{\sqrt{N}}\,\sum_{{{m}} = 0}^{N-1} \omega_N^{-{{m}}l}\,\tilde{T}_{{m}}.
\end{equation}
The previous definition can be interpreted as a decomposition of the original field in Fourier modes, i.e. rotations in the complex plane with wavenumber ${{m}}$ from $0$ to $N-1$.

At this point, it is possible to introduce a matrix notation, which is more convenient for solving the linear system of equations for heat conduction. Let us introduce the FT operator $\hat{U}_{FT}$, where the generic component at the ${{m}}$-th row and at the $l$-th column is given by $1/\sqrt{N}\,\omega_N^{{{m}}l}$, namely
\begin{equation}\label{FT-matrix}
    \hat{U}_{FT} := \frac{1}{\sqrt{N}}\,\begin{bmatrix}
1 & 1 & 1 & 1 & \dots & 1 \\
1 & \omega_N & \omega_N^2 & \omega_N^3 & \dots & \omega_N^{N-1} \\
1 & \omega_N^2 & \omega_N^4 & \omega_N^6 & \dots & \omega_N^{2(N-1)} \\
1 & \omega_N^3 & \omega_N^6 & \omega_N^9 & \dots & \omega_N^{3(N-1)} \\
\dots & \dots & \dots & \dots & \dots & \dots \\
1 & \omega_N^{N-1} & \omega_N^{2(N-1)} & \omega_N^{3(N-1)} & \dots & \omega_N^{(N-1)(N-1)} \\
\end{bmatrix}.
\end{equation}
In this way, the vector $\vec{\tilde{T}}$ given by Eq.~(\ref{tildeT_field}) and Fourier coefficients given by Eq.~(\ref{FT}) can be expressed as
\begin{equation}\label{FT-by-matrix}
    \vec{\tilde{T}} = \hat{U}_{FT}\,\vec{T}.
\end{equation}
It is worth to highlight that the adopted definition of the matrix $\hat{U}_{FT}$ makes it a unitary transformation, i.e. $\hat{U}_{FT}^{\dagger}\,\hat{U}_{FT} = \hat{U}_{FT}\,\hat{U}_{FT}^{\dagger} = I$, where $\hat{U}_{FT}^{\dagger}$ denotes the conjugate transpose of $\hat{U}_{FT}$, (namely Hermitian transpose). 
The latter transpose is relatively simple to be computed and it allows one to express the initial temperature profile of the heat conduction problem as
\begin{equation}\label{antiFT-by-matrix}
    \vec{T} = \hat{U}_{FT}^{\dagger}\,\vec{\tilde{T}},
\end{equation}
Introducing the previous definition into Eq.~(\ref{matrix-heat-conduction}) yields
\begin{equation}\label{FT-matrix-heat-conduction}
\hat{D}\,\vec{\tilde{T}}^+ = \vec{\tilde{T}},
\end{equation}
where $\hat{D} = \hat{U}_{FT}\,\hat{C}\,\hat{U}_{FT}^{\dagger}$ is a diagonal operator and the elements on the diagonal are the eigenvalues of the operator $\hat{C}$. In order to find these eigenvalues, let us apply the definition given by Eq.~(\ref{FT}) to the finite-difference formula given by Eq.~(\ref{line-heat-conduction}) and let us use the same nomenclature adopted in Eq.~(\ref{matrix-heat-conduction}),  namely
\begin{equation}\label{ft-line-heat-conduction}
    -r\,\mathcal{F}\left(T_{l-1}^+\right)+\left(1+2\,r\right)\,\tilde{T}_{{m}}^+
    -r\,\mathcal{F}\left(T_{l+1}^+\right) = \tilde{T_{{m}}},
\end{equation}
where $\mathcal{F}\left(\cdot\right)$ means the linear transform defined by Eq.~(\ref{FT}), i.e.
\begin{equation}\label{ft-W}
    \mathcal{F}\left(T_{l-1}^+\right) = \mathcal{F}^W_{{m}} = \frac{1}{\sqrt{N}}\,\sum_{l = 0}^{N-1} \omega_N^{{{m}}l}\,T_{(l-1)\bmod N},
\end{equation}
\begin{equation}\label{ft-E}
    \mathcal{F}\left(T_{l+1}^+\right) = \mathcal{F}^E_{{m}} = \frac{1}{\sqrt{N}}\,\sum_{l = 0}^{N-1} \omega_N^{{{m}}l}\,T_{(l+1)\bmod N},
\end{equation}
where $\bmod$ is the modulo operation, which returns the remainder of a division. 
It is important to highlight that this $\bmod$ operation is essential because the adopted labeling based on $l$ goes only from $0$ to $N-1$. Simplifying $\mathcal{F}^W_{{m}}$ yields 
\begin{equation}\label{ft-W2}
    \mathcal{F}^W_{{m}} = \frac{1}{\sqrt{N}}\,\sum_{l' = 0}^{N-1} \omega_N^{{{{m}}+{m}}l'}\,T_{l'} = \omega_N^{{m}}\,\tilde{T}_{{m}},
\end{equation}
where we set $l'=(l-1)\bmod N$ which implies $l=(l'+1)\bmod N$ {and we used the property $\omega_N^{x \bmod N}=\omega_N^{x}$ because $\omega_N^N = 1$}. Proceeding similarly for $\mathcal{F}^E_{{m}}$ and substituting these results in Eq.~(\ref{ft-line-heat-conduction}) yields
\begin{equation}\label{ft-line-heat-conduction2}
    \left[1+2\,r
    -r\,\omega_N^{{m}}
    -r\,\omega_N^{-{{m}}}\right]
    \,\tilde{T}_{{m}}^+ = \tilde{T_{{m}}}.
\end{equation}
Recalling the Euler's formula yields
\begin{equation}\label{ft-line-heat-conduction3}
    \left[1+2\,r
    -2\,r\,
    \cos{\left(\frac{2\pi\,{{m}}}{N}\right)}\right]
    \,\tilde{T}_{{m}}^+ = \tilde{T_{{m}}}.
\end{equation}
and consequently
\begin{equation}\label{ft-line-heat-conduction4}
    \left[1
    +4\,r\,
    \sin^2{\left(\frac{\pi\,{{m}}}{N}\right)}\right]
    \,\tilde{T}_{{m}}^+ = \tilde{T_{{m}}}.
\end{equation}
Comparing Eq.~(\ref{ft-line-heat-conduction4}) with Eq.~(\ref{FT-matrix-heat-conduction}), it is clear that $\hat{D}$ is a diagonal matrix with the diagonal elements equal to
\begin{equation}\label{diagonal-elements}
    \hat{D}_{{mm}} = {\lambda^C_m} = 1
    +4\,r\,
    \sin^2{\left(\frac{\pi\,{{m}}}{N}\right)},
\end{equation}
{where ${\lambda^C_m}$ is the $m$-th eigenvalue of the operator $\hat{C}$ given by Eq.~(\ref{conduction-matrix}). It is evident from the previous formula that $1\leq\lambda^C_m\leq(1+4\,r)$, as predicted by the Gershgorin circle theorem (1931), with $m=0$ and $m=N/2$ for the extreme values of the interval.} It is easy to compute the inverse matrix by replacing the main diagonal elements of the matrix $\hat{D}$ with their reciprocals, namely $\hat{R} = \hat{D}^{-1}$. The latter can be used to express the solution of the heat conduction problem as
\begin{equation}\label{one-step-matrix-heat-conduction-byFT}
    \vec{T}^+ = \hat{U}_{FT}^{\dagger}\,\hat{R}\,\hat{U}_{FT}\,\vec{T},
\end{equation}
which is an alternative route to Eq.~(\ref{one-step-matrix-heat-conduction}).

\section{Variational Quantum Eigensolver (VQE)}
\label{VQE}

Quantum computing is intimately connected with quantum information and deeply rooted in quantum physics. 
The rapid rate of progress in this field and its cross-disciplinary nature have made it difficult for newcomers to obtain a broad overview of the most important techniques and results \citep{nielsen_quantum_2010}. 
There is still a lot of work to do in hardware and software development before demonstrating any quantum computing supremacy.

Here we focus on solving a linear system of equations as a paradigmatic task for many engineering problems. 
There are many computational strategies for solving a linear system of equations by quantum computing, which are still an active field of research. Let us start with the variational quantum eigensolver (VQE) {in this section. See section \ref{HHL} for an alternative approach}. VQE is a hybrid algorithm that uses both classical operations and quantum operations to find the ground state (i.e. the stationary state of lowest energy) of a quantum system, which is designed to produce relevant information for the original problem of interest. In our case, the first step is to design a quantum system which allows one to derive the solution of the linear system of equations of interest, i.e. Eq.~(\ref{matrix-heat-conduction}). Variational quantum algorithms are promising candidates for observing quantum computation utility on noisy near-term devices. For this reason, VQE is implemented in most of the open-source software development kit, e.g. Qiskit (Quantum Information Software Kit) by IBM \citep{javadi-abhari_quantum_2024}.

\subsection{Quantum data structure}
\label{data}

\subsubsection{Two-qubit system}
\label{data:basics}

For readers who are not familiar with the basics of quantum computing, it is useful to introduce a few fundamental concepts, often explained through analogies with their classical counterparts. Readers already familiar with these basics may safely skip ahead to Section \ref{data:advanced}.

First of all, before familiarizing with the data structure of a quantum computer, let us recall some basics of binary coding, which is useful for both classical and quantum computing systems. To represent a number in binary, every digit has to be either 0 or 1 (as opposed to decimal numbers where a digit can take any value from 0 to 9). {In both cases, the mathematical notation used to write binary numbers or quantum states usually implies that the least significant bit (LSB) is written at the rightmost position. In other words, the rightmost qubit typically represents the LSB. For example, in a 4-qubit system, $0001$ in binary corresponds to the decimal value $1$, because the ``1'' is the LSB, representing $2^0$ instead of $2^3$ (if it were the most significant bit, MSB).} A decimal integer $l_{(10)}$, e.g. the already-mentioned index $l$ which identifies the mesh node, can be decomposed in terms of a binary number, which is represented by an equivalent bit string {$j_{(2)}=\beta_{n-1}\beta_{n-2}\dots\beta_1\beta_0$}, where
{
\begin{equation}\label{multibit-expansion}
    l \equiv l_{(10)} = \sum_{b=(n-1)}^{0}\beta_b\,2^b = \sum_{b=0}^{n-1}\beta_b\,2^b,
\end{equation}
}
and $\beta_b \in\{0,1\}$ is the value of the $b$-th computational unit in a binary computational system with $n$ bits. For example, the decimal number $6$, sometimes indicated as $6_{10}$ in order to emphasize the basis number equal to $10$, corresponds to the binary number $110$, or even better $6_{(10)} = 110_{(2)}$. This means that each decimal integer $d$ can be converted into an equivalent sequence of bits $\beta_b$ in the binary numeral system, i.e. {$\beta_{n-1}\dots\beta_1\beta_0$}. In the following, we will use equivalently the decimal integer $l$ or the corresponding bit string {$j_{(2)}:= j = \beta_{n-1}\dots\beta_1\beta_0$}. 

Let us move now to the data structure of a quantum computer. The building block of a quantum computer is a qubit. A qubit is a two-level quantum-mechanical system, with many states which are linear combinations (often called superpositions) of the fundamental basis states, corresponding to the states $0$ and $1$ for a classical bit. Let us indicate the quantum states by the Dirac notation $\ket{\cdot}$, which stands for the standard notation for normalized vectors in quantum mechanics. Consequently the computational basis states of each qubit, or simply the computational basis, are $\ket{0}$ and $\ket{1}$. The state vector for the generic $q$-th qubit in a system can be expressed as 
\begin{equation}\label{qubit}
    \ket{\psi_q} = \delta_q^{\ket{0}}\ket{0}+\delta_q^{\ket{1}}\ket{1},
\end{equation}
where $\delta_q^{\ket{0}}\in\mathbb{C}$ and $\delta_q^{\ket{1}}\in\mathbb{C}$ are complex numbers that represent the weight of $\ket{0}$ and $\ket{1}$ states of the superposition, and are called complex probability amplitudes. In principle, these two complex numbers may suggest that there are four degrees of freedom in each qubit, but there are also two physical constraints to be considered. First of all, if these complex numbers are presented in polar form, it is possible to realize that their global phases can be disregarded, because only the relative phase matters with regards to the expectation value of any observable \citep{nielsen_quantum_2010}. Secondly, the corresponding probabilities are normalized such that $\left|\delta_q^{\ket{0}}\right|^2+\left|\delta_q^{\ket{1}}\right|^2 = 1$\footnote{Please note that the square of a complex number may itself be complex. To obtain the amplitude probabilities, one must take the square of the modulus (absolute value) of the complex number.}. Taking into account these two constraints, the state of each qubit can be described by two angles $\varphi_q$ and $\zeta_q$, and the state vector can be expressed as:
\begin{equation}\label{qubit2}
    \ket{\psi_q} = \cos\left(\varphi_q/2\right)\ket{0}+e^{i\,\zeta_q}\sin\left(\varphi_q/2\right)\ket{1},
\end{equation}
which can be visualized by means of the so-called Bloch sphere (see Fig.~\ref{fig:bloch_sphere} and more details in Appendix \ref{hilbert-bloch})~\citep{nielsen_quantum_2010}. Comparing Eq.~(\ref{qubit}) and Eq.~(\ref{qubit2}) yields 
\begin{equation}\label{bloch}
    \delta_q^{\ket{0}} = \cos(\varphi_q/2)
    \;\;\;\text{and}\;\;\;
    \delta_q^{\ket{1}} = e^{i\,\zeta_q}\sin(\varphi_q/2) = 
    e^{i\,\zeta_q}\cos(\varphi_q/2-\pi/2).
\end{equation}
Therefore, if one considers $n$ qubits separately, i.e. isolated from each other, they could be used to store $N_{\text{sep}} = 2\,n$ real values, which are not many (usually $N_{\text{sep}}\ll N = 2^n$).
In many applications, assuming real probability amplitudes, i.e. $\zeta_q=0$, yields to a further contraction of the representable numbers, namely $N_{\text{sep}}^{\text{real}} = n$. 
\begin{figure}[htbp]
    \centering
    \includegraphics[scale=0.95]{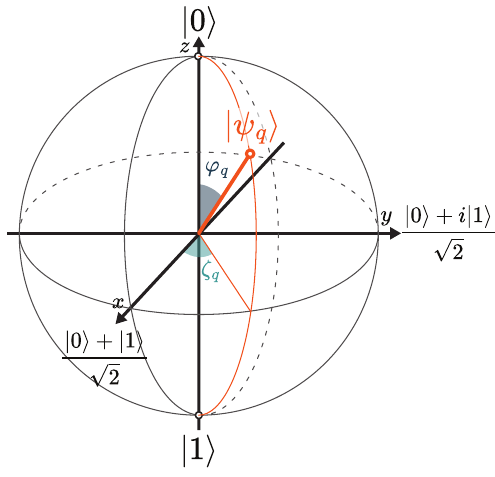}
    \caption{Bloch sphere representation of the qubit state $\ket{\psi_q}$. More details about the construction of the Bloch sphere representation of a single qubit are reported in Appendix \ref{hilbert-bloch}.
    }\label{fig:bloch_sphere}
\end{figure}

A quantum computer involves more than one qubit and therefore we need to familiarize also with the multi-qubit representation. For the sake of simplicity, let us consider first the very special case where $n$ qubits are isolated from each other, i.e. the quantum computer is in a separable quantum state. The fundamental tool for combining formally multiple qubits isolated from each other into a large single state vector is the tensor product, indicated by the $\otimes$ symbol. As an example, let us consider a composite system made of two separable qubits $\ket{\psi_0}$ and $\ket{\psi_1}$ (hence without entanglement). {Please note that, when composing physical systems, the sequential labeling of their components (e.g., $\ket{\psi_0}, \ket{\psi_1}, \dots, \ket{\psi_{n-1}}$) may differ from the mathematical notation used to represent the bit strings, i.e., $\beta_{n-1}\dots\beta_1\beta_0$.} The state vector of the composite system is expressed by $\ket{\psi_{\text{sep}}}$ and can be computed as
\begin{eqnarray}\label{two-qubits}
    \ket{\psi_{\text{sep}}} &=& 
    \ket{\psi_0} \otimes \ket{\psi_1} = \left(\delta_0^{\ket{0}}\ket{0}+\delta_0^{\ket{1}}\ket{1}\right) \otimes \left(\delta_1^{\ket{0}}\ket{0}+\delta_1^{\ket{1}}\ket{1}\right) = \nonumber\\
    &=& \delta_0^{\ket{0}}\delta_1^{\ket{0}}\ket{0}\ket{0}
    +\delta_0^{\ket{0}}\delta_1^{\ket{1}}\ket{0}\ket{1}
    +\delta_0^{\ket{1}}\delta_1^{\ket{0}}\ket{1}\ket{0}
    +\delta_0^{\ket{1}}\delta_1^{\ket{1}}\ket{1}\ket{1} = \nonumber\\
    &=& \delta_0^{\ket{0}}\delta_1^{\ket{0}}\ket{00}
    +\delta_0^{\ket{0}}\delta_1^{\ket{1}}\ket{01}
    +\delta_0^{\ket{1}}\delta_1^{\ket{0}}\ket{10}
    +\delta_0^{\ket{1}}\delta_1^{\ket{1}}\ket{11}.
\end{eqnarray}
In the previous derivation, the abbreviated notation for tensor product has been adopted, namely {$\ket{\beta_1}\otimes\ket{\beta_0}=\ket{\beta_1}\ket{\beta_0}=\ket{\beta_1\beta_0}$} (it is possible to assume that {$\ket{\beta_1}\ket{\beta_0}=\ket{\beta_1\beta_0}$} is always valid because the computational basis is always separable \citep{nielsen_quantum_2010}), where $\beta_q\in\{0,1\}$ and $\ket{\beta_q}$ is one of the computational basis states of the $q$-th qubit. Let us link Eq.~(\ref{two-qubits}) with the computational mesh. 
Let us convert the already-mentioned index $l$, which identifies a mesh node, to a binary number represented by a string {$\beta_1\beta_0$} such that
{
\begin{equation}\label{two-qubit-expansion}
    l = \beta_1\,2^1 + \beta_0\,2^0.
\end{equation}
}
Let $j$ be this string, namely {$j = \beta_1\beta_0$}, which is useful to identify both the mesh nodes by $j_{(2)} = l_{(10)}$ but also the corresponding computational basis states $\ket{j}$, namely
{
\begin{equation}\label{two-qubit-binary}
    \ket{j} = \ket{\beta_1\beta_0}.
\end{equation}
}
The corresponding complex probability amplitude of the computational basis state $\ket{j}$, in case of a composite system made of two (separable) qubits, can be expressed as
{
\begin{equation}\label{two-qubit-amplitude-sep}
    \alpha_j^{\text{sep}} = \delta_0^{\ket{\beta_1}}\delta_1^{\ket{\beta_0}},
\end{equation}
}
where $\alpha_j^{\text{sep}}\in\mathbb{C}$ in general. Therefore, the state vector of this composite separable system can be expressed as
\begin{equation}\label{two-qubits-compact-sep}
\ket{\psi_{\text{sep}}} = \ket{\psi_0}\ket{\psi_1} = 
\sum_{j_{(2)} = 00}^{11} \alpha_{j}^{\text{sep}}\ket{j} =
\alpha_{00}^{\text{sep}}\ket{00}+
\alpha_{01}^{\text{sep}}\ket{01}+
\alpha_{10}^{\text{sep}}\ket{10}+
\alpha_{11}^{\text{sep}}\ket{11},
\end{equation}
where the same abbreviated notation of the tensor product has been used also for the state vector of the composite system. A column vector representation is also useful for understanding how the tensor product works, namely
\begin{equation}\label{two-qubits-columns}
\begin{pmatrix}
    \delta_0^{\ket{0}} \\
    \delta_0^{\ket{1}}
\end{pmatrix}
\otimes
\begin{pmatrix}
    \delta_1^{\ket{0}} \\
    \delta_1^{\ket{1}}
\end{pmatrix}
=
\begin{pmatrix}
    \delta_0^{\ket{0}}\delta_1^{\ket{0}} \\
    \delta_0^{\ket{0}}\delta_1^{\ket{1}} \\
    \delta_0^{\ket{1}}\delta_1^{\ket{0}} \\
    \delta_0^{\ket{1}}\delta_1^{\ket{1}}
\end{pmatrix}
=
\begin{pmatrix}
    \alpha_{00}^{\text{sep}} \\
    \alpha_{01}^{\text{sep}} \\
    \alpha_{10}^{\text{sep}} \\
    \alpha_{11}^{\text{sep}}
\end{pmatrix}.
\end{equation}
The tensor product between two vectors produces a larger vector, and it should not be confused with the dyadic product in fluid-dynamics, which would lead to a second order tensor instead\footnote{The tensor product $\otimes$ of two vectors $\ket{\psi_A}$ and $\ket{\psi_B}$ creates a new vector $\ket{\psi_A}\otimes\ket{\psi_B}$ in a space with dimension $d_A \times d_B$, which is larger than the individual spaces having dimensions $d_A$ and $d_B$, respectively. On the other hand, the dyadic product (sometimes indicated by the same symbol in fluid-dynamics), produces a matrix (operator), not a vector, and it is indicated here by $\ket{\psi_A}\bra{\psi_B}$.}.
It is worth noting that the number of elements in the set $\{\alpha_{00}^{\text{sep}},\alpha_{01}^{\text{sep}},\alpha_{10}^{\text{sep}},\alpha_{11}^{\text{sep}}\}$ grows exponentially with $n$, but all these terms depend on amplitudes $\delta_0^{\ket{0}}$, $\delta_0^{\ket{1}}$, $\delta_1^{\ket{0}}$ and $\delta_1^{\ket{1}}$: there are $2\,n$ such terms, meaning their number grows linearly with $n$.
Therefore, in separable states, the representable numbers grow only linearly with $n$.

The tensor product can be used to represent only separable states, i.e. states without entanglement. 
{\it Entanglement} is a fundamental property of quantum mechanics where the quantum state of a system composed of multiple subsystems cannot be described as a simple product of the states of its individual parts \citep{nielsen_quantum_2010}. 
On the contrary, the system exists in a superposition of correlated states, such that the measurement of one subsystem instantaneously affects the state of the other, no matter how far apart they are \citep{nielsen_quantum_2010}. 
An intuitive example of entanglement for non-experts is reported in Appendix \ref{strategy}. 
In order to understand that entanglement allows one to represent (many more) new correlated states, which are unreachable by Eq.~(\ref{two-qubits-columns}), let us consider the following example based on only two qubits. 
Let us limit ourselves to only real probability amplitudes, i.e. $\zeta_q = 0$ for $q\in\{0,1\}$.
A well-known entangled state in the computational basis is the Bell state (see also Appendix \ref{dilemma} for an intuitive example), which is a maximally entangled state:
\begin{equation}\label{entangled}
\begin{pmatrix}
    \alpha_{00} \\
    \alpha_{01} \\
    \alpha_{10} \\
    \alpha_{11}
\end{pmatrix} = 
\begin{pmatrix}
    1/\sqrt{2} \\
    0 \\
    0 \\
    1/\sqrt{2}
\end{pmatrix}.
\end{equation}
This state cannot be factorized into a product of two individual qubit states, i.e., recalling that we assumed $\zeta_0 = \zeta_1 = 0$ for simplicity, there are no $\varphi_0$ and $\varphi_1$ in Eq.~(\ref{bloch}) which can be combined in Eq.~(\ref{two-qubits-columns}) to represent the previous entangled (correlated) state. 
This is clear because $\alpha_{00}$ and $\alpha_{11}$ require non-zero {$\delta_0^{\ket{\beta_1}}$} and {$\delta_1^{\ket{\beta_0}}$}, but this is contradictory with the conditions derived by $\alpha_{01}$ and $\alpha_{10}$. The Bell state can be included instead, by generalizing Eq.~(\ref{two-qubits-compact-sep}) for the unknown vector state
\begin{equation}\label{two-qubits-compact}
\ket{\psi} = \sum_{j = 00}^{11} \alpha_{j}\ket{j} =
\alpha_{00}\ket{00}+
\alpha_{01}\ket{01}+
\alpha_{10}\ket{10}+
\alpha_{11}\ket{11},
\end{equation}
where now the generic $\alpha_j\in\{\alpha_{00},\alpha_{01},\alpha_{10},\alpha_{11}\}$ can be any complex number, only fulfilling the normalization condition $\sum_j |\alpha_j|^2 = 1$.
The previous generalization is useful to realize that the number of elements in the set $\{\alpha_{00},\alpha_{01},\alpha_{10},\alpha_{11}\}$ grows exponentially with $n$, and these terms are now independent from each other: Therefore, in generic states, the representable numbers grow exponentially with $n$.
\begin{figure}[htbp]
    \centering
    \includegraphics[scale=0.95]{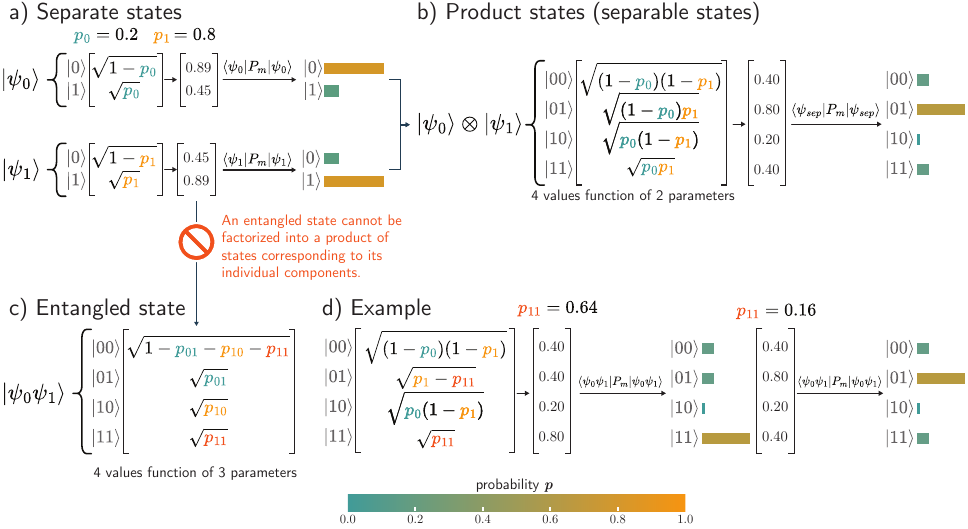}
    \caption{
        Schematic of a composite system of two qubits. (a) Separate states, (b) tensor product of the two individual states, (c) entangled state, and (d) an example where two marginal probabilities are fixed ($p_{00}$, $p_{10} $) and consistent with the previous case, while varying $p_{11} $. The measurement probabilities for each basis state are indicated using colored bars and computed using the projector $ P_m = \ket{m}\bra{m} $, where $ \ket{m} \in \{\ket{0}, \ket{1}\} $ for single-qubit states or $ \ket{m} \in \{\ket{00}, \ket{01}, \ket{10}, \ket{11}\} $ for the two-qubit case.
    }\label{fig:entanglement}    
\end{figure}
This exponential increase of the number of computational basis states can be understood by introducing some correlation among qubits due to entanglement. The fundamental difference between a separable two-qubit system (i.e. a system obtained by the tensor product of two individual states) and an entangled two-qubit system is depicted in Fig. \ref{fig:entanglement}. This example clearly shows that entanglement enables the exploration of a much broader space of states compared to what is achievable using only separable states. 
Hence, the entanglement is the key to exploit fully the tremendous potential of all computational basis states. It is the entanglement or -- with other words -- the presence of correlated states, which makes the microscopic scenario discussed here  substantially different from other microscopic theories, e.g. the kinetic theory of gases. In the latter theory, the assumption of molecular chaos (also known as the \textit{Stosszahlansatz}), is a key assumption in the derivation of the Boltzmann equation for dilute gases. The assumption of molecular chaos states that before a collision occurs, the velocities of two colliding particles are uncorrelated.

\subsubsection{Multiple-qubit system}
\label{data:advanced}

After becoming more familiar with the quantum basics, let us come back to a more realistic quantum computer. 
A quantum computer with $n$ qubit has typically a much larger computational basis than the previous example. 
In the following, the argument above will be generalized to any number $n$ of qubits. 
An $n$ qubit system has $N = 2^n$ computational basis states, analogously to a classical system, but, unlike classical bits that exist in one state at a time, qubits can also exist in superpositions of all these states, according to some probabilities, which can also be correlated and hence can be freely explored, thanks to the entanglement.
Therefore, a quantum computer can store (and process) an amount of real numbers which is at least equal to the number of computational basis states (see next). 
In principle, the total number of real degrees of freedom in a general $n$-qubit quantum state is $2^{n+1}-2$, accounting for normalization and the fact that the absolute phase of the state has no physical significance.
If we impose the condition that the amplitudes are real, the normalization constraint still applies, but the global phase is restricted to $\pm 1$, which does not introduce a continuous degree of freedom. 
As a result, the number of available degrees of freedom reduces to $2^n-1$.
In the following, we will sometimes simplify this expression by stating that a quantum computer can store (and process) at least $2^n$ real numbers, leveraging the imaginary parts of certain complex amplitudes to compensate for the missing degree of freedom.
This proves that, in practice, we have significant control over operating with $2^n$ real numbers by choosing appropriate gate sets or adding auxiliary qubits. Moreover, if we purposely restrict our gate set to only real-valued operations, the resulting quantum state will remain in the space of real amplitudes.

Again, for the sake of simplicity, let us first consider the discrete system state vector representing $n$ separable qubits (without entanglement), namely  
\begin{equation}\label{state-vector}
    \ket{\psi_{\text{sep}}} = \ket{\psi_0}\otimes\ket{\psi_1}\otimes\dots\ket{\psi_{n-1}} = 
    \ket{\psi_0}\ket{\psi_1}\dots\ket{\psi_{n-1}},
\end{equation}
where the last expression is an abbreviated notation for the tensor product among individual qubits \citep{nielsen_quantum_2010}.
More precisely, the system quantum state involves associating a complex coefficient $\alpha_j^{\text{sep}}\in \mathbb{C}$ (called an amplitude) with each computational basis state $\ket{j}$. 
The amplitude for separable states is actually a composite amplitude which comes from multiplying the qubit individual amplitudes $\delta_q^{\ket{\beta_q}}$ with each other, namely
{
\begin{equation}\label{multiqubit-amplitude}
    \alpha_j^{\text{sep}} = \delta_0^{\ket{\beta_{n-1}}}\dots\delta_1^{\ket{\beta_1}}\delta_{n-1}^{\ket{\beta_0}} = \prod_{q=0} ^{n-1}\delta_q^{\ket{\beta_q}},
\end{equation}
}
which generalizes Eq.~(\ref{two-qubit-amplitude-sep}). 
Using the relations given by Eqs. (\ref{bloch}) yields
\begin{equation}\label{multiqubit-amplitude-sep-bloch}
    \alpha_j^{\text{sep}} = \prod_{q=0}^{n-1}\delta_q^{\ket{\beta_q}} =
    \prod_{q=0}^{n-1} e^{i\,\zeta_q\,\beta_q}\cos\left[\varphi_q/2-\beta_q\,(\pi/2)\right],
\end{equation}
where we took advantage of the property $\cos\left(\varphi_q/2-\pi/2\right) = \sin\left(\varphi_q/2\right)$ and $\beta_q\in\{0,1\}$ as usual, depending on the considered $q$-th qubit.
Again, let us convert the already-mentioned index $l$, which identifies a mesh node, to a binary number represented by a string {$j = \beta_{n-1}\dots\beta_1\beta_0$} by the following formula: 
{
\begin{equation}\label{multiqubit-expansion}
    l = \sum_{q=0}^{n-1}\beta_q\,2^q,
\end{equation}
}
which ensures that $j_{(2)} = l_{(10)}$. 
The computational basis state corresponding to $j$ is indicated by $\ket{j}$, which generalizes the binary representation given by Eq.~(\ref{two-qubit-binary}), defined as
{
\begin{equation}\label{multiqubit-binary}
    \ket{j} = \ket{\beta_{n-1}\dots\beta_1\beta_0},
\end{equation}
}
where $\ket{j}\in\{\ket{0},\ket{1}\}^{\otimes n}$, which means that $\ket{j}$ can be $\ket{00\dots0}$, $\ket{00\dots1}$, $\dots$, $\ket{11\dots1}$ (the computational basis is always separable). Clearly there are $N=2^n$ elements in the set $\{\ket{0},\ket{1}\}^{\otimes n}$. As done in the previous example, it is possible to exploit the full capability of the previous set by including also correlated states by entanglement, which are many more (by far!), in the formula for the state vector of the system in order to ensure a general validity now, namely
\begin{equation}\label{multiqubit-compact}
\ket{\psi} = \ket{\psi_0\,\psi_1\,\dots\,\psi_{n-1}} = 
\sum_{{j\in\{0,\,1\}^n}} \alpha_j\ket{j}, 
\end{equation}
which generalizes Eq.~(\ref{two-qubits-compact}). Please note that, in general, $\ket{\psi_0\,\psi_1\,\dots\,\psi_{n-1}}\neq\ket{\psi_0}\ket{\psi_1}\dots\ket{\psi_{n-1}}$, namely the state vector of the composite system is usually not separable. The probability $p_j$ of finding the system state in the computational basis state $\ket{j}$ is given by $p_j = |\alpha_j|^2$, where $\sum_j p_j = \sum_j |\alpha_j|^2 = 1$. In other words, one can say that the discrete vector quantum state $\ket{\psi}$ is normalized, namely $\braket{\psi|\psi}=1$ where $\braket{\cdot|\cdot}$ denotes the inner product. 

For the sake of simplicity, let us suppose to construct a quantum circuit such that the output state is characterized by real amplitudes $\alpha_j = a_j \in \mathbb{R}$. In the output of such quantum circuit, the non-trivial (real) amplitudes $a_j$ are $2^n$ real numbers (where typically $2^n\gg n$), which can be used to store a huge number of relevant information for the problem of interest. On the other hand, in a classical computer, real numbers are typically stored as floating-point approximations rather than exact values (e.g. according to the IEEE 754 floating-point standard). In particular, $n$ classical bits can store a fixed-precision binary representation of a real number, which is equivalent to store one coded state among all available computational states (which are $2^n$). If one wants to express the same concept by using a classical probability distribution, it would be like the classical probability distribution is equal to one only for the coded state and zero otherwise (Dirac delta distribution). Hence the difference between a quantum computer and a classical computer is that we can load $2^n$ real numbers in the corresponding amplitudes of a quantum circuit thanks to superposition (non-trivial probability distribution of states), while we can code only one discrete state at the time in a classical computer among all possible states (Dirac delta distribution of states). 
The (potentially) tremendous advantage is clear ($2^n\gg \left\lfloor n_{\text{classic}} / 64 \right\rfloor$) and it can be represented in a non-rigorous way by the following expression (which is meaningful at the current stage of development of the quantum technology)
\begin{equation}\label{non-rigorous}
N_{\text{sep}} \ll N = 2^n. 
\end{equation}

Next, we need to understand how a quantum circuit can be used to perform the desired calculations. There are five main steps:
\begin{enumerate}
    \item Identification of the quantum state (Normalization);
    \item Design of the quantum system (Observable); 
    \item Selection of the quantum parametrized trial solution (Quantum ansatz);
    \item Minimization of the loss function (Optimization);
    \item Extraction of useful results (De-normalization).
\end{enumerate}

\subsection{Normalization}\label{normalization}
The first step is to identify the quantum state where to store the relevant information by normalizing Eq.~(\ref{matrix-heat-conduction}). Because a discrete quantum state $\ket{\psi}$ is normalized, namely $\braket{\psi|\psi}=1$, let us divide Eq.~(\ref{matrix-heat-conduction}) by a factor such that it becomes possible to identify a quantum state which depends on the temperature profile. In particular, let us choose this factor as follows:
\begin{equation}\label{unitary-matrix-heat-conduction}
    \frac{1}{(\vec{T}^+\cdot\vec{T}^+)(\vec{T}\cdot\vec{T})}\,\hat{C}\,\vec{T}^+ = 
    \frac{1}{(\vec{T}^+\cdot\vec{T}^+)(\vec{T}\cdot\vec{T})}\,\vec{T},
\end{equation}
where $\vec{T} \cdot \vec{T}$ denotes the inner product (i.e., $\vec{T}^\dagger \cdot \vec{T}$, and since $\vec{T}$ is real-valued, $\vec{T}^\dagger$ reduces to the transpose of $\vec{T}$); similarly for $\vec{T}^+ \cdot \vec{T}^+$. Let us define the quantum state $\ket{b}$ for mapping the initial temperature profile, namely
\begin{equation}\label{b}
    \ket{b} = \frac{1}{\sqrt{\vec{T}\cdot\vec{T}}}\,\vec{T},
\end{equation}
where, by construction, $\braket{b|b} = 1$. Before proceeding, let us be sure to appreciate the true meaning of the previous relation. Essentially it implies that each node $z_l$ of the computational mesh is associated with a quantum computational basis state $\ket{j}$ (where $\ket{j}\in\{0,\,1\}^{\otimes n}$ involves the binary representation of integer $l$). Similarly let us proceed with the quantum state $\ket{x}$ for mapping the updated temperature profile at the new time step, namely
\begin{equation}\label{x}
    \ket{x} = \frac{1}{\sqrt{\vec{T}^+\cdot\vec{T}^+}}\,\vec{T}^+,
\end{equation}
where the same normalization holds. Introducing the definitions given by Eq.~(\ref{b}) and Eq.~(\ref{x}) into Eq.~(\ref{unitary-matrix-heat-conduction}) yields
\begin{equation}\label{norm-matrix-heat-conduction}
    \sqrt{\frac{\vec{T}^+\cdot\vec{T}^+}{\vec{T}\cdot\vec{T}}}\,\hat{C}\,\ket{x} = 
    \ket{b}.
\end{equation}
It is also possible to define a normalization factor $f$ given by 
\begin{equation}\label{f}
    f := \sqrt{\frac{\vec{T}^+\cdot\vec{T}^+}{\vec{T}\cdot\vec{T}}},
\end{equation}
and to derive the linear system of target equations as
\begin{equation}\label{quantum-system}
    \hat{A}\ket{x} = 
    \ket{b},
\end{equation}
where $\hat{A} = f\,\hat{C}$. The problem is that the normalization factor is not known at the beginning of the numerical procedure because it depends on the solution $\vec{T}^+$, which derives from solving the linear system of equations. 
This means that $\hat{A}$ can be used to discuss the theoretical setup, but the practical numerical procedure must involve $\hat{C}$, because the latter depends only on the adopted FD formula.

\subsection{Observable}\label{step2}

The second step is to derive a quantum system, which provides information relevant to solve the target problem. Let us follow the strategy proposed in Ref. \citep{xu_variational_2021} to construct a Hamiltonian, which admits the quantum state $\ket{x}$ as the ground state. 
Although heat conduction is intrinsically a dissipative process and thus not Hamiltonian in the strict sense, we introduce a `Hamiltonian' here to refer to the matrix arising from the finite-difference discretization, which structurally resembles a Hamiltonian operator. This formalism facilitates the following analysis (see section \ref{philosophical} for more details).
Applying the methodology described in Ref. \citep{xu_variational_2021} to Eq.~(\ref{quantum-system}) yields the following Hamiltonian
\begin{equation}\label{Hp}
    \hat{H}' = 
    \hat{A}^{\dagger}\left(I-\ket{b}\bra{b}\right)
    \hat{A},
\end{equation}
where $\ket{\cdot}\bra{\cdot}$ denotes the outer product and $\hat{A}^{\dagger}$ is the conjugate transpose of $\hat{A}$ in general, while, in this case, $\hat{A}^{\dagger} = \hat{A}^T$, because $\hat{A}$ is real. In this case, $\hat{A}$ is also symmetric and therefore $\hat{A}^T = \hat{A}$. As pointed out before, let us formulate the quantum algorithm in terms of the practical operator $\hat{C}$ which does not involve the normalization factor. Since $\hat{A}^T=\hat{A} = f\,\hat{C}$, the previous Hamiltonian can be computed as $\hat{H}' = f^2\,\hat{H}$ where
\begin{equation}\label{observable}
    \hat{H} := \hat{O} = \hat{C}^T\left(I-\ket{b}\bra{b}\right)\hat{C}.
\end{equation}
In the previous formula, we implicitly remind that the Hamiltonian is just a special case of quantum observable (where energy is the actual observed quantity): therefore, it makes sense to use instead the symbol $\hat{O}$ from now on for making the procedure as universal as possible. Let us decompose this operator as $\hat{O} = \hat{C}^T\,\hat{M}\,\hat{C}$, where $\hat{M} = I-\ket{b}\bra{b}$ is a projector operator. It is possible to prove that $\hat{M}$ is a projector operator because $\hat{M}^2 = I-2 \ket{b}\bra{b} + \ket{b}\bra{b} = \hat{M}$. It projects onto the orthogonal complement of $\ket{b}$. That is, it removes the component of a vector along $\ket{b}$, leaving only the part orthogonal to $\ket{b}$.
First of all, let us analyze the eigenvalues of $\hat{M}$, by calling ${\lambda^M_m}$ the ${{m}}$-th eigenvalue and $\ket{\phi_{{m}}^{{M}}}$ the corresponding eigenstate, namely $\hat{M}\ket{\phi_{{m}}^{{M}}} = {\lambda^M_m}\ket{\phi_{{m}}^{{M}}}$. The ${{m}}$-th eigenvalue can be computed as
\begin{equation}\label{eigen-observable-core}
    {\lambda^M_m} = \braket{\phi_{{m}}^{{M}}|\hat{M}|\phi_{{m}}^{{M}}} = 
    1 - \braket{\phi_{{m}}^{{M}}|b}^2.
\end{equation}
Recalling the Cauchy–Schwarz inequality, namely
\begin{equation}\label{cauchy–schwarz}
    \braket{\phi_{{m}}^{{M}}|b} \leq
    \sqrt{\braket{\phi_{{m}}^{{M}}|\phi_{{m}}^{{M}}}}
    \sqrt{\braket{b|b}} = 1,
\end{equation}
it is possible to find out that
\begin{equation}\label{eigen-ocservable-core2}
    {\lambda^M_m} = 1 - \braket{\phi_{{m}}^{{M}}|b}^2 
    \geq 0.
\end{equation}
This means that all eigenvalues of $\hat{M}$ are larger than or equal to zero, i.e. the core operator $\hat{M}$ is positive semi-definite. Using the same orthonormal basis, it is possible to express $\hat{M}$ by spectral decomposition
\begin{equation}\label{observable-core}
    \hat{M} = 
    \sum_{{{m}}=0}^{N-1} {\lambda^M_m} \ket{\phi_{{m}}^{{M}}}\bra{\phi_{{m}}^{{M}}}.
\end{equation}
Now, coming back to the main observable $\hat{O}$, let us consider the expectation value of the observable $\hat{O}$ with regards to the generic state vector $\ket{\psi}$, namely
\begin{equation}\label{observable2}
    \bra{\psi}\hat{O}\ket{\psi} = \bra{\psi}\hat{C}^T\left(\sum_{{{m}} = 0}^{N-1} {\lambda^M_m}\,
    \ket{\phi_{{m}}^{{M}}}\bra{\phi_{{m}}^{{M}}}\right)\hat{C}\ket{\psi} 
    = \sum_{{{m}} = 0}^{N-1} {\lambda^M_m} 
    \braket{\phi_{{m}}^{{M}}|\hat{C}|\psi}^2,
\end{equation}
where we used $\hat{C}^T=\hat{C}$. Using the result given by Eq.~(\ref{eigen-ocservable-core2}) yields
\begin{equation}\label{observable3}
    \bra{\psi}\hat{O}\ket{\psi} \geq 0,
\end{equation}
which proves that the main observable $\hat{O}$ is also positive semi-definite, i.e. all eigenvalues of $\hat{O}$ are positive or equal to zero. Clearly $\ket{x}$ is the eigenvector of $\hat{O}$ corresponding to the zero eigenvalue, namely
\begin{equation}\label{minimum}
    \hat{O}\ket{x} = \hat{O}\,\hat{A}^{-1}\ket{b}=
    f^{-1}\,\hat{C}^T\,\left(I-\ket{b}\bra{b}\right) 
    \ket{b} = 0,
\end{equation}
which proves that $\ket{x}$ is the ground state of the operator $\hat{O}$, as expected by design. The linear algebra task given by Eq.~(\ref{quantum-system}) is converted to the task of finding the ground state of the Hamiltonian $\hat{H} := \hat{O}$ \citep{xu_variational_2021}.

\subsection{Quantum circuit ansatz}\label{step3}

The third step is the quantum circuit ansatz, i.e. a parameterized trial solution which should be able to approximate the ground state $\ket{x}$. The key point is that, in order to prove the quantum supremacy, there are too many elements in the vector $\ket{x}$ to work on them directly by a classical computer. Let us recall that VQE is a hybrid algorithm, where the optimization is supposed to be done by a classical computer \citep{nielsen_quantum_2010}. Hence let us introduce a vector of parameters $\vec{\theta}$, which are fewer such that they can be handled by a classical computer. The quantum circuit ansatz enforces a parameterized state $\ket{x(\vec{\theta}\,)}$ which makes possible to map these parameters on a generic quantum state. The parameterized state is the output of a unitary transformation (quantum gate), namely 
\begin{equation}\label{ansatz}
    \ket{x(\vec{\theta}\,)} = U(\vec{\theta}\,)\,\ket{0}^{\otimes\,n},
\end{equation}
where $\ket{0}^{\otimes\,n}$ stands for $\ket{0}\otimes\ket{0}\otimes\dots\ket{0} = \ket{0}\ket{0}\dots\ket{0} = \ket{00\dots0}$ (the computational basis is always separable) and $\vec{\theta}$ is a vector of tunable parameters. The parameters $\vec{\theta}$ are typically generic `rotations' of qubits which are optimized during the minimization step of the VQE algorithm (see next section \ref{step4}). 
In order to preserve the quantum advantage, the number of parameters to optimize over in the ansatz circuit must be much less than the size of the computational basis of the quantum states, because the former is handled by a classical optimizer/minimizer, while the latter exploits the full capability of the quantum computer. With other words, the number of parameters must grow as a polynomial in the number $n$ of qubits, while the size of the full vector $\ket{x}$ is exponential in the number of qubits \citep{javadi-abhari_quantum_2024}. It is not important which polynomial describes the growth of the number of parameters, because any polynomial cannot compete with the growth of the exponential function $2^n$ for large $n$. In the following sections, for example, we will see  $8\,n$ parameters in the ansatz depicted in Fig. (\ref{ansatz-3qubits}) and $4\,n$ parameters in the simplified ansatz depicted in Fig. (\ref{ansatz-3qubits-qrisp}). 

\subsection{Optimization}\label{step4}

The fourth step is the actual minimization of the loss function. A loss function quantifies the difference (``loss”) between a quantum state predicted by the ansatz for a given input and the ground state. Taking into account Eq.~(\ref{observable3}) and the ansatz given by Eq.~(\ref{ansatz}), the loss function can be defined as 
\begin{equation}\label{loss}
    L(\vec{\theta}\,) := \bra{x(\vec{\theta}\,)}\hat{O}\ket{x(\vec{\theta}\,)} \geq 0.
\end{equation}
The optimal set of parameters can be formally defined by the argument of the minimization problem with regards to the loss function, namely
\begin{equation}\label{vqe}
    \vec{\theta}_\text{min} := 
    \arg\underset{\vec{\theta}}{\min}\,\,
    L(\vec{\theta}\,) = 
    \arg\underset{\vec{\theta}}{\min}\, \bra{x(\vec{\theta}\,)}\hat{O}\ket{x(\vec{\theta}\,)}.
\end{equation}
Let us define $\ket{x_\text{min}}$ as
\begin{equation}\label{x_min}
    \ket{x_\text{min}} = \ket{x(\vec{\theta}_\text{min}\,)}.
\end{equation}
Because of numerical errors, $\ket{x_\text{min}}$ is different from the theoretical ground state, i.e. $\ket{x_\text{min}} \neq \ket{x}$, but it is usually close enough. 

\subsection{De-normalization}

The final step is to de-normalize the numerical quantum approximation $\ket{x_\text{min}}$ for coming back to the original quantity of interest, i.e. the temperature. Let us start with the initial temperature profile. Let us define the auxiliary quantity
\begin{equation}\label{theta}
    \theta := \sqrt{\vec{T}\cdot\vec{T}},
\end{equation}
which can be used to express Eq.~(\ref{b}) as $\vec{T} = \theta\ket{b}$. Let us compute this auxiliary quantity $\theta$ by the spatial average of the initial temperature profile, namely
\begin{equation}\label{scale2}
    \theta = 
    \frac{\sum_l T_l} 
    {\sum_j \braket{j|b}}.
\end{equation}
where $\sum_j \braket{j|b}$ is the sum of all real amplitudes in $\ket{b}$. Please remember that $\ket{j}\in\{0,\,1\}^{\otimes n}$ involves the binary representation of integer $l$. Similarly, recalling Eq.~(\ref{x}), let us define
\begin{equation}\label{theta+}
    \theta^+ := \sqrt{\vec{T}^+\cdot\vec{T}^+}.
\end{equation}
which can now be computed by using the quantum numerical approximation $\ket{x_\text{min}}$. Taking advantage of the energy conservation, which implies 
\begin{equation}\label{energy-conservation}
    \sum_l T_l^+ = \sum_l T_l,
\end{equation}
the quantity $\theta^+$ can be computed by the following formula
\begin{equation}\label{scale}
    \theta^+ = 
    \frac{\sum_l T_l}
    {\sum_j \braket{j|x_\text{min}}}.
\end{equation}
It is worth to note that, in case of accurate minimization, all terms in the summation at the denominator in Eq.~(\ref{scale}) are positive because they correspond to the nodal values of the normalized new temperatures. 
Clearly $f=\theta^+/\theta$. The quantity $\theta^+$ is essential to de-normalize the quantum solution and to come back to the updated temperature profile, namely
\begin{equation}\label{result}
    \vec{T}^+ = \theta^+\ket{x_\text{min}}.
\end{equation}

So far we presented the straightforward implementation of the VQE approach which demonstrates a fundamental
possibility to solve linear algebra problems, and in
particular the discretized conduction equation on a quantum processor. However, it has one essential disadvantage: this algorithm requires to decompose the observable $\hat{O}$ in terms of a sequence of Pauli matrices (see next section for details). Usually the number of Pauli matrices in this decomposition is exponential in the number of qubits and hence it spoils the potential quantum speedup \citep{guseynov_depth_2023}. More sophisticated variational methods have been already proposed in the literature, which are more promising from a practical point of view \citep{guseynov_depth_2023}. One possibility consists in evaluating the loss function by an adaption of a fundamental quantum circuit, the so-called Hadamard test \citep{ingelmann_two_2024}. An even more effective implementation consists in combining the Hadamard test with the quantum Fourier transform \citep{guseynov_depth_2023}. A more advanced approach for near-term algorithms, namely algorithms suitable for near-term quantum hardware, is represented by the so-called ansatz tree \citep{huang_near-term_2021}, which has been already applied to the discretized conduction equation \citep{guseynov_depth_2023}. These techniques will be explored and compared in a future work. 

\subsection{Practical details of implementation}

In this section, we need to complete the algorithm presented in previous section by adding more details about the actual implementation of the algorithm in a quantum computer. Even though these details are general, we will focus on Qiskit \citep{javadi-abhari_quantum_2024} by IBM as an example open-source software development kit. Qiskit \citep{javadi-abhari_quantum_2024} is an open-source framework for quantum computing that allows users to design, simulate, and run quantum programs on real hardware. It provides an intuitive way to build quantum circuits, optimize them for execution, and simulate their behavior before running on actual quantum processors. Qiskit also includes tools for error mitigation and circuit optimization, making it more practical for real-world use. 

\subsubsection {Decomposition in Pauli matrices}
\label{step2qiskit}

In the presented algorithm, the loss function to be minimized $L(\vec{\theta})$ is defined by the expectation value of the observable $\hat{O}$ defined by Eq.~(\ref{observable}). To measure the observable $\hat{O}$ given by Eq.~(\ref{observable}) on a quantum computer by Qiskit \citep{javadi-abhari_quantum_2024}, one must represent it as a sum of tensor products of Pauli matrices, that is
\begin{equation}\label{Pauli-decomposition}
    \hat{H} \equiv \hat{O} = \sum_{p = 0}^{N_p-1}
    \gamma_p\,\hat{P}_p,
\end{equation}
where $N_p$ is the number of terms in the Pauli decomposition of the Hamiltonian, $\gamma_p\in \mathbb{R}$ because $\hat{O} = \hat{O}^\dagger$ is Hermitian (actually it is real and symmetric in this case), $\hat{P}_p\in\{I,X,Y,Z\}^{\otimes n}$, and the Pauli matrices are
\begin{equation}\label{Pauli-operators}
    {I} := 
    \begin{pmatrix}
        1 & 0 \\
        0 & 1 
    \end{pmatrix},\;\;
    {X} := 
    \begin{pmatrix}
        0 & 1 \\
        1 & 0 
    \end{pmatrix},\;\;
    {Y} := 
    \begin{pmatrix}
        0 & -i \\
        i & 0 
    \end{pmatrix},\;\;
    {Z} := 
    \begin{pmatrix}
        1 & 0 \\
        0 & -1 
    \end{pmatrix}.
\end{equation}
We will clarify soon the physical meaning of this decomposition in terms of Pauli matrices, but it is important to first understand the tensor product between matrices. As an example, let us consider again a composite system made of two qubits. In this case, the generic $p$-th element of the decomposition looks like
\begin{eqnarray}\label{two-qubit-Pauli-element}
\hat{P}_p &=& \sigma^{p0}\otimes\sigma^{p1}=
\begin{bmatrix}
\sigma^{p0}_{11}\,\sigma^{p1} & \sigma^{p0}_{12}\,\sigma^{p1} \\
\sigma^{p0}_{21}\,\sigma^{p1} & \sigma^{p0}_{22}\,\sigma^{p1}
\end{bmatrix}=\nonumber\\
&=& \begin{bmatrix}
\sigma^{p0}_{11}\,
\begin{pmatrix}
\sigma^{p1}_{11} & \sigma^{p1}_{12} \\
\sigma^{p1}_{21} & \sigma^{p1}_{22}
\end{pmatrix}
&
\sigma^{p0}_{12}\,
\begin{pmatrix}
\sigma^{p1}_{11} & \sigma^{p1}_{12} \\
\sigma^{p1}_{21} & \sigma^{p1}_{22}
\end{pmatrix}
\\
\sigma^{p0}_{21}\,
\begin{pmatrix}
\sigma^{p1}_{11} & \sigma^{p1}_{12} \\
\sigma^{p1}_{21} & \sigma^{p1}_{22}
\end{pmatrix}
&
\sigma^{p0}_{22}\,
\begin{pmatrix}
\sigma^{p1}_{11} & \sigma^{p1}_{12} \\
\sigma^{p1}_{21} & \sigma^{p1}_{22}
\end{pmatrix}
\end{bmatrix} = \nonumber\\
&=& \begin{bmatrix}
\sigma^{p0}_{11}\,\sigma^{p1}_{11} & \sigma^{p0}_{11}\,\sigma^{p1}_{12} & 
\sigma^{p0}_{12}\,\sigma^{p1}_{11} & \sigma^{p0}_{12}\,\sigma^{p1}_{12} \\
\sigma^{p0}_{11}\,\sigma^{p1}_{21} & \sigma^{p0}_{11}\,\sigma^{p1}_{22} & 
\sigma^{p0}_{12}\,\sigma^{p1}_{21} & \sigma^{p0}_{12}\,\sigma^{p1}_{22} \\
\sigma^{p0}_{21}\,\sigma^{p1}_{11} & \sigma^{p0}_{21}\,\sigma^{p1}_{12} & 
\sigma^{p0}_{22}\,\sigma^{p1}_{11} & \sigma^{p0}_{22}\,\sigma^{p1}_{12} \\
\sigma^{p0}_{21}\,\sigma^{p1}_{21} & \sigma^{p0}_{21}\,\sigma^{p1}_{22} & 
\sigma^{p0}_{22}\,\sigma^{p1}_{21} & \sigma^{p0}_{22}\,\sigma^{p1}_{22}
\end{bmatrix},
\end{eqnarray}
where $\sigma^{p0}$ and $\sigma^{p1}$ are matrices which can be $I$, $X$, $Y$ or $Z$ and they refer to the first and the second qubit, respectively. 
After becoming more familiar with this nomenclature, let us come back to the case with $n$ qubits, which can become pretty complicated, namely
\begin{equation}\label{Pauli-element}
\hat{P}_p = \sigma^{p0}\otimes\sigma^{p1}\otimes\dots\sigma^{p(n-1)},
\end{equation}
where the detailed expressions are omitted for the sake of simplicity. Fortunately, the tensor product among matrices has a fundamental property \citep{nielsen_quantum_2010}, which only applies to separable states but can help in understanding this decomposition. Let us suppose to apply the $p$-th element of the decomposition $\hat{P}_p$ to a separable vector state $\ket{\psi_{\text{sep}}} = \ket{\psi_0}\otimes\ket{\psi_1}\otimes\dots\ket{\psi_{n-1}}$, which yields
\begin{eqnarray}\label{property}
\hat{P}_p\ket{\psi_{\text{sep}}} &=&
\left[\sigma^{p0}\otimes\sigma^{p1}\otimes\dots\sigma^{p(n-1)}\right]
\left(\ket{\psi_0}\otimes\ket{\psi_1}\otimes\dots\ket{\psi_{n-1}}\right) = \nonumber\\
&=& \sigma^{p0}\ket{\psi_0}\otimes
\sigma^{p1}\ket{\psi_1}\otimes\dots
\sigma^{p(n-1)}\ket{\psi_{n-1}}.
\end{eqnarray}
The previous formula means that the result of $\hat{P}_p\ket{\psi_{\text{sep}}}$ is simply the tensor product of the individual calculation $\sigma^{pq}\ket{\psi_q}$ for all the qubits. This implies
\begin{eqnarray}\label{property2}
\bra{\psi_{\text{sep}}}\hat{P}_p\ket{\psi_{\text{sep}}} =
\prod_{q=0}^{n-1}
\bra{\psi_q}\sigma^{pq}\ket{\psi_q}.
\end{eqnarray}
and consequently 
\begin{equation}\label{Pauli-decomposition2}
    \bra{\psi_{\text{sep}}}\hat{O}\ket{\psi_{\text{sep}}} = \sum_{p = 0}^{N_p-1}
    \gamma_p\,\prod_{q=0}^{n-1}
\bra{\psi_q}\sigma^{pq}\ket{\psi_q},
\end{equation}
which means that the expectation value of the observable $\hat{O}$ with regards to separable states can be computed by a sequence of measurements on one qubit at a time (but it is crucial to change the measurement basis for the $q$-th qubit corresponding the $\sigma^{pq}$ matrix). In case of non separable states, i.e. in case of entanglement, the previous simplification does not hold. 

Even though the previous formula is a special case, it allows one to appreciate that there is a computational problem at this point, namely $N_p$ grows pretty fast with $n$ (exponentially, see next). Let us do an example. In case of a system with $3$ qubits, the decomposition in Pauli matrices given by Eq.~(\ref{Pauli-decomposition}) is the following
\begin{eqnarray}\label{three-qubit-Pauli-decomposition}
    \hat{O} &=&\gamma_0 III + \gamma_1 IIX + \gamma_2 IXI +\gamma_3 IXX + \gamma_4 IXZ + \gamma_5 IYY + \nonumber\\
    &&\gamma_6 IZI + \gamma_7 IZX + \gamma_8 IZZ +
    \gamma_9 XII + \gamma_{10} XIX + \gamma_{11} XXI +\nonumber\\
    &&\gamma_{12} XXX + \gamma_{13} XXZ + \gamma_{14} XYY + \gamma_{15} XZI + \gamma_{16} XZX +\gamma_{17} XZZ + \nonumber\\
    &&\gamma_{18} YIY + \gamma_{19} YXY + \gamma_{20} YYI + \gamma_{21} YYX + \gamma_{22} YYZ + \gamma_{23} YZY + \nonumber\\
    &&\gamma_{24} ZII + \gamma_{25} ZIX + \gamma_{26} ZIZ + \gamma_{27} ZXI + \gamma_{28} ZXX + \gamma_{29} ZXZ + \nonumber\\
    &&\gamma_{30} ZYY + \gamma_{31} ZZI + \gamma_{32} ZZX + \gamma_{33} ZZZ,
\end{eqnarray}
where, for example, $III$ means $I \otimes I \otimes I$ and similarly for all remaining terms. In this example, for $n=3$ the Pauli decomposition requires $N_p=34$. For the sake of comparison, for $n=4$ the Pauli decomposition requires $N_p = 120 \sim \exp(k_p\,n)$, where $k_p$ is a proper constant. Therefore, the present application appears to be a case where the number of Pauli products in the Hamiltonian decomposition grows exponentially with the number of qubits, as suggested in \citep{guseynov_depth_2023}. However, the reference does not explicitly analyze this scaling, and as noted, leveraging the distribution of Pauli string weights can potentially reduce the complexity. While a brute-force or naive approach would indeed be impractical, the literature suggests alternative methods -- such as truncation, grouping of Pauli strings, and other techniques -- that might be applicable in this context. It remains unclear whether these approaches could be effectively employed in this specific case. More details about the Pauli decomposition can be found in Ref. \citep{hantzko_tensorized_2024}.

\subsubsection{Efficient circuit ansatz}
\label{step3qiskit}

In the presented algorithm, the loss function to be minimized $L(\vec{\theta})$ is defined with regards to a parameterized trial solution $\ket{x(\vec{\theta}\,)}$, which is called the ansatz. The ansatz given by Eq.~(\ref{ansatz}) is a parameterized trial solution $\ket{x(\vec{\theta}\,)}$ which should be able to approximate the ground state $\ket{x}$. The parametrized solution is the output of a unitary transformation (quantum gate) $U(\vec{\theta}\,)$, which depends on a vector $\vec{\theta}$ of $N_\theta$ tunable parameters. Naively one would like to have a procedure for correlating the parameters in $\vec{\theta}$ with the real amplitudes in $\ket{x(\vec{\theta}\,)}$ by means of some analytical formulas. This approach is usually called real data loading or better encoding, and some algorithms have been proposed in literature \citep{araujo_divide-and-conquer_2021}. For optimization problems -- and for VQE in particular -- real data loading/encoding is not strictly necessary and it will be omitted here in favor of a more efficient approach, namely an approach with less tunable parameters (see also Appendix \ref{appedix-real-data-loading}).

In Qiskit \citep{javadi-abhari_quantum_2024}, let us consider a unitary transformation $U(\vec{\theta}\,)$ which consists of two ingredients: (i) four layers of single-qubit operations, (ii) spanned by controlled $NOT$ gates (also called controlled-$X$ gates) for ensuring some degree of entanglement \citep{nielsen_quantum_2010}. This is a heuristic pattern that can be used to prepare trial states for variational quantum algorithms or classification circuit for machine learning \citep{javadi-abhari_quantum_2024}. The single-qubit operations consist of the sequential application $R_YR_Z$ (in this case, there is no tensor product implied because both apply to the same qubit) of a $R_Y$ gate and a $R_Z$ gate, defined as
\begin{equation}\label{RY}
    R_Y(\theta_Y) := \exp{\left(-i\,\frac{\theta_Y}{2}\,Y\right)} = 
    \begin{pmatrix}
        \cos{(\theta_Y/2)} & -\sin{(\theta_Y/2)} \\
        \sin{(\theta_Y/2)} & \cos{(\theta_Y/2)} 
    \end{pmatrix},
\end{equation}
and
\begin{equation}\label{RZ}
    R_Z(\theta_Z) := \exp{\left(-i\,\frac{\theta_Z}{2}\,Z\right)} = 
    \begin{pmatrix}
        \exp{(-i\,\theta_Z/2)} & 0 \\
        0 & \exp{(i\,\theta_Z/2)} 
    \end{pmatrix}.
\end{equation}
The previous definitions can be thought as derived from the same generic formula
\begin{equation}\label{lie}
    \exp{\left(-i\,\theta_\sigma\,\sigma\right)} = 
    \cos{\theta_\sigma}\,I - 
    i\,\sin{\theta_\sigma}\,\sigma,
\end{equation}
where $\theta_\sigma$ is a parameter and $\sigma$ is a matrix which can be the Pauli matrix $Y$ or $Z$. The previous generic formula derives from the property of Pauli matrices (after multiplication by $i=\sqrt{-1}$ to make them anti-Hermitian) to generate transformations in the sense of Lie algebras \citep{nielsen_quantum_2010}. This formula is analogous for Pauli matrices to the Euler's formula of complex analysis. 

\begin{figure}
  \includegraphics[width=1.0\linewidth]{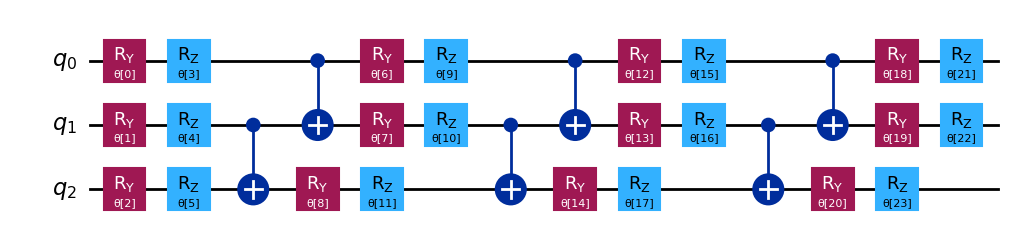}
  \caption{Efficient ansatz (3 qubit, 24 parameters = four layers with six parameters each or, equivalently, eight parameters per qubit). For clarity, horizontal lines represent quantum wires which correspond to qubits in the circuit, red squares are the $R_Y$ gates (see Eq.~(\ref{RY})), blue squares are the $R_Z$ gates (see Eq.~(\ref{RZ})), blue dots represent control points in controlled gates, $\oplus$ symbol is used for a controlled-$X$ ($CNOT$) gate. The latter gate explicitly guarantees the desired entanglement.}
  \label{ansatz-3qubits}
\end{figure}

This ansatz is called \say{\textit{EfficientSU2}} circuit in Qiskit \citep{javadi-abhari_quantum_2024} and it is plotted in Fig. (\ref{ansatz-3qubits}) for a system with $3$ qubits. {We have already discussed the basics of coding binary numbers. When dealing with computer memory, however, the endianness must be specified — that is, how bits or qubits are stored in memory. In this paper, the little-endian convention is used \citep{nielsen_quantum_2010}, meaning that the LSB (the bit representing the smallest place value, $2^0$) is stored at the lowest memory address. Consequently, in circuit diagrams, as the one reported in Fig. (\ref{ansatz-3qubits}), the topmost qubit represents the LSB and the bottom qubit represents the MSB.} For a system with $n=3$ qubits, $N_\theta = 24$ because there are four layers with six parameters each (two gates $R_Y$ and $R_Z$ for each qubit) or, equivalently, eight parameters per qubit. For the sake of comparison, for $n=4$ the number of parameters in this ansatz becomes $N_\theta = 8\,n = 32$. It is essential that the number of ansatz parameters to optimize over is linear as in this case (or polynomial at worst) in the number of qubits, in order to ensure a potential quantum supremacy (because $N_\theta \sim k_\theta\,n \ll 2^n=N$, where $k_\theta$ is a proper constant). 

\subsubsection{Minimization of the loss function}
\label{step4qiskit}

VQE is a hybrid algorithm that combines (i) classical operations for the converging iterations and (ii) loss function evaluations by quantum operations, to find the ground state of the target quantum system, which is designed in our case to update the one-dimensional temperature profile consistently with the heat conduction equation. 

For the converging iterations by classical operations, one can use the \say{\textit{minimize}} function of the \say{\textit{scipy.optimize}} library in the SciPy platform \citep{virtanen_scipy_2020}. It is recommended to focus on Jacobian-free methods: for example, the \say{\textit{COBYLA}} solver and \say{\textit{L-BFGS-B}} solver, which produce similar performance according to our preliminary experiments. The goal is to minimize the number of evaluations of loss functions, which requires limiting the tolerance for termination in the range $1$--$10\times10^{-3}$. 

\subsection{Simulated results}

\subsubsection{Simulated results by Qiskit}

For quantum computers in \href{https://en.wikipedia.org/wiki/Noisy_intermediate-scale_quantum_era}{the NISQ era}, the discussed algorithm for real applications is still very challenging, mainly because of qubit decoherence. For this reason, in order to perform some preliminary experiments, let us use the \say{\textit{BaseEstimatorV2}} simulator available in Qiskit \citep{javadi-abhari_quantum_2024}, which estimates expectation values for provided quantum circuit and observable combinations. An example implementation of the VQE in Qiskit is reported in Appendix \ref{example-codes}.

\begin{figure}[htbp]
    \centering
    \includegraphics[width=0.7\linewidth]{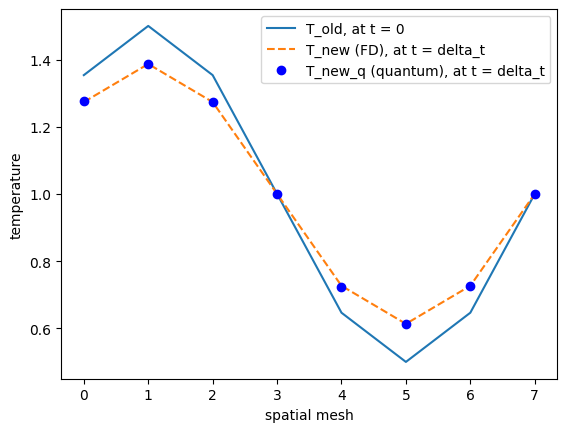}
    \caption{One time-step update of the temperature profile according to heat conduction equation by quantum computing ($3$ qubits, \textit{BaseEstimatorV2} quantum simulator, \textit{COBYLA} classical minimizer with tolerance for termination $1\times10^{-3}$). The blue line is the initial temperature profile (with mean equal to $1$), the orange dashed line is the new temperature profile at time $\Delta t$, computed by finite-difference method. The blue dots are the mesh node temperatures computed by the quantum simulator.}
    \label{conduction-3qubits}
\end{figure}

\begin{figure}[htbp]
    \centering
    \includegraphics[width=.7\linewidth]{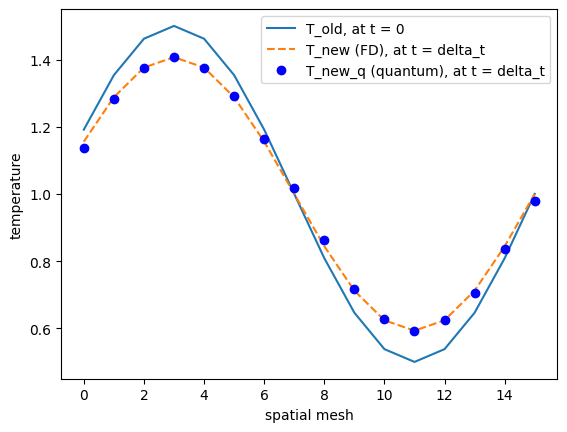}
    \caption{One time-step update of the temperature profile according to heat conduction equation by quantum computing ($4$ qubits, \textit{BaseEstimatorV2} quantum simulator, \textit{COBYLA} classical minimizer with tolerance for termination $1\times10^{-3}$). The blue line is the initial temperature profile (with mean equal to $1$), the orange dashed line is the new temperature profile at time $\Delta t$, computed by finite-difference method. The blue dots are the mesh node temperatures computed by the quantum simulator.}
    \label{conduction-4qubits}
\end{figure}

Let us compute the outcome of applying the time-progress operator $\hat{C}^{-1}$ given by Eq.~(\ref{one-step-matrix-heat-conduction}) to an initial temperature profile. With other words, let us perform one time step to update the temperature profile of our target problem. In Fig. (\ref{conduction-3qubits}), the results for one time-step update of the temperature profile according to the heat conduction equation are reported in case of $n=3$ qubits ($N=8$), \textit{BaseEstimatorV2} quantum simulator and \textit{COBYLA} classical minimizer with tolerance for termination $1\times10^{-3}$). This minimization required $839$ evaluations of the loss function, which are still too many for most existing quantum computers to compete with classical computers. Similarly, in Fig. (\ref{conduction-4qubits}), the results for the same problem in case of $n=4$ qubits ($N=16$) are reported (the quantum simulator and the classical minimizer are the same as before). In this second case, even though the results look acceptable, we performed $10^6$ evaluations of the loss function, hitting the upper maximum limit which we set in advance. 

\subsubsection{Simulated results by Qrisp}

While the physics side of quantum computing makes significant progress, the support for high-level quantum programming abstractions is still in its infancy compared to modern classical languages and frameworks \citep{seidel_qrisp_2022}. An interesting example is provided by Qrisp, which is a high-level programming language developed by Fraunhofer for creating and compiling quantum algorithms \citep{seidel_qrisp_2022}. Its structured programming model enables scalable development and maintenance \citep{seidel_qrisp_2022}. An example implementation of the VQE in Qrisp is reported in Appendix \ref{example-codes}.

\begin{figure}[htbp]
  \includegraphics[width=1.0\linewidth]{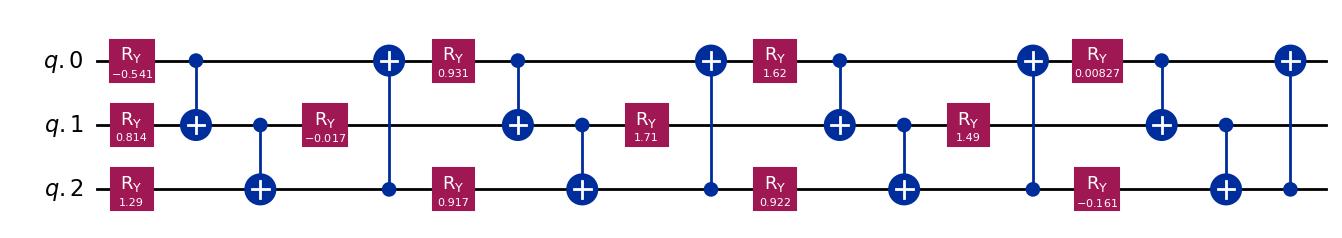}
  \caption{Simplified ansatz (3 qubit, 12 parameters = four layers with three parameters each or, equivalently, four parameters per qubit). For clarity, horizontal lines represent quantum wires which correspond to qubits in the circuit, red squares are the $R_Y$ gates (see Eq.~(\ref{RY})), blue dots represent control points in controlled gates, $\oplus$ symbol is used for a controlled-$X$ ($CNOT$) gate.}
  \label{ansatz-3qubits-qrisp}
\end{figure}

\begin{figure}[htbp]
    \centering
\begin{subfigure}{.48\textwidth}
  \centering
  \includegraphics[width=1.0\linewidth]{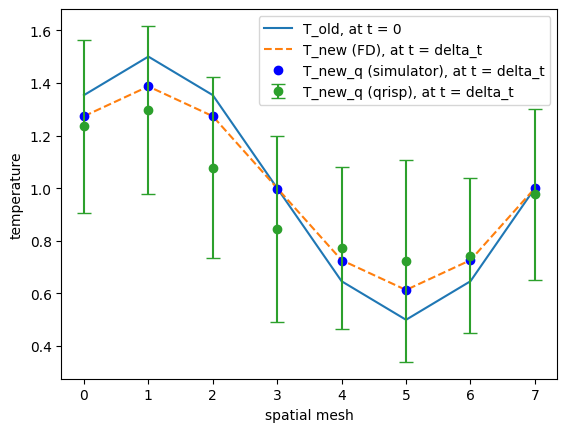}
  \caption{Parameter \textit{`precision': 0.025}}
  \label{repetitions:sfig1}
\end{subfigure}
\begin{subfigure}{.48\textwidth}
  \centering
  \includegraphics[width=1.0\linewidth]{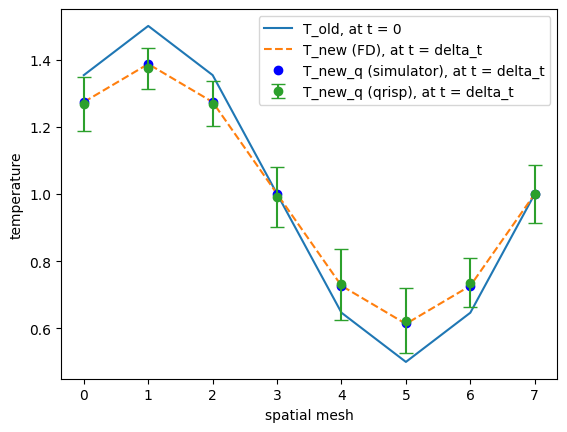}
  \caption{Parameter \textit{`precision': 0.001}}
  \label{repetitions:sfig2}
\end{subfigure}
\caption{Simulated results obtained by Qrisp example code for implementing the VQE (see Appendix \ref{example-codes}). In both cases, $30$ repetitions are considered for statistics (mean and standard deviation), but with \textit{`precision': 0.025} and \textit{`precision': 0.001} of the performed measurements, respectively. In the first case, the uncertainty is larger than the temperature differences due to the heat conduction time step.}
\label{repetitions-qrisp}
\end{figure}

For simplicity, the Qrisp example code employs the simplified ansatz shown in Fig. \ref{ansatz-3qubits-qrisp}, which consists solely of $R_Y$ and $CNOT$ gates. Among the relevant parameters for the \textit{vqe.run} method, used to compute the system's energy in the Qrisp example code (see Appendix \ref{example-codes}), the most important is `\textit{precision}', as it determines the number of shots during execution on real hardware. In quantum computing, `shots' refer to the number of times a quantum circuit is executed to collect measurement statistics. Since quantum measurements are probabilistic, multiple shots are required to estimate expectation values with sufficient accuracy. Hence precision refers to how accurately the Hamiltonian is evaluated. The number of shots the real quantum hardware performs per iteration scales quadratically with the inverse precision. Therefore it is important to estimate properly the required precision in order to assess the feasibility of running a VQE algorithm on real quantum hardware.

In Fig. \ref{repetitions-qrisp} the impact of parameter `\textit{precision}' on simulated results is investigated by 30 repetitions of the VQE algorithm for collecting some relevant statistics (mean and standard deviation) of the performed measurements. In particular, \textit{`precision': 0.025} and \textit{`precision': 0.001} are considered. In the first case, the uncertainty is larger than the temperature differences due to the heat conduction time step, making the simulation practically useless. This proves that actual precision, or equivalently the maximum number of shots, are limiting factors for successfully implementing VQE algorithms on real quantum hardware.

\subsection{VQE based on diagonalizing the measurement}

As already mentioned, the key for modeling an irreversible phenomenon by an ideal quantum computer is to properly design the measurement. Unfortunately, the naive VQE approach discussed in the previous section presents a challenge: in the general case, the observable $\hat{O}$ may require an expansion involving an exponential number of Pauli matrices \citep{guseynov_depth_2023}. See the expansion given by Eq.~(\ref{three-qubit-Pauli-decomposition}) for $n=3$ and the discussion afterwards. For this reason, we discuss here a better approach based on diagonalizing the measurement \citep{guseynov_depth_2023}. The key idea is to simplify the observable by transferring relevant information about the problem in the preparation of the state by a proper circuit. In order to do so, we need to introduce first the Quantum Fourier Transform, which is a unitary transformation by construction.

\subsubsection{Quantum Fourier Transform (QFT)}

Before deriving the QFT, let us generalize our nomenclature about the normalization needed to pass from a temperature vector and the corresponding quantum state. A generic quantum state $\ket{\psi}$ is related to the corresponding temperature vector by a proper scaling factor. For example, according to Eq.~(\ref{b}) and Eq.~(\ref{theta}), the following scaling holds $\ket{b} = (1/\theta)\,\vec{T}$, which means that state $\ket{b}$ is obtained by normalizing $\vec{T}$ by the scaling factor $\theta$. Similarly, $\ket{x} = (1/\theta^+)\,\vec{T}^+$, where it is important to highlight that $\theta^+\neq \theta$. More specifically, the scaling factor $N/\theta^2$ changes during the simulation. In general, let us define the linear mapping as $\ket{\psi} = (1/\theta_\psi)\,\vec{T}$. This generalized mapping will be used in the rest of this section. 

The Fourier transformation is defined in this document in such a way so as to realize a unitary transformation by construction. Consequently $\hat{U}_{FT}$ is a unitary matrix, which can be automatically implemented by means of a unitary quantum circuit \citep{nielsen_quantum_2010}. For the sake of simplicity, let us use in the following $\hat{U}_{QFT}$, where $\hat{U}_{QFT}:= \hat{U}_{FT}$, {with $\hat{U}_{FT}$ given by Eq.~(\ref{FT-matrix})}. Because the FT is a linear transformation, $\ket{\tilde{\psi}} = \hat{U}_{QFT}\,\ket{\psi} = (1/\theta_\psi)\,\vec{\tilde{T}}$ holds too. Consequently, 
\begin{equation}\label{QFT-by-matrix}
    \ket{\tilde{\psi}} = \hat{U}_{QFT}\,\ket{\psi},
\end{equation}
where, for example, $\ket{\psi}$ can be $\ket{b}$ or $\ket{x}$. The quantum states $\ket{\psi}$ and $\ket{\tilde{\psi}}$ are defined as
\begin{equation}\label{b-basis}
    \ket{\psi} = \sum_{{j\in\{0,\,1\}^n}} \psi_j\,\ket{j},
\end{equation}
\begin{equation}\label{tildeb-basis}
    \ket{\tilde{\psi}} = \sum_{{j\in\{0,\,1\}^n}} \tilde{\psi}_j\,\ket{j},
\end{equation}
where the computational basis is the same, as it happens also for the classical case given by Eq.~(\ref{T_field}) and Eq.~(\ref{tildeT_field}), {and hence we used the same binary index $j$}. Taking into account Eq.~(\ref{FT}), the quantum FT of the generic ${\tilde{j}}$-th amplitude of the transformed state is given by
\begin{equation}\label{QFT}
    \tilde{{\psi}}_{{\tilde{j}}} = \frac{1}{\sqrt{N}}\,\sum_{{j\in\{0,\,1\}^n}} \omega_N^{{\tilde{j}}j}\,{\psi}_j.
\end{equation}
The problem arises with Eq.~(\ref{spectrum}) because the wavenumber spectrum is not linear with regards to the transformed field. This means that the wavenumber spectrum of a quantum state can not be a quantum state. Therefore, let us define the vector $\vec{p}$ as the wavenumber spectrum, namely 
\begin{equation}\label{spectrum-qft}
\vec{p}_\psi := \operatorname{Diag}(\ket{\tilde{\psi}}\bra{\tilde{\psi}}),
\end{equation}
where $\operatorname{Diag}(\cdot)$ is the diagonal extraction operator and $\ket{\tilde{\psi}}\bra{\tilde{\psi}}$ is the density matrix, which -- in case of pure states -- represents the classical probability distribution over measurement outcomes in the standard basis. It is possible to prove that the wavenumber spectrum defined by Eq.~(\ref{spectrum-qft}) is automatically normalized, namely
\begin{equation}\label{spectrum-qft-sum}
\sum_{{\tilde{j}\in\{0,\,1\}^n}}
p_{{\tilde{j}}} = \braket{\tilde{\psi}|\tilde{\psi}} = 
\bra{\tilde{\psi}}\hat{U}_{QFT}^{\dagger}\,\hat{U}_{QFT}\ket{\tilde{\psi}} = \braket{\psi|\psi} = 1.
\end{equation}
By comparing Eq.~(\ref{spectrum-qft}) with Eq.~(\ref{spectrum}), it is easy to prove that $\vec{p}_\psi = \ket{\tilde{\psi}} \odot \ket{\tilde{\psi}^*} = (N/\theta_\psi^2)\,\vec{p}^{\;c}$, where the superscript $^*$ means the complex conjugate. The scaling factor $N/\theta^2$ changes during the simulation. For example, for the temperature profile reported in Appendix \ref{appedix-discrete-FT-works}, $N/\theta^2 = 8/9$ initially, but it tends to unity when the solution approaches the steady-state temperature profile.

\begin{figure}
\begin{subfigure}{\textwidth}
  \centering
  \includegraphics[width=1.0\linewidth]{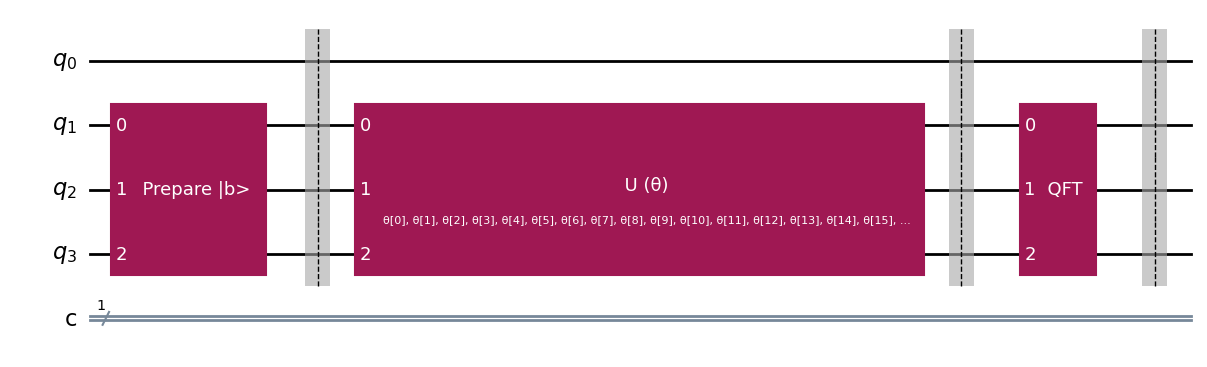}
  \caption{Circuit, which can be described as $\hat{U}_{QFT}\,U\,U_b$, for preparing the quantum state $\ket{\tilde{x}}$ (with $n$ qubits), which is used by the measurement protocol $\mathcal{M}_{\hat{D}^2}$ given by Eq.~(\ref{measurement-first}).}
  \label{fig-diagonal-measure-U-3qubits}
\end{subfigure}
\begin{subfigure}{\textwidth}
  \centering
  \includegraphics[width=1.0\linewidth]{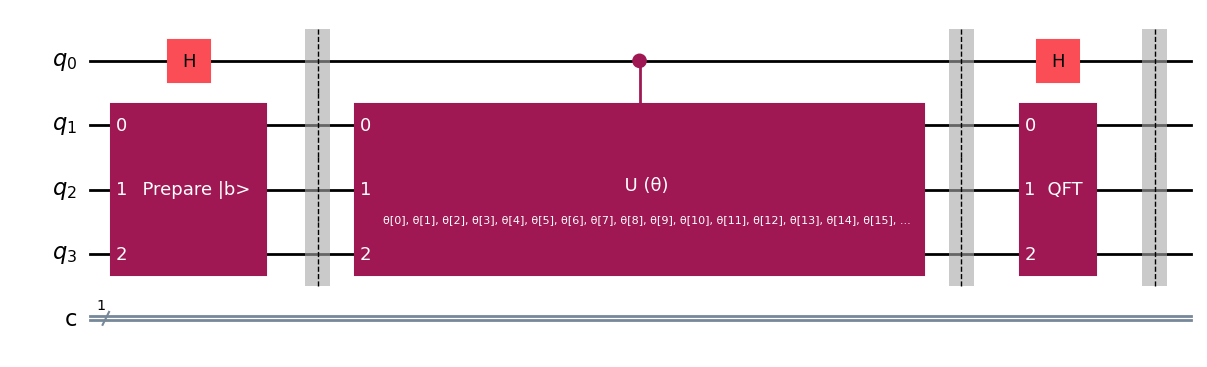}
  \caption{Circuit, which can be described as $U_{HI}$, for preparing the quantum state $\ket{\xi}$ (with $n+1$ qubits), which is used by the measurement protocol $\mathcal{M}_{\hat{D},\,\operatorname{Re}}$ given by Eq.~(\ref{measurement-second-Re}).}
  \label{fig-diagonal-measure-UHI-3+aqubits}
\end{subfigure}
\begin{subfigure}{\textwidth}
  \centering
  \includegraphics[width=1.0\linewidth]{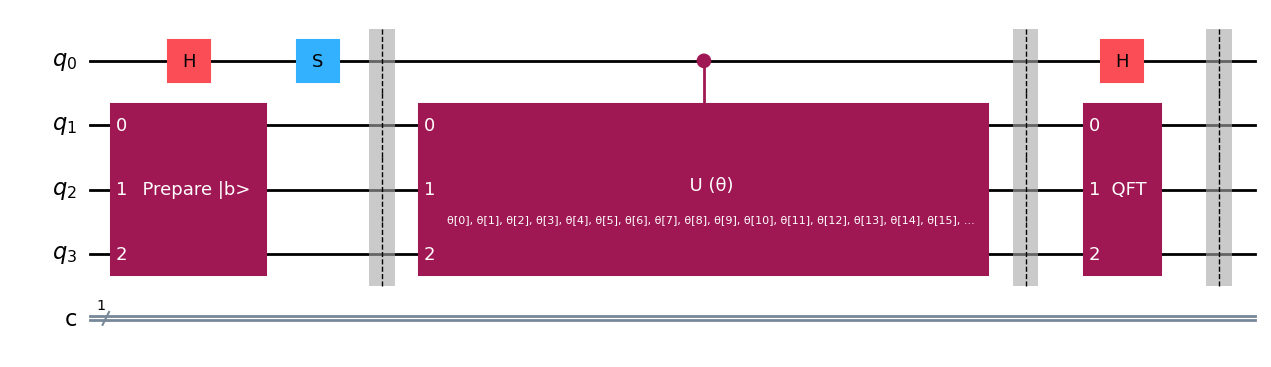}
  \caption{Circuit, which can be described as $U_{HS}$, for preparing the quantum state $\ket{\xi'}$ (with $n+1$ qubits), which is used by the measurement protocol $\mathcal{M}_{\hat{D},\,\operatorname{Im}}$ given by Eq.~(\ref{measurement-second-Im}).}
  \label{fig-diagonal-measure-UHS-3+aqubits}
\end{subfigure}
\caption{Quantum circuits used by the measurement protocols $\mathcal{M}_{\hat{D}^2}$, $\mathcal{M}_{\hat{D},\,\operatorname{Re}}$ and $\mathcal{M}_{\hat{D},\,\operatorname{Im}}$, needed for computing the loss function according to Eq.~(\ref{loss4}). Note that unitary transformations are graphically represented from left to right, but they apply in the reverse order in the computational formulas \citep{nielsen_quantum_2010}.}
\label{fig-xi}
\end{figure}

\subsubsection{Hadamard test approach}

Before proceeding with the methodology proposed in Ref. \citep{guseynov_depth_2023}, let us first clarify how the loss function given by Eq.~(\ref{loss}) is actually computed by a quantum algorithm. First of all, the observable $\hat{O}$ is represented as a sum of $N_p$ tensor products of Pauli matrices, as reported in Eq.~(\ref{Pauli-decomposition}). Secondly, Eq.~(\ref{ansatz}) means that the parametrized solution $\ket{x(\vec{\theta}\,)}$ is actually computed by means of a unitary transformation $U(\vec{\theta}\,)$, coded by the selected ansatz. Let us highlight these implementation features in the definition of the loss function given by Eq.~(\ref{loss}), namely
\begin{equation}\label{loss2}
    L(\vec{\theta}\,) = \bra{x(\vec{\theta}\,)}\hat{O}\ket{x(\vec{\theta}\,)}
    = \mathcal{M}\left(\hat{O}, U(\vec{\theta}\,)\right),
\end{equation}
where $\mathcal{M}(\hat{O},U)$ is the measurement protocol for estimating the expectation value of the observable $\hat{O}$ by means of the circuit $U$. Eq.~(\ref{loss2}) means that the naive implementation, discussed in the previous section, consists in computing the loss function by a direct measurement protocol. Unfortunately, the latter requires large $N_p$, i.e. too many tensor products to represent $\hat{O}$.

On the other hand, the key idea here is to use the Quantum Fourier Transform (QFT) to simplify the observable $\hat{O}$ by encoding the relevant information about the problem into some state preparations via appropriate quantum circuits. This approach avoids the issue of exponential growth in the number of Pauli matrices required for the decomposition of the observable. Let us diagonalize the loss function by recalling the definition of observable given by Eq.~(\ref{observable}) and by using the definition of QFT given by Eq.~(\ref{QFT-by-matrix}), namely,
\begin{eqnarray}\label{loss3}
    L(\vec{\theta}\,) &=& \bra{x(\vec{\theta}\,)}\hat{C}^T\left(I-\ket{b}\bra{b}\right)\hat{C}\ket{x(\vec{\theta}\,)} \nonumber \\ &=& \bra{x}\hat{C}^T\hat{C}\ket{x}-\bra{x}\hat{C}^T\ket{b}\bra{b}\hat{C}\ket{x}\nonumber\\
    &=& \bra{\tilde{x}}\hat{D}^2\ket{\tilde{x}}-\bra{x}\hat{C}^T\ket{b}\left(\bra{x}\hat{C}^T\ket{b}\right)^* \nonumber \\ &=&
    \bra{\tilde{x}}\hat{D}^2\ket{\tilde{x}}-\left|\bra{\tilde{x}}\hat{D}\ket{\tilde{b}}\right|^2
\end{eqnarray}
where the dependence of $\ket{x(\vec{\theta}\,)}$ on $\vec{\theta}$ was dropped for the sake of simplicity and $\hat{D} = \hat{U}_{QFT}\,\hat{C}\,\hat{U}_{QFT}^{\dagger}$, which means that QFT diagonalizes the conduction operator $\hat{C}$, as already discussed in the previous sections. In deriving the last formula, the following property was used: $\hat{U}_{QFT}\,\hat{C}^T\,\hat{U}_{QFT}^\dagger = \hat{D}^T = \hat{D}$. Moreover, it is worth to recall that $\bra{b} \hat{C} \ket{x}$ is a complex number and that $\bra{b} \hat{C} \ket{x} = (\bra{x} \hat{C}^T \ket{b})^*$, where $(\cdot)^*$ is the complex conjugate. The QFT clearly simplifies the observables which are now $\hat{D}$ and $\hat{D}^2$, but it introduces the challenge of efficiently preparing non-trivial states necessary for measuring the terms involved in computing the loss function \citep{guseynov_depth_2023}. 

Let us identify the measurement protocols for estimating both terms in Eq.~(\ref{loss3}). Let us start with the first term $\bra{\tilde{x}}\hat{D}^2\ket{\tilde{x}}$. Since $\hat{D}^2$ has real matrix elements and is Hermitian (as it arises from a symmetry), its expectation value is necessarily real \citep{nielsen_quantum_2010}. We need to use the selected ansatz in a way which is different from what was done before. We use it to pass from state $\ket{b}$ to state $\ket{x}$ for some parameter vector $\vec{\theta}$, namely 
\begin{equation}\label{ansatz-new}
    \ket{x(\vec{\theta}\,)} = U(\vec{\theta}\,)\,\ket{b}.
\end{equation}
The previous formula may appear incompatible with Eq.~(\ref{ansatz}). Actually there is no contradiction between Eq.~(\ref{ansatz}) and Eq.~(\ref{ansatz-new}) because the ansatz is just a sort-of numerical spline, which can realize different transformations by different parameter vectors $\vec{\theta}$. Consequently, another transformation must be used to prepare the state $\ket{b}$, namely $\ket{b} = U_b\,\ket{0}^{\otimes\,n}$, which leads to
\begin{equation}\label{ansatz-new2}
    \ket{x(\vec{\theta}\,)} = U(\vec{\theta}\,)\,\ket{b} =
    U(\vec{\theta}\,)\,U_b\,\ket{0}^{\otimes\,n}.
\end{equation}
Finally
\begin{equation}\label{ansatz-new3}
    \ket{\tilde{x}} = 
    \ket{\tilde{x}(\vec{\theta}\,)} = \hat{U}_{QFT}\,U(\vec{\theta}\,)\,\ket{b} =
\hat{U}_{QFT}\,U(\vec{\theta}\,)\,U_b\,\ket{0}^{\otimes\,n}.
\end{equation}
Consequently the measurement protocol for estimating the first term in Eq.~(\ref{loss3}) becomes
\begin{equation}\label{measurement-first}
    \bra{\tilde{x}}\hat{D}^2\ket{\tilde{x}} 
    = \mathcal{M}\left(\hat{D}^2, \hat{U}_{QFT}\,U\,U_b\right) = \mathcal{M}_{\hat{D}^2},
\end{equation}
which is depicted in Fig. (\ref{fig-diagonal-measure-U-3qubits}). 

The remaining term $\bra{\tilde{x}}\hat{D}\ket{\tilde{b}}$ in Eq.~(\ref{loss3}) is more difficult to compute because it is asymmetric with regards to the states which must collapse on the observable $\hat{D}$. In this case, even though $\hat{D}$ is Hermitian, $\bra{\tilde{x}}\hat{D}\ket{\tilde{b}}$ is not necessarily real. Therefore, we can use two circuits for computing its real and imaginary parts. Following the procedure suggested in Ref. \citep{guseynov_depth_2023}, an additional ancilla qubit is added, and some modifications of the standard Hadamard test \citep{nielsen_quantum_2010} are properly designed. In particular, the modified Hadamard tests, depicted in Fig. \ref{fig-diagonal-measure-UHI-3+aqubits} and in Fig. \ref{fig-diagonal-measure-UHS-3+aqubits}, are used to prepare two states $\ket{\xi}$ and $\ket{\xi'}$, respectively. These quantum states are defined with regards to an enlarged system made of $n+1$ qubits, where the ancilla qubit is added to the original $n$ qubits. {When composing physical systems, like adding an ancilla qubit in this case, the sequential labeling of their components (e.g., $\ket{\psi_0}, \ket{\psi_1}, \dots, \ket{\psi_{n-1}}$) may differ from the mathematical notation used to represent the bit strings, i.e., $\beta_{n-1}\dots\beta_1\beta_0$. In this case, we will conventionally list the ancilla qubit first.} The quantum circuits preparing these states act as unitary gates, namely
\begin{equation}\label{circuit-xi-Re}
    \ket{\xi} = \ket{\xi(\vec{\theta}\,)} = U_{HI}(\vec{\theta}\,)\,\ket{0}^{\otimes\,(n+1)},
\end{equation}
\begin{equation}\label{circuit-xi-Im}
    \ket{\xi'} = \ket{\xi'(\vec{\theta}\,)} = U_{HS}(\vec{\theta}\,)\,\ket{0}^{\otimes\,(n+1)}.
\end{equation}
These quantum states are used in the following measurement protocols (which are proved in the following):
\begin{equation}\label{measurement-second-Re}
\operatorname{Re}\left(\bra{\tilde{x}}\hat{D}\ket{\tilde{b}}\right) = \bra{\xi}Z\otimes\hat{D}\ket{\xi} = \mathcal{M}\left(Z\otimes\hat{D}, U_{HI}\right) = \mathcal{M}_{\hat{D},\,\operatorname{Re}},
\end{equation}
\begin{equation}\label{measurement-second-Im}
\operatorname{Im}\left(\bra{\tilde{x}}\hat{D}\ket{\tilde{b}}\right) = \bra{\xi'}Z\otimes\hat{D}\ket{\xi'} = \mathcal{M}\left(Z\otimes\hat{D}, U_{HS}\right) = \mathcal{M}_{\hat{D},\,\operatorname{Im}},
\end{equation}
where $Z$ is one of the Pauli matrices reported in Eq.~(\ref{Pauli-operators}). Before proving the above measurement protocols, it is worth to realize that they can be used to compute the loss function given by Eq.~(\ref{loss3}), namely
\begin{eqnarray}\label{loss4}
    L(\vec{\theta}\,) &=& 
    \bra{\tilde{x}}\hat{D}^2\ket{\tilde{x}}-\left|\bra{\tilde{x}}\hat{D}\ket{\tilde{b}}\right|^2 \nonumber\\
    &=&\bra{\tilde{x}}\hat{D}^2\ket{\tilde{x}}-\left[\operatorname{Re}\left(\bra{\tilde{x}}\hat{D}\ket{\tilde{b}}\right)\right]^2 -\left[\operatorname{Im}\left(\bra{\tilde{x}}\hat{D}\ket{\tilde{b}}\right)\right]^2\nonumber\\
    &=&\mathcal{M}_{\hat{D}^2} 
    -\mathcal{M}_{\hat{D},\,\operatorname{Re}}^2
    -\mathcal{M}_{\hat{D},\,\operatorname{Im}}^2.
\end{eqnarray}
This is the novel methodology proposed in Ref. \citep{guseynov_depth_2023}, which reduces significantly the number of Pauli matrices required for the decomposition of the observable, and hence substitutes the naive measurement protocol reported in Eq.~(\ref{loss2}). The key idea is to simplify the observables, at the price of making the circuits for generating the measurement states more complex. 

The novel methodology given by Eq.~(\ref{loss4}) is based essentially on the measurement protocols given by Eq.~(\ref{measurement-second-Re}) and Eq.~(\ref{measurement-second-Im}). In order to prove them, one needs to derive explicitly the states $\ket{\xi}$ and $\ket{\xi'}$. Let us start with $\ket{\xi}$, which is prepared by the circuit reported in Fig. \ref{fig-diagonal-measure-UHI-3+aqubits}. Recalling that the Hadamard gate \citep{nielsen_quantum_2010} is given by
\begin{equation}\label{Hadamard-gate}
    {H} := \frac{1}{\sqrt{2}}
    \begin{pmatrix}
        \,1 & 1\, \\
        \,1 & -1\, 
    \end{pmatrix},
\end{equation}
$H \ket{0} = (\ket{0}+\ket{1})/\sqrt{2}$ and $H \ket{1} = (\ket{0}-\ket{1})/\sqrt{2}$ because, by convention, $\ket{0} = (1, 0)^T$ and $\ket{1} = (0, 1)^T$. Analyzing the circuit depicted in Fig. \ref{fig-diagonal-measure-UHI-3+aqubits} yields
\begin{eqnarray}\label{xi}
\ket{\xi}|_{\text{1st barrier}} &=& \frac{1}{\sqrt{2}} \left(\ket{0}+ \ket{1}\right) \otimes \ket{\tilde{b}}.
\end{eqnarray}
In this case, the controlled U gate is only applied to the target if the controlled qubit(s) is in the $|1\rangle$ state, namely
\begin{equation}\label{controlled-U}
    {U}^c := {
    \ket{0}\bra{0}\otimes I+
    \ket{1}\bra{1}\otimes U}=
    \begin{pmatrix}
        1 & 0 & 0 & 0 \\
        0 & 1 & 0 & 0 \\
        0 & 0 & U_{11} & U_{12} \\
         0 & 0 & U_{21} & U_{22}
    \end{pmatrix}.
\end{equation}
Consequently, according to equation \ref{ansatz-new}, the state at the second barrier in this case becomes:
\begin{eqnarray}
\ket{\xi}|_{\text{2nd barrier}} = \frac{1}{\sqrt{2}} \left(\ket{0}\otimes \ket{\tilde{b}}+ \ket{1}\otimes \ket{\tilde{x}}\right).
\end{eqnarray}
Finally, applying the Hadamard gate again to the previous state yields:
\begin{eqnarray}
\ket{\xi}|_{\text{3rd barrier}} = \ket{\xi} &=& \frac{1}{\sqrt{2}} \left[\frac{1}{\sqrt{2}}\left(\ket{0}+ \ket{1}\right)\otimes \ket{\tilde{b}}+ \frac{1}{\sqrt{2}}\left(\ket{0} - \ket{1}\right)\otimes \ket{\tilde{x}}\right] = \nonumber\\ 
&=& \frac{1}{2} \ket{0} \otimes (\ket{\tilde{b}} + \ket{\tilde{x}}) + \frac{1}{2} \ket{1} \otimes (\ket{\tilde{b}} - \ket{\tilde{x}}).
\end{eqnarray}
This is a particularly interesting quantum state: (i) because it represents a superposition between the vector of known terms of the linear system $\ket{\tilde{b}}$ and the solution vector $\ket{\tilde{x}}$; (ii) furthermore, the presence of the ancilla qubit allows us to distinguish between two linear combinations, $\ket{\tilde{b}} + \ket{\tilde{x}}$ and $\ket{\tilde{b}} - \ket{\tilde{x}}$, thereby broadening the range of computations that can be performed with this state. The state exhibits quantum entanglement between the ancilla qubit and the register containing $\ket{\tilde{b}}$ and $\ket{\tilde{x}}$, meaning that measurement of the ancilla directly affects the state of the second register. Having simultaneous access to both $\ket{\tilde{b}} + \ket{\tilde{x}}$ and $\ket{\tilde{b}} - \ket{\tilde{x}}$ enables the use of quantum interference to extract global features of the solution, such as inner products or similarity tests. Recalling that $Z\ket{0} = \ket{0}$ and $Z\ket{1} = -\ket{1}$, applying the observable $Z \otimes \hat{D}$ to $\ket{\xi}$ yields
\begin{eqnarray}\label{ZoD-ketxi}
(Z \otimes \hat{D}) \ket{\xi} &=& \frac{1}{2} Z\ket{0} \otimes \hat{D}(\ket{\tilde{b}} + \ket{\tilde{x}}) + \frac{1}{2} Z\ket{1} \otimes \hat{D}(\ket{\tilde{b}} - \ket{\tilde{x}})\nonumber\\
&=& \frac{1}{2} \ket{0} \otimes \hat{D}(\ket{\tilde{b}} + \ket{\tilde{x}}) - \frac{1}{2} \ket{1} \otimes \hat{D}(\ket{\tilde{b}} - \ket{\tilde{x}}).
\end{eqnarray}
Next, we want to compute $\bra{\xi} (Z \otimes \hat{D}) \ket{\xi}$. The complex conjugate state $\bra{\xi}$ is given by
\begin{equation}\label{conjugate-xi}
\bra{\xi} = \frac{1}{2} \bra{0} \otimes (\bra{\tilde{b}} + \bra{\tilde{x}}) + \frac{1}{2} \bra{1} \otimes (\bra{\tilde{b}} - \bra{\tilde{x}}).
\end{equation}
Consequently the expectation value $\bra{\xi} (Z \otimes \hat{D}) \ket{\xi}$ is given by
\begin{eqnarray}\label{braxi-ZoD-ketxi}
\bra{\xi} (Z \otimes \hat{D}) \ket{\xi} = &&
\frac{1}{4} \left[\bra{0} \otimes (\bra{\tilde{b}} + \bra{\tilde{x}})\right] \left[\ket{0} \otimes \hat{D}(\ket{\tilde{b}} + \ket{\tilde{x}})\right]\nonumber\\
&&- \frac{1}{4} \left[\bra{0} \otimes (\bra{\tilde{b}} + \bra{\tilde{x}})\right] \left[\ket{1} \otimes \hat{D}(\ket{\tilde{b}} - \ket{\tilde{x}})\right]\nonumber\\
&&+ \frac{1}{4} \left[\bra{1} \otimes (\bra{\tilde{b}} - \bra{\tilde{x}})\right] \left[\ket{0} \otimes \hat{D}(\ket{\tilde{b}} + \ket{\tilde{x}})\right]\nonumber\\
&&- \frac{1}{4} \left[\bra{1} \otimes (\bra{\tilde{b}} - \bra{\tilde{x}})\right] \left[\ket{1} \otimes \hat{D}(\ket{\tilde{b}} - \ket{\tilde{x}})\right].
\end{eqnarray}
The second and the third term in the previous expression are null, because $\langle 0 | 1 \rangle = 0$ and $\langle 1 | 0 \rangle = 0$. Taking into account that $\langle 0 | 0 \rangle = 1$ and $\langle 1 | 1 \rangle = 1$ yields
\begin{eqnarray}\label{braxi-ZoD-ketxi2}
\bra{\xi} (Z \otimes \hat{D}) \ket{\xi} = &&
\frac{1}{4} \left( \bra{\tilde{b}} \hat{D} \ket{\tilde{b}} + \bra{\tilde{b}} \hat{D} \ket{\tilde{x}} + \bra{\tilde{x}} \hat{D} \ket{\tilde{b}} + \bra{\tilde{x}} \hat{D} \ket{\tilde{x}} \right)\nonumber\\
&&- \frac{1}{4} \left( \bra{\tilde{b}} \hat{D} \ket{\tilde{b}} - \bra{\tilde{b}} \hat{D} \ket{\tilde{x}} - \bra{\tilde{x}} \hat{D} \ket{\tilde{b}} + \bra{\tilde{x}} \hat{D} \ket{\tilde{x}} \right)\nonumber\\
= &&\frac{1}{2} \left(\bra{\tilde{b}} \hat{D} \ket{\tilde{x}} + \bra{\tilde{x}} \hat{D} \ket{\tilde{b}} \right).
\end{eqnarray}
Taking into account that
\begin{equation}\label{inner-product-property}
    (\bra{\tilde{x}} \hat{D} \ket{\tilde{b}})^* = \bra{\tilde{b}} \hat{D}^\dagger \ket{\tilde{x}} = 
    \bra{\tilde{b}} \hat{D} \ket{\tilde{x}},
\end{equation}
it is possible to derive the following property
\begin{equation}\label{inner-product-property2}
\bra{\tilde{b}} \hat{D} \ket{\tilde{x}} + \bra{\tilde{x}} \hat{D} \ket{\tilde{b}} = 2 \operatorname{Re} \left(\bra{\tilde{x}} \hat{D} \ket{\tilde{b}}\right).
\end{equation}
Consequently, using Eq.~(\ref{inner-product-property2}) into Eq.~(\ref{braxi-ZoD-ketxi2}) yields
\begin{equation}\label{braxi-ZoD-ketxi3}
\bra{\xi} (Z \otimes \hat{D}) \ket{\xi} = \operatorname{Re} \left(\bra{\tilde{x}} \hat{D} \ket{\tilde{b}}\right) = \mathcal{M}_{\hat{D},\,\operatorname{Re}},
\end{equation}
which is the desired result for the measurement protocol $\mathcal{M}_{\hat{D},\,\operatorname{Re}}$. 

On the other hand, let us proceed with $\ket{\xi'}$, which is prepared by the circuit reported in Fig. \ref{fig-diagonal-measure-UHS-3+aqubits}. Recalling that the phase gate \citep{nielsen_quantum_2010} is given by
\begin{equation}\label{Shift-gate}
    {S} := \begin{pmatrix}
        \,1 & 0\, \\
        \,0 & i\, 
    \end{pmatrix}.
\end{equation}
Hence, $S \ket{0} = \ket{0}$ and $S \ket{1} = i\ket{1}$.
Analyzing the circuit depicted in Fig. \ref{fig-diagonal-measure-UHS-3+aqubits} yields
\begin{eqnarray}\label{xi-prime}
\ket{\xi'}|_{\text{1st barrier}} &=& \frac{1}{\sqrt{2}} \left(\ket{0}+ i\,\ket{1}\right) \otimes \ket{\tilde{b}},\\
\ket{\xi'}|_{\text{2nd barrier}} &=& \frac{1}{\sqrt{2}} \left(\ket{0}\otimes \ket{\tilde{b}}+ i\,\ket{1}\otimes \ket{\tilde{x}}\right),\\
\ket{\xi'}|_{\text{3rd barrier}} = \ket{\xi'} &=& \frac{1}{\sqrt{2}} \left[\frac{1}{\sqrt{2}}\left(\ket{0}+ \ket{1}\right)\otimes \ket{\tilde{b}}+ \frac{i}{\sqrt{2}}\left(\ket{0} - \ket{1}\right)\otimes \ket{\tilde{x}}\right] = \nonumber\\ 
&=& \frac{1}{2} \ket{0} \otimes (\ket{\tilde{b}} + i\,\ket{\tilde{x}}) + \frac{1}{2} \ket{1} \otimes (\ket{\tilde{b}} - i\,\ket{\tilde{x}}).
\end{eqnarray}
The same considerations about the unique properties of the state $\ket{\xi}$ hold as well for the state $\ket{\xi'}$. Proceeding in the same way discussed above, the expectation value $\bra{\xi'} Z \otimes \hat{D} \ket{\xi'}$ is given by
\begin{equation}\label{braxip-ZoD-ketxip}
    \bra{\xi'} Z \otimes \hat{D} \ket{\xi'} = \frac{i}{2} \left( \bra{\tilde{b}} \hat{D} \ket{\tilde{x}} - \bra{\tilde{x}} \hat{D} \ket{\tilde{b}} \right).
\end{equation}
Taking into account again Eq.~(\ref{inner-product-property}), it is possible to derive the following property
\begin{equation}\label{inner-product-property3}
\bra{\tilde{b}} \hat{D} \ket{\tilde{x}} - \bra{\tilde{x}} \hat{D} \ket{\tilde{b}} = \left( \bra{\tilde{x}} \hat{D} \ket{\tilde{b}} \right)^* - \bra{\tilde{x}} \hat{D} \ket{\tilde{b}} = -2i \, \operatorname{Im} \left( \bra{\tilde{x}} \hat{D} \ket{\tilde{b}} \right).
\end{equation}
Consequently, using Eq.~(\ref{inner-product-property3}) into Eq.~(\ref{braxip-ZoD-ketxip}) yields
\begin{equation}\label{braxip-ZoD-ketxip2}
    \bra{\xi'} Z \otimes \hat{D} \ket{\xi'} = \operatorname{Im} \left( \bra{\tilde{x}} \hat{D} \ket{\tilde{b}} \right) = \mathcal{M}_{\hat{D},\,\operatorname{Im}},
\end{equation}
which is again the desired result for the measurement protocol $\mathcal{M}_{\hat{D},\,\operatorname{Im}}$.

\section{Harrow–Hassidim–Lloyd (HHL) algorithm}
\label{HHL}

The VQE approach described in Section~\ref{VQE} is certainly a good starting point, due to the intuitive analogy between solving a linear system of equations and finding the ground state of a quantum system, essentially captured by Eq.~(\ref{observable}). 
Moreover, VQE is considered more robust when dealing with current noisy intermediate-scale quantum (NISQ) devices. 
However, several issues may hinder the ability of VQE to effectively scale quantum simulations, even on future, hypothetically ideal quantum computers. 
First, in some cases the parametrized observable may be affected by the barren plateau (BP) phenomenon, in which the optimization landscape of the ansatz becomes exponentially flat and featureless as the number of qubits increases \citep{larocca_barren_2025}. 
To mitigate its negative impact on trainability, one may adopt local cost functions \citep{Bravo-Prieto2023} or alternating-layered ansatz circuits \citep{Cerezo2021}, although there is no guarantee that these techniques fully solve the problem. Furthermore, as the number of qubits — and consequently the number of parameters in the ansatz — increases, the optimization space becomes high-dimensional, which can be challenging for classical optimizers.

To this respect, let us consider in this section the Harrow–Hassidim–Lloyd (HHL) algorithm \citep{PhysRevLett.103.150502} which, to date, can be considered one of the most promising quantum algorithms for solving linear systems on future, fault-tolerant quantum computers.
HHL is theoretically appealing because it offers an exponential quantum speedup under well-defined assumptions—specifically, when the matrix is sparse, well-conditioned, and efficiently representable \citep{PhysRevLett.103.150502}, with potential impact also on industrial applications \citep{leonardo2025}. 
In spite of the promising features, the practical implementation of HHL on current noisy intermediate-scale quantum (NISQ) devices remains severely limited. 
The algorithm requires deep circuits involving controlled rotations, quantum phase estimation, and accurate eigenvalue inversion—operations highly sensitive to gate noise, decoherence, and restricted circuit depth. 
Thus, while HHL stands as a theoretically powerful algorithm with strong asymptotic promises, its practical use is postponed to the era of large-scale, error-corrected quantum computers. 
Meanwhile, it is worth the effort to investigate the applicability of the HHL algorithm in solving practical problems by using classical High Performance Computing (HPC) facilities. 
Moreover, HHL has stimulated the development of useful tutorials (e.g., Ref. \citep{hhl_tutorial}) that help readers learn basic concepts in quantum computing. The main goal of this document is to provide a rigorous analysis of the HHL algorithm and to discuss its details in a clear, step-by-step manner.

\begin{figure}[htbp]
    \centering
    \includegraphics[width=1.0\linewidth]{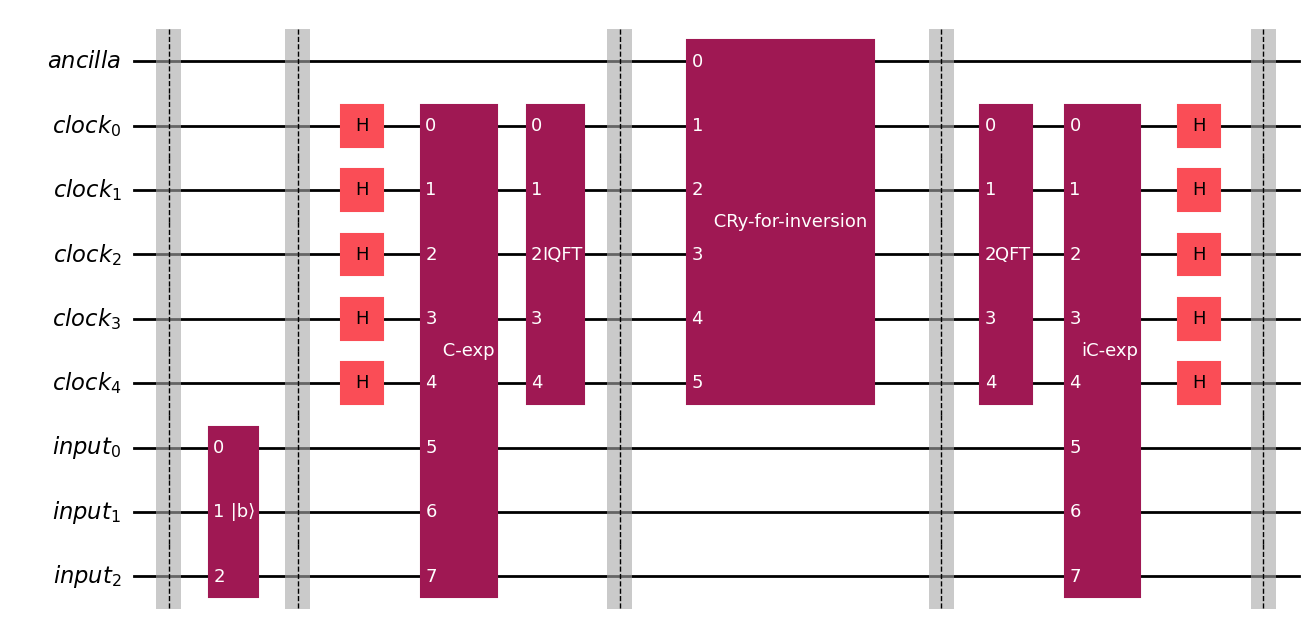}
    \caption{Circuit ideally implementing HHL algorithm with $n=3$ input/output qubits for solving the same problem discussed in the previous sections, $n_c=5$ clock qubits and $1$ ancilla qubit. ``C-exp" stands for controlled exponentiation (or controlled Hamiltonian evolution), while ``iC-exp" stands for its logical inverse. ``CRy-for-inversion" stands for controlled rotation for eigenvalue inversion. Because this circuit does not include any measurements, it is suitable only for ideal (statevector) simulations.   
    }\label{fig:circuit-hhl-3q5c}
\end{figure}

The overall HHL circuit is shown in Fig. (\ref{fig:circuit-hhl-3q5c}) for the same problem discussed in the previous sections. 
The main difference between HHL and VQE is that the former requires extra qubits for approximating the solution. The HHL algorithm requires extra qubits because it must temporarily store and process the eigenvalues of the matrix of the target problem and perform controlled operations that depend on those eigenvalues. These extra qubits hold intermediate quantum information, which are essential for implementing the inverse of the matrix of the linear system coherently on a quantum state.
Therefore, in addition to the $n$ qubits that encode the numerical solution at the mesh nodes (with $N = 2^n$), HHL requires additional $n_c$ “clock” qubits and one extra ancilla qubit. 
Therefore, the total number of qubits required by the algorithm is $n + n_c + 1$. Clearly, there is a noticeable overhead for small systems when $n \sim (n_c + 1)$, but this becomes negligible when $n \gg (n_c + 1)$. We can imagine the $n$ qubits used for the computational mesh as the input (and output) register. A quantum register is a collection of qubits that together form the basic unit of quantum memory or state representation in a quantum computer. Similarly, the $n_c$ qubits form the clock register and the ancilla qubit constitutes the ancilla register. Hence, the HHL algorithm uses three quantum registers, which will be conventionally listed in the above order. Within each register, the corresponding computational basis is defined by a binary representation, which uses the usual mathematical notation to represent the bit strings, i.e., $\beta_{n-1}\dots\beta_1\beta_0$. In order to make clearer the meaning of the qubits, it is also possible to add a superscript which specifies the register they belong to (omitting the one for input/output consistently with the rest of the document), namely
\begin{equation}\label{overall-state}
\ket{\beta_{n-1}\dots\beta_1\beta_0\,
\beta_{n_c-1}^c\dots\beta_1^c\beta_0^c\,
\beta_0^a},
\end{equation}
where the three registers are separable at inlet and output, but not necessarily in the intermediate steps of the algorithm when they exchange/share information. 

Coming back to the main algorithm, HHL is essentially made of two components: (i) the Quantum Phase Estimation (QPE) and (ii) the binarized inversion module, which are described in the following subsections \ref{QPE} and \ref{inv}, before combining them in subsection \ref{hhl-all}. In this section, only ideal (statevector) simulations are reported, because the practical implementation of HHL on current noisy intermediate-scale quantum (NISQ) devices remains severely limited.

\subsection{Quantum Phase Estimation (QPE)}
\label{QPE}

The non-trivial starting point of the HHL algorithm is the Quantum Phase Estimation (QPE) algorithm \citep{nielsen_quantum_2010}. The Quantum Phase Estimation (QPE) algorithm is a quantum algorithm used to estimate the phase associated with an eigenvalue of a given unitary operator. 

In order to understand what is the quantum phase, let us recall the main result about the eigenvalues of the numerical finite-difference procedure for the heat conduction equation reported in section \ref{hce}, namely $1\leq\lambda^C_m\leq(1+4\,r)$, where $\lambda^C_m$ is the $m$-th eigenvalue of the operator $\hat{C}$ given by Eq.~(\ref{conduction-matrix}). The parameter $r = D\,\Delta t / \Delta z^{2}$ is the (dimensionless) numerical Fourier number. It should increase during mesh refinement in order to keep $\Delta t$ constant; namely, $r \sim 1/\Delta z^{2}$. Otherwise, time stepping would require an impractically large number of iterations on a quantum computer with a significant number of qubits. Therefore it makes sense to normalize the evolution matrix as 
\begin{equation}\label{Gamma-def}
\hat{\Gamma} := \frac{1}{1+4\,r}\,\hat{C},
\end{equation}
which can be used to reformulate Eq.~(\ref{matrix-heat-conduction}) as $\hat{\Gamma}\,\vec{\xi} = \ket{b}$, where the unknown vector $\vec{\xi}$ is defined as
\begin{equation}\label{xi-def}
\vec{\xi} = \frac{1+4\,r}{\theta}\,\vec{T}^+ = 
\frac{\theta^+}{\theta}\,(1+4\,r)\,\ket{x} = g\,\ket{x},
\end{equation}
while $g := f\,(1+4\,r)$, $\theta$ is given by Eq. (\ref{theta}) and $\theta^+$ by Eq. (\ref{theta+}). 
The normalized evolution matrix $\hat{\Gamma}$ is real and symmetric. Hence it can be expressed by spectral decomposition, using its eigenvectors $\ket{\phi_m^\Gamma}$ and its eigenvalues $\lambda_m^\Gamma$, namely
\begin{equation}\label{Gamma-def2}
\hat{\Gamma} = \sum_{m=0}^{N-1} \lambda_m^\Gamma\,\ket{\phi_m^\Gamma}\bra{\phi_m^\Gamma}
= \sum_{j\in\{0,\,1\}^n} \lambda_j^\Gamma\,\ket{\phi_j^\Gamma}\bra{\phi_j^\Gamma},
\end{equation}
where $j$ is the bit string corresponding to $m$ as usual. Because of the previous normalization, now the following relation holds
\begin{equation}\label{Gamma-eigenvalues-limits}
\frac{1}{1+4\,r}\leq\lambda_j^\Gamma\leq 1.
\end{equation}
The input $\ket{b}$ can be also expressed in the basis formed by the eigenvectors of $\hat{\Gamma}$, such that
\begin{equation}\label{b-def}
\ket{b} = \sum_{j\in\{0,\,1\}^n} b_j \ket{j}
 = \sum_{j\in\{0,\,1\}^n} b_j^\Gamma\ket{\phi_j^\Gamma},
\end{equation}
as well as the solution
\begin{equation}\label{x-def}
\ket{x} = \sum_{j\in\{0,\,1\}^n} x_j^\Gamma\ket{\phi_j^\Gamma}.
\end{equation}
Recalling that $g\,\hat{\Gamma}\,\ket{x}=\ket{b}$ and that the eigenvectors of this matrix form an orthonormal basis yields
\begin{equation}\label{x-def2}
x_j^\Gamma = \frac{b_j^\Gamma}{g\,\lambda_j^\Gamma}.
\end{equation}
Because $\braket{x|x}=1$ by construction, as already discussed in section \ref{VQE}, then $\sum_{j} |x_j^\Gamma|^2=1$ holds. The last relations allows one to compute $g$, namely
\begin{equation}\label{g-def}
g=\sqrt{\sum_{j} \left|\frac{b_j^\Gamma}{\lambda_j^\Gamma}\right|^2},
\end{equation}
and consequently
\begin{equation}\label{x-def3}
x_j^\Gamma = \frac{1}{\sqrt{\sum_{j} \left|{b_j^\Gamma}/{\lambda_j^\Gamma}\right|^2}}\,\frac{b_j^\Gamma}{\lambda_j^\Gamma},
\end{equation}

In essence, the HHL algorithm is a step-by-step procedure for computing the coefficients $x_j^\Gamma$ (and hence $\ket{x}$), by computing the expansion coefficients $b_j^\Gamma$ and inverting the eigenvalues $\lambda_j^\Gamma$. 
Generalizing the pedagogical approach proposed in Ref. (\citep{hhl_tutorial}) for the present case, it is possible to derive the system quantum states at every computational step in the following.
\begin{itemize}
    \item Step \#1
\end{itemize}
The input register must be prepared to have the amplitudes corresponding to the coefficients of $\ket{b}$. 
This can be done by many techniques designed for data loading/encoding in quantum circuits. In practical applications, the cost of loading classical information into a quantum device can dominate the overall asymptotic complexity of a quantum algorithm. Hence, data encoding remains an active and challenging research area in quantum computing \citep{Agliardi2025}. 
Several approaches aim to mitigate this bottleneck. 
Some algorithms rely on divide-and-conquer strategies for efficiently preparing amplitude-encoded states, exploiting hierarchical structures such as segment trees to reduce state-preparation costs \citep{araujo_divide-and-conquer_2021}. 
Some details about the latter approach are reported in Appendix \ref{appedix-real-data-loading}. 
Starting with $\ket{\Psi_0} := \ket{0}^{\otimes n}\,\ket{0}^{\otimes n_c}_c\,\ket{0}_a$ and applying a proper procedure for data encoding yields
\begin{equation}\label{Psi1}
\ket{\Psi_1} := \ket{b}\,\ket{0}^{\otimes n_c}_c\,\ket{0}_a.
\end{equation}

\begin{itemize}
    \item Step \#2
\end{itemize}

This is the proper beginning of the QPE algorithm. First of all, Hadamard gates are applied to the clock qubits to create a superposition of the clock qubits, namely
\begin{equation}\label{Psi2a}
\ket{\Psi_2} := \ket{b}\,
\frac{1}{\sqrt{N_c}}\left(\ket{0}+\ket{1}\right)^{\otimes n_c}\,\ket{0}_a,
\end{equation}
where $N_c = 2^{n_c}$. This superposition can be further expanded explicitly as
\begin{eqnarray}\label{Psi2b}
\ket{\Psi_2} &=& \ket{b}\,
\frac{1}{\sqrt{N_c}}
\left(\ket{0}+\ket{1}\right)\otimes\dots
\left(\ket{0}+\ket{1}\right)\otimes
\left(\ket{0}+\ket{1}\right)
\,\ket{0}_a\nonumber\\
&=& \ket{b}\,
\frac{1}{\sqrt{N_c}}
\sum_{k\in\{0,\,1\}^{n_c}}\ket{k}_c\,\ket{0}_a,
\end{eqnarray}
where $k$ is again a binary number and $\ket{k}$ is a generic state of the computational basis of the clock register. Please note that the number $N_c$ of the states of the clock register typically differs from the number $N$ of the states of the input register. The previous system quantum state is still separable in its constituent registers, namely
\begin{equation}\label{Psi2}
\ket{\Psi_2} = 
\left(
\sum_{j\in\{0,\,1\}^n} b_j^{\Gamma}\ket{\phi_j^{\Gamma}}
\right)\otimes
\left(
\frac{1}{\sqrt{N_c}}
\sum_{k\in\{0,\,1\}^{n_c}}\ket{k}_c
\right)
\otimes\ket{0}_a,
\end{equation}
where we used Eq.~(\ref{b-def}). In the following, we will sometimes decompose the binary number $k$ in a string of bits, namely
\begin{equation}\label{state-binary-clock}
    k := k_{(2)} = \beta^c_{n_c-1}\dots\beta^c_1\beta^c_0,
\end{equation}
where the generic clock bit is $\beta^c_q$. The corresponding decimal number is
\begin{equation}\label{multiqubit-expansion-clock}
    k_{(10)} = \sum_{q=0}^{n_c-1}\beta^c_q\,2^q.
\end{equation}

\begin{figure}[htbp]
    \centering
    \includegraphics[width=0.6\linewidth]{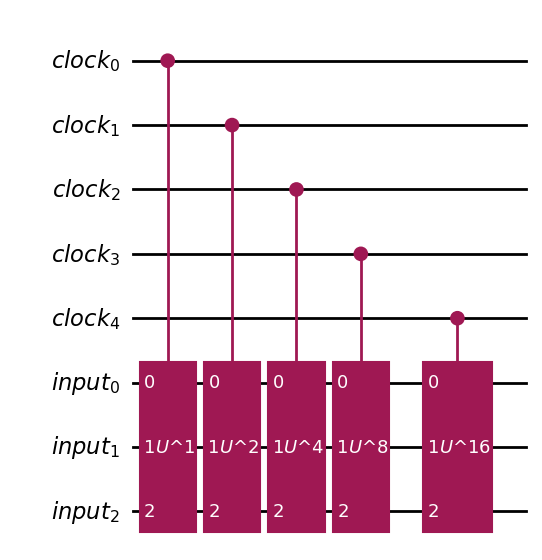}
    \caption{Circuit implementing controlled exponentiation (or controlled Hamiltonian evolution) in the HHL algorithm with $n=3$ input qubits for solving the same problem discussed in the previous sections and $n_c=5$ clock qubits. This circuit shows what is inside the block called ``C-exp" in Fig. (\ref{fig:circuit-hhl-3q5c}). 
    }\label{fig:circuit-hhl-exponentiation-3q5c}
\end{figure}

\begin{itemize}
    \item Step \#3
\end{itemize}

The next step is the controlled exponentiation (or controlled Hamiltonian evolution) shown in Fig. (\ref{fig:circuit-hhl-exponentiation-3q5c}). 
The QPE extracts eigenvalues (as quantum phases defined below), but only of a unitary operator. 
We can define the unitary operator $\hat{U}$ as the evolution determined by the Hamiltonian $\hat{\Gamma}$ we are interested in, namely
\begin{equation}\label{U-def}
\hat{U} = e^{i\,\hat{\Gamma}\,\varphi} = \sum_{s = 0}^{\infty}\frac{1}{s!}\,(i\,\hat{\Gamma}\,\varphi)^s = 
\sum_{j\in\{0,\,1\}^n} 
e^{i\,\lambda_j^\Gamma\,\varphi}
\,\ket{\phi_j^\Gamma}\bra{\phi_j^\Gamma},
\end{equation}
where $\varphi$ is a fictitious evolution time (which also indirectly explains why we called ``clock" the second register). 
For reasons that will be clarified later, we assume
\begin{equation}\label{varphi}
\varphi = 2\pi\left(\frac{N_c-1}{N_c}\right).
\end{equation}
A unitary operator has eigenvalues of modulus $1$, which can be written as $e^{i2\pi\phi_j}$, where $\phi_j$ is the $j$-th quantum phase. 
Quantum phases are elusive concepts because multiplying an entire quantum state by $e^{i2\pi\phi_j}$ does not change measurement outcomes. 
However HHL algorithm, and QPE in particular, uses quantum phases of the input register to compute the eigenvalues we are interested in. 
It is worth noting that $\hat{U}$ is also diagonal in the basis defined by the eigenvalues of $\hat{\Gamma}$, namely
\begin{equation}\label{U-def2}
\hat{U} = \sum_{j\in\{0,\,1\}^n} e^{i2\pi\phi_j}
\ket{\phi_j^{\Gamma}}\bra{\phi_j^{\Gamma}}.
\end{equation}
Comparing the previous expression with Eq.~(\ref{U-def}) allows to derive the relation between quantum phases and original eigenvalues as $2\pi\phi_j = \lambda_j^{\Gamma}\varphi$. Using the assumption given by Eq.~(\ref{varphi}) implies
\begin{equation}\label{phi}
\phi_j = \lambda_j^{\Gamma}\,\left(\frac{N_c-1}{N_c}\right).
\end{equation}
It is also trivial to compute the eigenvalues of the previous operator because
\begin{equation}\label{U-def3}
\hat{U}\ket{\phi_j^{\Gamma}} = e^{i 2\pi\phi_j}
\ket{\phi_j^{\Gamma}} = e^{i \lambda_j^{\Gamma}\varphi}
\ket{\phi_j^{\Gamma}},
\end{equation}
%
%
%

For the powers of the unitary operator $\hat{U}$, similar relations hold. The powers of the unitary operator prove to be very useful. The idea is to use different powers of $\hat{U}$ in order to highlight the specific behavior of different quantum phases $\phi_j$ (and hence of different eigenvalues). We want to operate on the input register differently, depending on the clock register. Let us reformulate Eq.~(\ref{Psi2}) as
\begin{equation}\label{Psi2bis}
\ket{\Psi_2} = \frac{1}{\sqrt{N_c}}
\sum_{j\in\{0,\,1\}^n} b_j^{\Gamma}
\sum_{k\in\{0,\,1\}^{n_c}}
\ket{\phi_j^{\Gamma}}
\ket{k}_c
\ket{0}_a,
\end{equation}
because $\ket{\phi_j^{\Gamma}}$ does not depend on $k$. Let us proceed in the following way: (i) firstly, we factorize its binary representation as $k=\beta^c_{n_c-1}\dots\beta^c_1\beta^c_0$; (ii) secondly, we use $\beta^c_q$ to perform some operations on $\ket{\phi_j^{\Gamma}}$. In particular, if $\beta^c_q = 1$ then we apply $\hat{U}^{2^{q}}$ to $\ket{\phi_j^{\Gamma}}$, otherwise, if $\beta^c_q = 0$, then nothing happens. It is possible to simplify the last statement by saying that, for every bit of the string $k=\beta^c_{n_c-1}\dots\beta^c_1\beta^c_0$, one apply $\hat{U}^{\beta^c_q\,2^{q}}$ on $\ket{\phi_j^{\Gamma}}$, because $\hat{U}^{0} = I$. Because $\beta^c_q$ depends on $k$, we can imagine to define the following operator (of Hamiltonian evolution)
\begin{equation}\label{C-exp}
{E_V} = \prod_{q=0}^{n_c-1}\hat{U}^{\beta^c_q(k)\,2^{q}}
= \hat{U}^{\sum_{q=0}^{n_c-1}\beta^c_q(k)\,2^{q}} =
\hat{U}^{k_{(10)}}.
\end{equation}
Applying the previous operator to $\ket{\phi_j^{\Gamma}}$ yields
\begin{equation}\label{C-exp2}
{E_V}\ket{\phi_j^{\Gamma}} = \hat{U}^k\ket{\phi_j^{\Gamma}} =
e^{i2\pi\phi_j k_{(10)}}
\ket{\phi_j^{\Gamma}},
\end{equation}
where we used Eq.~(\ref{U-def3}) and Eq.~(\ref{multiqubit-expansion-clock}). 
%
%
%
The operator $E_V$ depends on the clock register $\ket{k}_c$ as it is clear in Eq.~(\ref{C-exp}). Let us define the following controlled operator for the whole quantum system
\begin{equation}\label{C-exp-system}
{CE_V} = \left(\sum_{k\in\{0,\,1\}^{n_c}}
\hat{U}^k\otimes \ket{k}_c\bra{k}_c
\right)\otimes I_a,
\end{equation}
and let us apply it to the system quantum state given by Eq.~(\ref{Psi2bis}) which yields
\begin{equation}\label{Psi3c}
\ket{\Psi_3} := {CE_V} \ket{\Psi_2} =
\frac{1}{\sqrt{N_c}}
\sum_{j\in\{0,\,1\}^n} b_j^{\Gamma}
\sum_{k\in\{0,\,1\}^{n_c}}
e^{i2\pi\phi_j k_{(10)}}
\ket{\phi_j^{\Gamma}}
\ket{k}_c\ket{0}_a.
\end{equation}
Because $\ket{\phi_j^{\Gamma}}$ does not depend on $k$, it is convenient to move the accumulated phase to the clock register, leaving the input register $\ket{\phi_j^{\Gamma}}$ unchanged, namely
\begin{equation}\label{Psi3}
\ket{\Psi_3} = 
\sum_{j\in\{0,\,1\}^n} 
\left[
b_j^{\Gamma}
\ket{\phi_j^{\Gamma}}\otimes
\left(
\frac{1}{\sqrt{N_c}}
\sum_{k\in\{0,\,1\}^{n_c}}
e^{i2\pi\phi_j k_{(10)}}
\ket{k}_c
\right)\right]\otimes\ket{0}_a.
\end{equation}
This phenomenon is general and it is called the kickback effect, which modifies the control state but leaves the target state unchanged. With other words, after applying the operator $E_V$ to an eigenvector, the phase accumulates on the clock register. The phase kickback is a fundamental quantum phenomenon in which a phase acquired by a target quantum state during a controlled operation is effectively transferred back onto the control qubit. A simple example is reported in Appendix \ref{kickback}.

As a final remark of this step, it is important to note that the state given by Eq.~(\ref{Psi3}) entangles the clock register with the input register. In order to make it even more evident, let us reformulate the previous state as
\begin{equation}\label{Psi3bis}
\ket{\Psi_3} = 
\left(
\sum_{j\in\{0,\,1\}^n} 
b_j^{\Gamma}
\ket{\phi_j^{\Gamma}}\otimes
\ket{\upsilon_j}_c\right)\otimes\ket{0}_a,
\end{equation}
where
\begin{equation}\label{upsilon}
\ket{\upsilon_j}_c := 
\frac{1}{\sqrt{N_c}}
\sum_{k\in\{0,\,1\}^{n_c}}
e^{i2\pi\phi_j k_{(10)}}
\ket{k}_c.
\end{equation}
Because the phases $\phi_j$ make the control states $\ket{\upsilon_j}_c$ different for different $j$, the overall state becomes a sum of non-parallel product terms, which cannot be factored into a single tensor product -- hence it is generically entangled. Now the clock register stores information about quantum phases $\phi_j$, which we need to extract in the following step.

\begin{itemize}
    \item Step \#4
\end{itemize}

In this step, let us apply the Quantum Fourier Transformation (QFT) to the clock register of the quantum state given by Eq.~(\ref{Psi3bis}), which means to apply QFT to $\ket{\upsilon_j}_c$ given by Eq.~(\ref{upsilon}). The QFT is defined in this document in such a way so as to realize a unitary transformation by construction \citep{nielsen_quantum_2010}. Consequently $\hat{U}_{QFT}$ is analogous to the unitary matrix $\hat{U}_{FT}$ given by Eq.~(\ref{FT-matrix}), but adapted for the number of states $N_c=2^{n_c}$ of the clock register. This means that $\hat{U}_{QFT}^{\dagger}\,\hat{U}_{QFT} = \hat{U}_{QFT}\,\hat{U}_{QFT}^{\dagger} = I$, where $\hat{U}_{QFT}^{\dagger}$ denotes the conjugate transpose of $\hat{U}_{QFT}$ (namely Hermitian transpose). The quantum state of this step can be formally defined as
\begin{equation}\label{Psi4}
\ket{\Psi_4} := 
\left(
\sum_{j\in\{0,\,1\}^n} 
b_j^{\Gamma}
\ket{\phi_j^{\Gamma}}\otimes
\hat{U}_{QFT}^{\dagger}\ket{\upsilon_j}_c\right)\otimes\ket{0}_a.
\end{equation}
Let us elaborate on the previous expression by applying $\hat{U}_{QFT}^{\dagger}$ to $\ket{\upsilon_j}_c$ given by Eq.~(\ref{upsilon}), namely
\begin{equation}\label{iQFT-Phi}
\hat{U}_{QFT}^{\dagger}\ket{\upsilon_j}_c = 
\frac{1}{\sqrt{N_c}}
\hat{U}_{QFT}^{\dagger}\sum_{k\in\{0,\,1\}^{n_c}}
e^{i2\pi\phi_j k_{(10)}}
\ket{k}_c=
\frac{1}{\sqrt{N_c}}
\sum_{k\in\{0,\,1\}^{n_c}}
e^{i2\pi\phi_j k_{(10)}}\,
\hat{U}_{QFT}^{\dagger}\ket{k}_c.
\end{equation}
Applying $\hat{U}_{QFT}^{\dagger}$ to a basis vector $\ket{k}_c$ yields the ${k}$-th column of the matrix $\hat{U}_{QFT}^{\dagger}$\footnote{Let us consider a matrix $A$ and a vector $\vec{a}$. Let us do the usual matrix-vector multiplication, namely $A\cdot\vec{a}$, which gives a vector. The generic component of the latter vector is $(A\cdot\vec{a})_\alpha$ and it can be expressed as $\sum_\beta A_{\alpha\beta}\,a_\beta$. Using the unit vector $\vec{e}_\alpha$ is possible to express $A\cdot\vec{a} = \sum_\alpha\sum_\beta A_{\alpha\beta}\,a_\beta\,\vec{e}_\alpha$ or equivalently $A\cdot\vec{a} = \sum_\alpha c_\alpha\,\vec{e}_\alpha$ where $c_\alpha = \sum_\beta A_{\alpha\beta}\,a_\beta$. In case $\vec{a}$ is another unit vector $\vec{a}'=\vec{e}_\gamma$, then $c_\alpha' = \sum_\beta A_{\alpha\beta}\,\delta_{\gamma\beta} = A_{\alpha\gamma}$, which means that $\vec{c}\,'$ is the $\gamma$-th column of the matrix.}
\begin{equation}\label{iQFT-basis}
\hat{U}_{QFT}^{\dagger}\ket{k}_c = 
 \frac{1}{\sqrt{N_c}} \sum_{\tilde{k}\in\{0,\,1\}^{n_c}} \omega_{N_c}^{-k_{(10)}\tilde{k}_{(10)}}\ket{\tilde{k}}_c.
\end{equation}
Consequently
\begin{equation}\label{iQFT-Phi2}
\hat{U}_{QFT}^{\dagger}\ket{\upsilon_j}_c = 
\frac{1}{{N_c}}
\sum_{k\in\{0,\,1\}^{n_c}}
\sum_{\tilde{k}\in\{0,\,1\}^{n_c}}
e^{i2\pi k_{(10)}\left[\phi_j-\tilde{k}_{(10)}/N_c\right]}
\ket{\tilde{k}}_c.
\end{equation}
Using the last expression into Eq.~(\ref{Psi4}) and swapping $k$ with $\tilde{k}$ yield
\begin{equation}\label{Psi4b}
\ket{\Psi_4} =
\sum_{j\in\{0,\,1\}^n}
\left[
b_j^{\Gamma}
\ket{\phi_j^{\Gamma}}\otimes
\left(
\frac{1}{{N_c}}
\sum_{k\in\{0,\,1\}^{n_c}}
\sum_{\tilde{k}\in\{0,\,1\}^{n_c}}
e^{i2\pi \tilde{k}_{(10)}\left[\phi_j-{k}_{(10)}/N_c\right]}
\ket{{k}}_c
\right)\right]\otimes\ket{0}_a.
\end{equation}
The last expression can be reformulated as
\begin{equation}\label{Psi4c}
\ket{\Psi_4} =
\sum_{j\in\{0,\,1\}^n}
\left[
b_j^{\Gamma}
\ket{\phi_j^{\Gamma}}\otimes
\left(
\sum_{k\in\{0,\,1\}^{n_c}}
\alpha^{QPE}_{jk}
\ket{{k}}_c
\right)\right]\otimes\ket{0}_a,
\end{equation}
%
%
where the complex amplitude matrix $\alpha^{QPE}_{jk}$ is defined as
\begin{equation}\label{alpha-QPE_jk}
\alpha^{QPE}_{jk}:=
\frac{1}{{N_c}}
\sum_{\tilde{k}\in\{0,\,1\}^{n_c}}
e^{i2\pi \tilde{k}_{(10)}\left[\phi_j-{k}_{(10)}/N_c\right]} =
\frac{1}{{N_c}}
\sum_{\tilde{k}\in\{0,\,1\}^{n_c}}
e^{i2\pi \tilde{k}_{(10)}\left[\lambda_j^\Gamma(N_c-1)-{k}_{(10)}\right]/N_c},
\end{equation}
which used the definition given by Eq.~(\ref{phi}). The last definition can be used to make more compact Eq.~(\ref{iQFT-Phi2}), namely
\begin{equation}\label{iQFT-Phi2-norm}
\hat{U}_{QFT}^{\dagger}\ket{\upsilon_j}_c = 
\sum_{k\in\{0,\,1\}^{n_c}}
\alpha^{QPE}_{jk}\,
\ket{{k}}_c.
\end{equation}
Because $\ket{\upsilon_j}_c$ is clearly normalized and the operator $\hat{U}_{QFT}^{\dagger}$ is unitary, then
\begin{equation}\label{iQFT-Phi2-norm2}
\sum_{k\in\{0,\,1\}^{n_c}}
\left|
\alpha^{QPE}_{jk}\,
\right|^2=1,
\end{equation}
which will be useful in the following. It is worth the understand better the term $\overline{m}_{jk} = \lambda_j^\Gamma(N_c-1)-{k}_{(10)}$ in the complex amplitude matrix. Eq.~(\ref{Gamma-eigenvalues-limits}) ensures that $0<\lambda_j^\Gamma\leq 1$ for all eigenvalues. Consequently $0<(N_c-1)\,\lambda_j^{\Gamma}\leq (N_c-1)$, which means that $(N_c-1)\,\lambda_j^{\Gamma}$ is included in the same numerical interval spanned by all the states of the clock register with $n_c$ qubits. With other words, given $k\in\{0,\,1\}^{n_c}$, then $k_{(10)}\in[0,\,N_c-1]$ and also $(N_c-1)\,\lambda_j^{\Gamma} \in [0,\,N_c-1]$. The difference is that $k_{(10)}$ corresponds to the binary integer $k$, while $(N_c-1)\,\lambda_j^{\Gamma}$ has no match with an integer in general. Consequently $\overline{m}_{jk}$ may be close to an integer, but it is not an integer in general. Eq.~(\ref{alpha-QPE_jk}) can be simplified by realizing that it involves a finite geometric series, namely
\begin{equation}\label{alpha_jk3}
\alpha^{QPE}_{jk} =
e^{if_{jk}} 
\frac{\sin(\pi\,\overline{m}_{jk})}{N_c\,\sin(\pi\,\overline{m}_{jk}/N_c)},
\end{equation}
where $f_{jk} = \pi(N_c-1)\,\overline{m}_{jk}/N_c$. Before passing to the following module, it is interesting to realize that also the quantum state given by Eq.~(\ref{Psi4c}) is entangled, because the clock register depends on the input register by the complex matrix $\alpha^{QPE}_{jk}$.

\begin{figure}[htbp]
    \centering
    \includegraphics[width=1\linewidth]{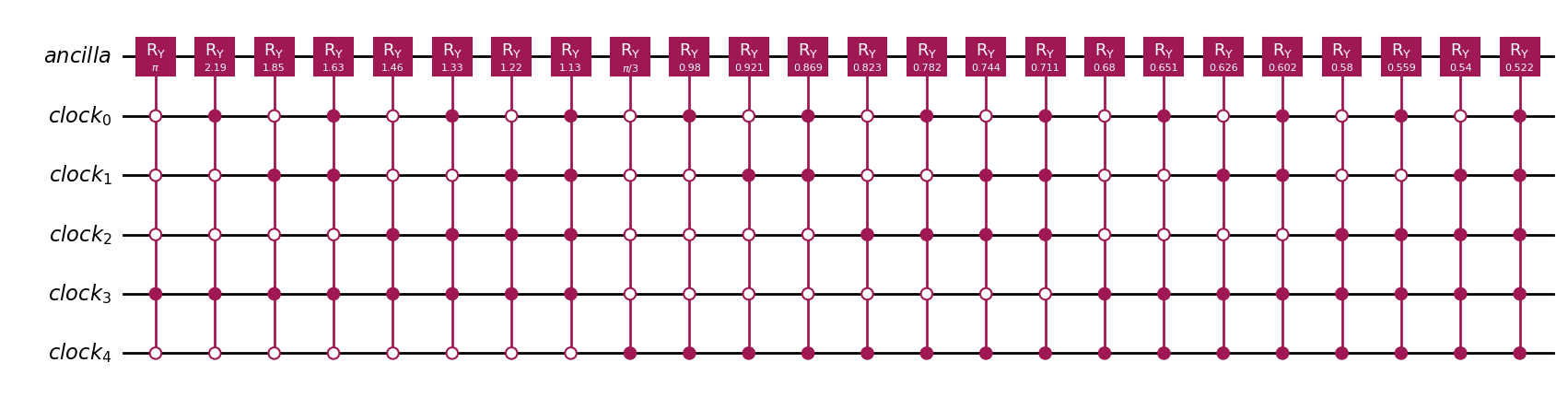}
    \caption{Circuit implementing controlled rotation on the ancilla qubit in the HHL algorithm with $n_c=5$ clock qubits. As it is clear from the circuit, $k_{(10)}^{\min}=8$, which corresponds to $k^{\min} = 1000$ and is the controller condition of the first rotation block on the left. This circuit shows what is inside the block called ``CR\_y-for-inversion" in Fig. (\ref{fig:circuit-hhl-3q5c}).   
    }\label{fig:circuit-hhl-rotation-3q5c}
\end{figure}

\subsection{Binarized inversion module}
\label{inv}

The second essential component of the HHL algorithm is the binarized inversion module. There are many advanced and efficient quantum circuits for inverting eigenvalues, e.g. see Ref. \citep{Wang2020}. Here we focus on a very simple straightforward implementation.

\begin{itemize}
    \item Step \#5
\end{itemize}

Before defining the operator for the inversion, let us estimate the minimum binarized eigenvalue which we have to invert. The binary clock register ${k}_{(10)}$ that we have to invert is very close to $\lambda_j^\Gamma(N_c-1)$. Taking into account Eq.~(\ref{Gamma-eigenvalues-limits}) yields
\begin{equation}\label{eigenvalue-approx2}
\frac{N_c-1}{1+4\,r} \leq 
\lambda_j^{\Gamma}\,(N_c-1) \leq (N_c-1).
\end{equation}
In order to take into account the round-off error and the measurement probability scattering, we can assume that the minimum binarized eigenvalue to be inverted can be expressed as
\begin{equation}\label{min-binary-eigen}
k_{(10)}^{\min} = \left\lfloor 
\frac{N_c-1}{1+4\,r}
\right\rfloor
-\Delta N_\epsilon,
\end{equation}
where $\Delta N_\epsilon$ is a tolerance integer (e.g. in Fig. (\ref{fig:circuit-hhl-rotation-3q5c}) we assumed $\Delta N_\epsilon = 2$ over $N_c = 2^5 = 32$). Knowing that $(k_j^\lambda)_{(10)} > k_{(10)}^{\min}$ allows to save some operations and to increase the sampling rate (as it will be clarified in the following). Therefore let us define the controlled rotation operator for inverting the eigenvalues as
\begin{equation}\label{C-Ry}
CR_Y = I \otimes 
\sum_{k\in\{0,\,1\}^{n_c}\geq k^{\min}} 
\ket{k}_c\bra{k}_c 
\otimes R_Y \left[2\arcsin(k_{(10)}^{\min}/k_{(10)})\right].
\end{equation}
This operator is shown in Fig. (\ref{fig:circuit-hhl-rotation-3q5c}) for the case with $n_c=5$ clock qubits (plus the ancilla qubit). As it is clear from the circuit in Fig. (\ref{fig:circuit-hhl-rotation-3q5c}), in this case $k_{(10)}^{\min}=8$, which corresponds to $k^{\min} = 1000$ and is the controller condition of the first rotation block on the left. Starting with $k_{(10)}^{\min}=8$ allows to save seven controlled rotations ($k_{(10)}>0$ for avoiding division by zero). In top of the saving, it is important to understand how this operator works. The portion $\ket{k}_c\bra{k}_c$ is the projection operator which selects in the clock register when the state $\ket{k}_c$ happens. If the state $\ket{k}_c$ happens, then the ancilla qubit is rotated by a angle which is 
\begin{equation}\label{conntrolled-rotation}
\theta_Y = 2\arcsin\left[\frac{k_{(10)}^{\min}}{k_{(10)}}\right] = 2\arcsin\left(\alpha_k\right),
\end{equation}
where $\alpha_k = k_{(10)}^{\min}/k_{(10)} < 1$. Clearly if $k = k^{\min}$, then $\theta_Y = \pi$, which is the leftmost rotation in Fig. (\ref{fig:circuit-hhl-rotation-3q5c}). Taking into account the definition given by Eq.~(\ref{RY}), applying the operator $CR_Y$ to the previous system quantum state yields
%
%
\begin{equation}\label{Psi5-adv}
\ket{\Psi_5} := CR_Y\ket{\Psi_4} =
\sum_{j\in\{0,\,1\}^n}
b_j^{\Gamma}
\ket{\phi_j^{\Gamma}}\otimes
\left[
\sum_{k\in\{0,\,1\}^{n_c}\geq k^{\min}}
\alpha^{QPE}_{jk}
\ket{{k}}_c
\otimes
\left(
\sqrt{1-\alpha_k^2}\,\ket{0}_a
+\alpha_k\,\ket{1}_a
\right)\right].
\end{equation}
In the previous formula, the inverse eigenvalues implicitly appear but we need the following step for isolating this result.

\begin{itemize}
    \item Step \#6
\end{itemize}

%
%

The key idea here is to measure the ancilla qubit, to discard the calculation if the result is $\ket{0}$ and to keep it for statistics if it is $\ket{1}$. Following the Born rule, quantum mechanics tells us that, after a measurement, the post-measurement state $\ket{\Psi_6}$ must be normalized by dividing by the square root of the probability of the observed outcome, namely 
%
%
\begin{equation}\label{Psi6-adv}
\ket{\Psi_6} :=
\frac{1}{\sqrt{N_M}}
\sum_{j\in\{0,\,1\}^n}
\left[
b_j^{\Gamma}\,\ket{\phi_j^{\Gamma}}\otimes
\sum_{k\in\{0,\,1\}^{n_c}\geq k^{\min}}
\left(
\alpha^{QPE}_{jk}
\alpha_k\,
\ket{{k}}_c
\right)\right]
\otimes
\ket{1}_a
\end{equation}
It is interesting to note that the measurement of the ancilla transferred the inverse of the eigenvalues from the ancilla qubit to the clock register. The normalization factor $N_M$ in the previous formula is required to ensure that $\ket{\Psi_6}$ is properly normalized (see next).

The previous state seems close to solving the problem, but the clock register is still entangled with the input register. Moreover the inverse eigenvalues in the clock register are unknown. The inverse of the operation that originally entangled them is required at this point, as discussed in the following section.

\begin{figure}[htbp]
    \centering
    \includegraphics[width=0.8\linewidth]{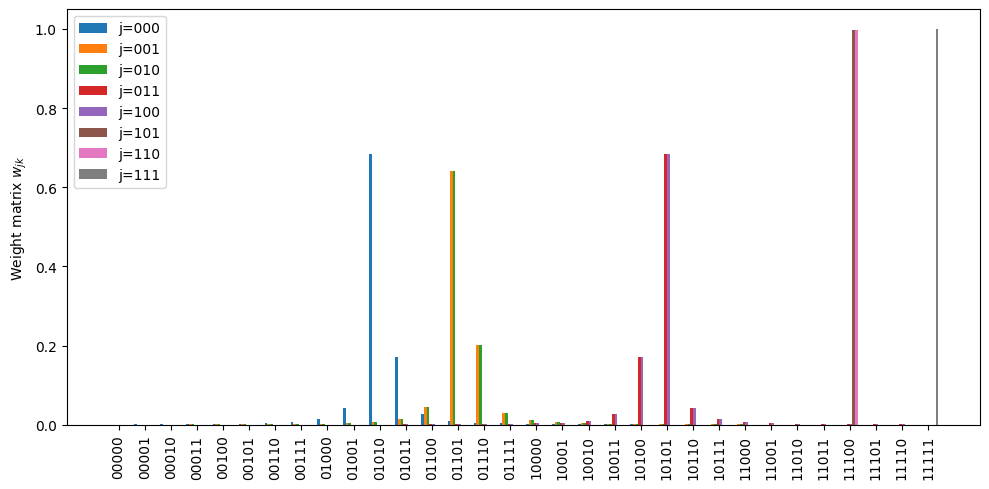}
    \caption{Normalized weights $w_{jk}$, where $\sum_{k} w_{jk} = 1$ for every $j$-th component of the spectral decomposition. It is worth noting that the binarization process implies $(N_c-1)\,\lambda_j^{\Gamma} \approx (k_j^\lambda)_{(10)}$, which introduces some spread in the spectral distribution of the binarized eigenvalues. For example, $(N_c-1)\,\lambda_0^{\Gamma}=10.33$, which falls in between $k=01010$ and $k'=01011$.
    }\label{fig:circuit-hhl-binarized-weights-3q5c}
\end{figure}

\subsection{Inverse QPE}
\label{hhl-all}

As it is clear from the previous steps, the clock register has become an eigenvalue register. To obtain the “solution’’ in the input/output register, one must quantum-erase (uncompute) the eigenvalue register, i.e., apply exactly the inverse of the operation that originally encoded it, namely the QPE.

\begin{itemize}
    \item Step \#7, \#8 and \#9 (for consistency with Ref. \citep{hhl_tutorial})
\end{itemize}

First of all, let us collect in one single unitary operation $\hat{U}_{QPE}$ some previous steps, i.e. Hadamard transformation (step \#2), Hamiltonian evolution (step \#3) and inverse QFT (step \#4), namely
\begin{equation}\label{U_QPE}
\hat{U}_{QPE} =
\left(I\otimes \hat{U}_{QFT}^\dagger \otimes I_a\right)
\,
CE_V
\,
\left(I\otimes H^{\otimes n_c} \otimes I_a\right).
\end{equation}
The logical order goes from right to left: step \#2 implies $\left(I\otimes H^{\otimes n_c} \otimes I_a\right)\ket{\Psi_1} = \ket{\Psi_2}$; step \#3 implies $CE_V\ket{\Psi_2} = \ket{\Psi_3}$ and, finally, step \#4 implies $\left(I\otimes \hat{U}_{QFT}^\dagger \otimes I_a\right)\ket{\Psi_3} = \ket{\Psi_4}$. Using the newly defined operator, these three steps, which are essential for the QPE algorithm, can be summarized as  $\hat{U}_{QPE}\ket{\Psi_1} = \ket{\Psi_4}$. The binarized inversion module allowed to compute $\ket{\Psi_6}$. Hence the already-mentioned quantum uncomputation can be formally realized by the inverse operator $\hat{U}_{QPE}^\dagger$, namely
\begin{equation}\label{Psi9-adv}
\ket{\Psi_9} := \hat{U}_{QPE}^\dagger\ket{\Psi_6} =
\frac{1}{\sqrt{N_M}}
\sum_{j\in\{0,\,1\}^n}
b_j^{\Gamma}\,
\sum_{k\in\{0,\,1\}^{n_c}\geq k^{\min}}
\alpha^{QPE}_{jk}
\alpha_k\,
\hat{U}_{QPE}^\dagger
\ket{\phi_j^{\Gamma}}
\ket{{k}}_c
\ket{1}_a.
\end{equation}
Although some simplifications are possible at the cost of accuracy, the optimal HHL algorithm requires an additional projection onto the ground state of the clock register, i.e. $\ket{0}_c^{\otimes n_c}$, which can be implemented through an appropriate measurement. For the generic $j$-th mesh node the probability of the corresponding state can be computed by the following projection, which can be conveniently written as  
\begin{equation}\label{clock-ground}
a_j = b_j^{\Gamma}\,g(\lambda_j^\Gamma) 
=
\bra{\phi_j^{\Gamma}}
\bra{0}_c^{\otimes n_c}\bra{1}_a
\ket{\Psi_9},
\end{equation}
where $g(\lambda_j^\Gamma)$ is sometimes called ``filter" and is a proper function which will be clarified in the following. Equivalently, we can express it as 
\begin{equation}\label{clock-ground2}
a_j = b_j^{\Gamma}\,g(\lambda_j^\Gamma) =
\sum_{j'\in\{0,\,1\}^n}
b_{j'}^{\Gamma}\,
\sum_{k\in\{0,\,1\}^{n_c}\geq k^{\min}}
\alpha^{QPE}_{j'k}
\alpha_k\,
\bra{\phi_j^{\Gamma}}
\bra{0}_c^{\otimes n_c}\bra{1}_a
\hat{U}_{QPE}^\dagger
\ket{\phi_{j'}^{\Gamma}}
\ket{{k}}_c
\ket{1}_a.
\end{equation}
Please note that, in order to avoid confusion with $j$ of the amplitude $a_j$ we are searching for, we introduced the index $j'$ in the summation. In order to compute the quantum measurement in the previous expression, let us take into account the following property
\begin{equation}\label{proof}
\bra{\phi_j^{\Gamma}}
\bra{0}_c^{\otimes n_c}\bra{1}_a
\hat{U}_{QPE}^\dagger
\ket{\phi_{j'}^{\Gamma}}
\ket{{k}}_c
\ket{1}_a=
\left(
\bra{\phi_{j'}^{\Gamma}}
\bra{k}_c\bra{1}_a
\hat{U}_{QPE}
\ket{\phi_{j}^{\Gamma}}
\ket{0}_c^{\otimes n_c}
\ket{1}_a
\right)^*,
\end{equation}
where the superscript $^*$ indicates the complex conjugate, as usual. The result $\hat{U}_{QPE} 
\ket{\phi_j^{\Gamma}}
\ket{0}_c^{\otimes n_c}\ket{1}_a$ can be computed as a particular case of Eq.~(\ref{Psi9-adv}), i.e. assuming $\ket{\Psi_9}=\ket{\phi_j^{\Gamma}}$, namely
\begin{equation}\label{U_QPE-eigenvector}
\hat{U}_{QPE} 
\ket{\phi_j^{\Gamma}}
\ket{0}_c^{\otimes n_c}\ket{1}_a
= 
\ket{\phi_j^{\Gamma}}\otimes
\sum_{k'\in\{0,\,1\}^{n_c}}
\left(
\alpha^{QPE}_{jk'}\,
\ket{{k'}}_c
\right)
\otimes\ket{1}_a,
\end{equation}
where again index $k'$ is introduced for avoiding confusion. Applying the result given by Eq.~(\ref{U_QPE-eigenvector}) into Eq.~(\ref{proof}) yields
\begin{equation}\label{proof-result}
\bra{\phi_j^{\Gamma}}
\bra{0}_c^{\otimes n_c}\bra{1}_a
\hat{U}_{QPE}^\dagger
\ket{\phi_{j'}^{\Gamma}}
\ket{{k}}_c
\ket{1}_a = 
\delta_{j'j}
\left(
\alpha^{QPE}_{jk}
\right)^*.
\end{equation}
Applying the last expression into Eq.~(\ref{clock-ground2}) yields
\begin{equation}\label{clock-ground3}
a_j = b_j^{\Gamma}\,g(\lambda_j^\Gamma) =
\sum_{j'\in\{0,\,1\}^n}
b_{j'}^{\Gamma}\,
\sum_{k\in\{0,\,1\}^{n_c}\geq k^{\min}}
\alpha^{QPE}_{j'k}
\alpha_k\,
\delta_{j'j}
\left(
\alpha^{QPE}_{jk}
\right)^*,
\end{equation}
or equivalently
\begin{equation}\label{clock-ground4}
g(\lambda_j^\Gamma) = a_j/b_j^\Gamma = 
\sum_{k\in\{0,\,1\}^{n_c}\geq k^{\min}}
w_{jk}
\,\alpha_k,
\end{equation}
where
\begin{equation}\label{weight}
w_{jk} = \left|\alpha^{QPE}_{jk}\right|^2 = \left|\frac{\sin(\pi\,\overline{m}_{jk})}{N_c\,\sin(\pi\,\overline{m}_{jk}/N_c)}\right|^2.
\end{equation}
Taking into into account Eq.~(\ref{iQFT-Phi2-norm2}), it is possible to realize that the previous terms are actually normalized weights, namely
\begin{equation}\label{iQFT-Phi2-norm3}
\sum_{k\in\{0,\,1\}^{n_c}}
w_{jk} = 1,
\end{equation}
for every $j$-th component. See also Fig. \ref{fig:circuit-hhl-binarized-weights-3q5c} for an example. After computing the amplitudes $a_j = b_j^{\Gamma}\,g(\lambda_j^\Gamma)$, we can use them to normalize the final state as
\begin{equation}\label{Psi9-adv2}
\ket{\Psi_9} =
\frac{1}{\sqrt{\sum_j \left|b_j^{\Gamma}\,g(\lambda_j^\Gamma)\right|^2}}
\sum_{j\in\{0,\,1\}^n}
b_j^{\Gamma}\,g(\lambda_j^\Gamma)
\ket{\phi_j^{\Gamma}}
\ket{0}_c^{\otimes n_c}
\ket{1}_a.
\end{equation}
In the previous final state, for the generic $j$-th component of the input decomposition, we can define the actual HHL eigenvalue as
\begin{equation}\label{HHL-eigenvalues}
\frac{1}{\lambda_j^{HHL}} := 
\frac{(N_c-1)}{k_{(10)}^{min}}\,
g(\lambda_j^\Gamma) =  
\frac{(N_c-1)}{k_{(10)}^{min}}\,
\sum_{k\in\{0,\,1\}^{n_c}\geq k^{\min}}
w_{jk}\,\alpha_k = 
\sum_{k\in\{0,\,1\}^{n_c}\geq k^{\min}}
w_{jk}
\frac{(N_c-1)}{k_{(10)}}.
\end{equation}
The HHL eigenvalues can be used to define the coefficients of the decomposition of the HHL solution, which appear in the pre-factors of the final quantum state, namely
\begin{equation}\label{x-def5}
\frac{b_j^{\Gamma}\,g(\lambda_j^\Gamma)}{\sqrt{\sum_{j} \left|b_j^{\Gamma}\,g(\lambda_j^\Gamma)\right|^2}} = \frac{b_j^\Gamma/\lambda_j^{HHL}}{\sqrt{\sum_{j} \left|b_j^\Gamma/\lambda_j^{HHL}\right|^2}} = x_j^{HHL}.
\end{equation}
Introducing these coefficients in Eq.~(\ref{Psi9-adv2}) yields
\begin{equation}\label{Psi9-adv3}
\ket{\Psi_9} =
\sum_{j\in\{0,\,1\}^n}
x_j^{HHL}\,
\ket{\phi_j^{\Gamma}}
\ket{0}_c^{\otimes n_c}
\ket{1}_a = 
\ket{x^{HHL}}\,
\ket{0}_c^{\otimes n_c}
\ket{1}_a,
\end{equation}
where the state $\ket{x^{HHL}}$ is the quantum state produced by the HHL algorithm. In the next section, we will use a simplified derivation of the HHL algorithm to show that ${\lambda_j^{HHL}} \approx {\lambda_j^{\Gamma}}$ and consequently that $\ket{x^{HHL}}\approx \ket{x}$.

\begin{figure}[htbp]
    \centering
    \includegraphics[width=0.8\linewidth]{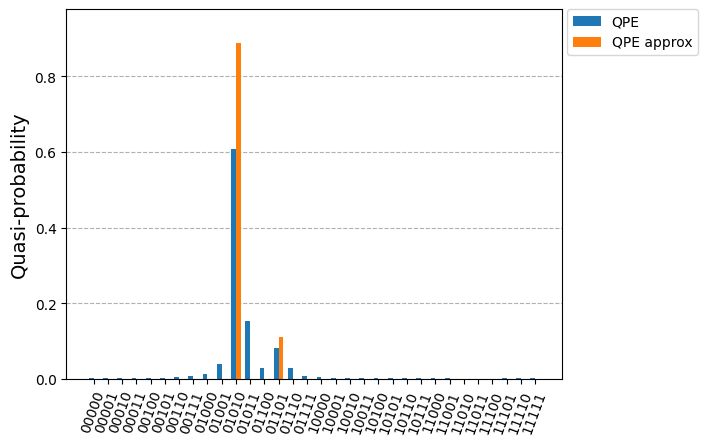}
    \caption{Probability distribution of the binarized approximations of the eigenvalues of the problem $\ket{k_j^\lambda}$ stored in the clock register, after completing QPE. Accessing this information is only possible in ideal (statevector) simulations (otherwise, the wave function would collapse in real computations). The two peaks correspond to (i) eigenvalue $k_0^\lambda = 01010$ or equivalently $(k_0^\lambda)_{(10)}=10$, and to (ii) eigenvalue $k_1^\lambda = 01101$ or equivalently $(k_1^\lambda)_{(10)}=13$. It is worth noting that the binarization process implies $(k_j^\lambda)_{(10)} \approx (N_c-1)\,\lambda_j^{\Gamma}$, which introduces some spread in the spectral distribution of the binarized eigenvalues (see Fig. \ref{fig:circuit-hhl-binarized-weights-3q5c}).
    }\label{fig:circuit-hhl-binarized-eigenvalues-3q5c}
\end{figure}

\begin{figure}[htbp]
    \centering
    \includegraphics[width=0.8\linewidth]{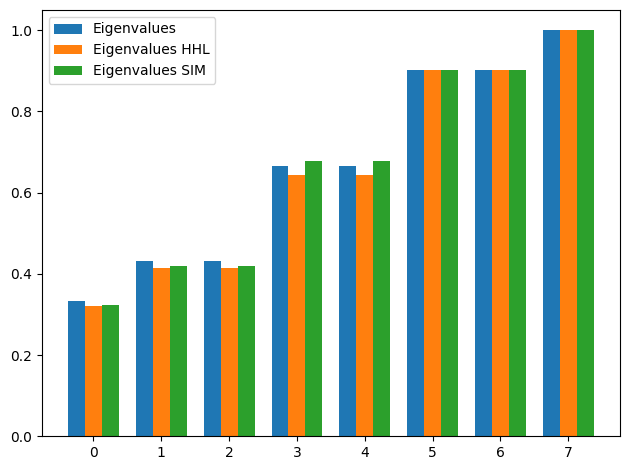}
    \caption{Comparison among eigenvalues: (i) original ones for the considered example; (ii) those approximated by the HHL algorithm; (iii) those computed by mimicking the HHL algorithm, but assuming the approximation given by Eq.~(\ref{hypothesis}). The abscissa represents the index labels ranging from $0$ to $N-1$, corresponding to the normalized eigenvalues of the matrix $\hat{\Gamma}$ given by Eq. (\ref{Gamma-def}) in increasing order. 
    By design, the maximum normalized eigenvalue is equal to one and it is exactly represented by the binary integer $\{1\}^{n_c}$. The approximated eigenvalues reported with label $5$ and $6$ differ from the exact values by less than $0.1\,\%$ and hence the discrepancies are not visible.
    }\label{fig:circuit-hhl-all-eigenvalues-3q5c}
\end{figure}

\subsection{Simplified derivation of eigenvalue inversion}
\label{simplified-inv}

In the previous sections, we have already pointed out that the term $\overline{m}_{jk} = \lambda_j^\Gamma(N_c-1)-{k}_{(10)}$ is not necessarily an integer in general. Eq.~(\ref{Gamma-eigenvalues-limits}) ensures that $0<\lambda_j^\Gamma\leq 1$ for all eigenvalues. Consequently $0<(N_c-1)\,\lambda_j^{\Gamma}\leq (N_c-1)$, which means that $(N_c-1)\,\lambda_j^{\Gamma}$ is included in the same numerical interval $[0,\,N_c-1]$, as it happens for $k_{(10)}$. The difference is that $k_{(10)}$ corresponds to the binary integer $k$, while $(N_c-1)\,\lambda_j^{\Gamma}$ has no match with an integer in general. Let us relax here the last condition, namely  
\begin{eqnarray}\label{hypothesis}
(N_c-1)\,\lambda_j^{\Gamma} \approx (k_j^\lambda)_{(10)},
\end{eqnarray}
where $k_j^\lambda\in \mathbb{Z}$ is a proper integer. In this case, the term $\overline{m}_{jk} \approx (k_j^\lambda)_{(10)}-{k}_{(10)} = m_{jk} \in \mathbb{Z}$ and consequently $\alpha^{QPE}_{jk} \approx \delta_{k\,k_j^\lambda}$, or equivalently $w_{jk} \approx \delta_{k\,k_j^\lambda}$. It is interesting to see the impact of the approximation given by Eq.~(\ref{hypothesis}) in the derivation of the previous sections. In particular, substituting it in Eq.~(\ref{iQFT-Phi2}) yields
\begin{equation}\label{iQFT-Phi3}
\hat{U}_{QFT}^{\dagger}\ket{\upsilon_j}_c \approx 
\frac{1}{{N_c}}
\sum_{k\in\{0,\,1\}^{n_c}}
\sum_{\tilde{k}\in\{0,\,1\}^{n_c}}
e^{i2\pi k_{(10)}\left(k_j^{\lambda}-\tilde{k}\right)_{(10)}/N_c}
\ket{\tilde{k}}_c.
\end{equation}
Because now $k_j^{\lambda}$ are binary integers, only the terms with $\tilde{k} = k_j^{\lambda}$ is non-zero, which simplifies the previous expression as
\begin{equation}\label{iQFT-Phi4}
\hat{U}_{QFT}^{\dagger}\ket{\upsilon_j}_c \approx 
\frac{1}{{N_c}}
\sum_{k\in\{0,\,1\}^{n_c}}
e^{0}
\ket{k_j^{\lambda}}_c = \ket{k_j^{\lambda}}_c,
\end{equation}
which is perfectly consistent with $\alpha^{QPE}_{jk} \approx \delta_{k\,k_j^\lambda}$ and Eq.~(\ref{iQFT-Phi2-norm}). Applying the last result to the generic quantum state given by Eq.~(\ref{Psi4}) yields
\begin{equation}\label{Psi4-approx}
\ket{\Psi_4} \approx \ket{\Psi_4'} :=
\sum_{j\in\{0,\,1\}^n} 
\left(
b_j^{\Gamma}
\ket{\phi_j^{\Gamma}}\otimes
\ket{k_j^\lambda}_c
\right)\otimes\ket{0}_a.
\end{equation}
In ideal (statevector) simulations (otherwise, the wave function would collapse in real computations), it possible to compute the probability distribution of the binarized approximations of the eigenvalues of the problem $\ket{k_j^\lambda}$ stored in the clock register.
See an example in Fig. (\ref{fig:circuit-hhl-binarized-eigenvalues-3q5c}). It is worth noting that the binarization process implies $(k_j^\lambda)_{(10)} \approx (N_c-1)\,\lambda_j^{\Gamma}$, which introduces some spread in the spectral distribution of the binarized eigenvalues (see Fig. \ref{fig:circuit-hhl-binarized-weights-3q5c}). 
However, in some cases, this spread may even result being beneficial for the accuracy of the numerical solution produced by the HHL algorithm.

Proceeding as described in the previous sections for inverting the binarized eigenvalues and taking advantage of the assumption given by $\alpha^{QPE}_{jk} \approx \delta_{k\,k_j^\lambda}$ in Eq.~(\ref{Psi6-adv}) yield
\begin{equation}\label{Psi6-approx}
\ket{\Psi_6} \approx \ket{\Psi_6'} :=
\frac{1}{\sqrt{N_M}}
\sum_{j\in\{0,\,1\}^n}
\alpha_j
b_j^{\Gamma}\,
\ket{\phi_j^{\Gamma}}
\ket{k_j^\lambda}_c
\ket{1}_a,
\end{equation}
where $\alpha_j = k_{(10)}^{\min}/(k_j^\lambda)_{(10)}$. 
In order to derive $N_M$, it is possible to focus on the input register and the clock register only. 
Using (i) the tensor-product property on inner products and (ii) the orthogonality of the eigenvectors yields
\begin{eqnarray}\label{Psi6-normalization}
N_M &=& \sum_{j\in\{0,\,1\}^n} 
\alpha_j\,b_j^{\Gamma}
\bra{\phi_j^{\Gamma}}
\bra{k_j^\lambda}_c
\sum_{j'\in\{0,\,1\}^n} 
\alpha_{j'}\,b_{j'}^{\Gamma}
\ket{\phi_{j'}^{\Gamma}}
\ket{k_{j'}^\lambda}_c\nonumber\\
&=&\sum_{j,j'\in\{0,\,1\}^n} 
\alpha_j\,b_j^{\Gamma}
\alpha_{j'}\,b_{j'}^{\Gamma}
\bra{\phi_j^{\Gamma}}
\bra{k_j^\lambda}_c
\ket{\phi_{j'}^{\Gamma}}
\ket{k_{j'}^\lambda}_c\nonumber\\
&=&\sum_{j,j'\in\{0,\,1\}^n} 
\alpha_j\,b_j^{\Gamma}
\alpha_{j'}\,b_{j'}^{\Gamma}
\braket{\phi_j^{\Gamma}|\phi_{j'}^{\Gamma}}
\braket{k_{j}^\lambda|k_{j'}^\lambda}
=\sum_{j,j'\in\{0,\,1\}^n} 
\alpha_j\,b_j^{\Gamma}
\alpha_{j'}\,b_{j'}^{\Gamma}
\,\delta_{jj'}
\braket{k_{j}^\lambda|k_{j'}^\lambda}\nonumber\\
&=&\sum_{j\in\{0,\,1\}^n} 
|\alpha_j\,b_j^{\Gamma}|^2
\braket{k_{j}^\lambda|k_{j}^\lambda}=
\sum_{j\in\{0,\,1\}^n}
|\alpha_j\,b_j^{\Gamma}|^2.
\end{eqnarray}
The previous expression completes the normalization factor appearing in Eq.~(\ref{Psi6-approx}). As discussed in the previous sections, we now need to project the state $\ket{\Psi_6'}$ onto the ground state of the clock register, namely
\begin{equation}\label{Psi9-approx}
\ket{\Psi_9} \approx \ket{\Psi_9'} := \hat{U}_{QPE}^\dagger\ket{\Psi_6'} =
\frac{1}{\sqrt{\sum_j \left|\alpha_j\,b_j^{\Gamma}\right|^2}}
\sum_{j\in\{0,\,1\}^n}
\alpha_j\,b_j^{\Gamma}\,
\ket{\phi_j^{\Gamma}}
\ket{0}_c^{\otimes n_c}
\ket{1}_a.
\end{equation}
Comparing Eq.~(\ref{Psi9-adv2}) and Eq.~(\ref{Psi9-approx}), it is clear that
\begin{equation}\label{clock-ground5-approx}
g(\lambda_j^\Gamma) \approx
\alpha_j = k_{(10)}^{\min}/(k_j^\lambda)_{(10)},
\end{equation}
which is consistent with the assumption $w_{jk} \approx \delta_{k\,k_j^\lambda}$. 
In the previous final state, for the generic $j$-th component of the input decomposition, we can define the generic simplified eigenvalue as
\begin{equation}\label{SIM-eigenvalues}
\frac{1}{\lambda_j^{SIM}} :=   
\frac{(N_c-1)}{(k_j^\lambda)_{(10)}}.
\end{equation}
The simplified eigenvalues can be used in the coefficients of the decomposition of the solution, which appear in the pre-factors of the final quantum state, namely 
\begin{equation}\label{x-def4}
\frac{\alpha_j\,b_j^\Gamma}{\sqrt{\sum_{j} \left|\alpha_j\,b_j^\Gamma\right|^2}} = \frac{b_j^\Gamma/(k_j^\lambda)_{(10)}}{\sqrt{\sum_{j} \left|b_j^\Gamma/(k_j^\lambda)_{(10)}\right|^2}} = 
\frac{1}{\sqrt{\sum_{j} \left|{b_j^\Gamma}/{\lambda_j^{SIM}}\right|^2}}\,\frac{b_j^\Gamma}{\lambda_j^{SIM}} = x_j^{SIM}.
\end{equation}
Substituting the previous result in Eq.~(\ref{Psi9-approx}) yields
\begin{equation}\label{Psi6-approx2}
\ket{\Psi_9'} =
\sum_{j\in\{0,\,1\}^n} 
x_j^{SIM}\,
\ket{\phi_j^{\Gamma}}
\ket{0}_c^{\otimes n_c}
\ket{1}_a =
\ket{x^{SIM}}\otimes
\ket{0}_c^{\otimes n_c}
\otimes
\ket{1}_a.
\end{equation}
Substituting the fundamental hypothesis given by Eq.~(\ref{hypothesis}) into Eq.~(\ref{SIM-eigenvalues}) yields
\begin{equation}\label{SIM-eigenvalues2}
\frac{1}{\lambda_j^{SIM}} =   
\frac{(N_c-1)}{(k_j^\lambda)_{(10)}} \approx
\frac{1}{\lambda_j^{\Gamma}}.
\end{equation}

\begin{figure}[htbp]
    \centering
    \includegraphics[width=0.7\linewidth]{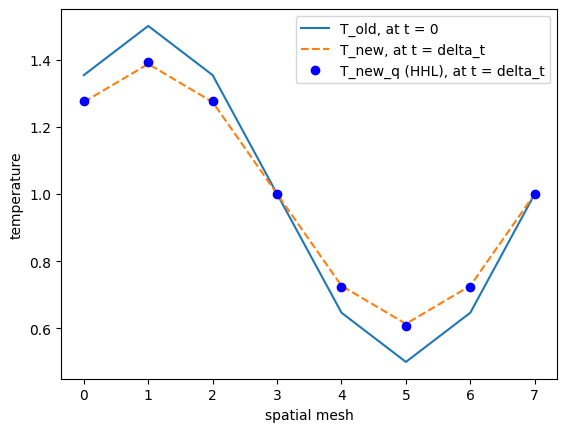}
    \caption{One time-step update of the temperature profile according to heat conduction equation by quantum computing ($3$ qubits) by HHL algorithm. The blue line is the initial temperature profile (with mean equal to $1$), the orange dashed line is the new temperature profile at time $\Delta t$, computed by finite-difference method. The blue dots are the mesh node temperatures computed by the HHL algorithm (ideal statevector simulator). The circuit implementing HHL algorithm with $n=3$ input qubits, $n_c=5$ clock qubits and $1$ ancilla qubit is shown in Fig. \ref{fig:circuit-hhl-3q5c}. In particular, the binarized inversion module with $n_c=5$ qubits in the clock register is shown in Fig. \ref{fig:circuit-hhl-rotation-3q5c}.}
    \label{conduction-3qubits-HHL}
\end{figure}

Interestingly, the simplified derivation of the HHL algorithm discussed in this section also provides an intuitive explanation of why the algorithm yields a good approximation of the solution to the original problem. 
In fact, QPE makes it possible to extract binarized distributions of the original eigenvalues, as shown in Fig. \ref{fig:circuit-hhl-binarized-eigenvalues-3q5c}. 
These distributions can then be used to compute an approximate estimate of the eigenvalues of the original problem, as illustrated in Fig. \ref{fig:circuit-hhl-all-eigenvalues-3q5c}. 
Therefore, it becomes clear that ${\lambda_j^{HHL}} \approx \lambda_j^{SIM} \approx {\lambda_j^{\Gamma}}$. 
Consequently, the HHL algorithm provides an approximate solution to the original problem, namely $\ket{x^{HHL}} \approx \ket{x^{SIM}} \approx \ket{x}$, as shown in Fig. \ref{conduction-3qubits-HHL} for the considered case. 
However, there is an important difference between the HHL algorithm and the VQE algorithm. The key idea behind HHL is to retain only those quantum states for which the ancilla qubit is in the state $\ket{1}_a$ and, simultaneously, the clock register is in the ground state, i.e. $\ket{0}_c^{\otimes n_c}$. 
In the present example, the states satisfying $\ket{0}_c^{\otimes n_c}\ket{1}_a$ account for only $55.5\,\%$ of the total number of states. 
This implies that, statistically, about half of the measurement outcomes in actual shot-based simulations must be discarded when implementing the HHL algorithm. 
This is generally not considered a significant issue once quantum computers reach full maturity.

\section{Conclusions}
\label{conclusions}

At the current stage of technological development, predicting the potential impact of quantum computing on Thermal Science remains extremely challenging, as it depends on future advancements. As a paradigmatic case, we focused on solving the heat conduction equation, with the starting point being the development of algorithms that leverage quantum computing most effectively for this application.

In these notes, we began by analyzing the Variational Quantum Eigensolver (VQE) algorithm in section \ref{VQE}, as it establishes a crucial connection between solving linear systems of equations -- common in Thermal Science -- and finding the ground state of quantum systems, a fundamental problem that provides deeper insight into quantum mechanics. While VQE faces practical challenges for implementation on real quantum computers, the complexity of decomposing the target observable into Pauli matrices depends on the specific problem. For instance, in molecular Hamiltonian functions, the number of Pauli strings typically scales as $\sim n^4$, which is still computationally demanding but not exponential. Despite these challenges, with appropriate techniques, VQE may still be applicable \citep{zimboras_myths_2025}.

Next, we moved to analyze the Harrow–Hassidim–Lloyd (HHL) algorithm in section \ref{HHL}, because it is considered one of the most promising quantum algorithms for solving linear systems on future, fault-tolerant quantum computers. HHL is theoretically appealing because it offers an exponential quantum speedup under well-defined assumptions—specifically, when the matrix is sparse, well-conditioned, and efficiently representable. This makes it a cornerstone example of quantum advantage for a classically hard problem. However, the practical implementation of HHL on current noisy intermediate-scale quantum (NISQ) devices remains severely limited. The algorithm requires deep circuits involving controlled rotations, quantum phase estimation, and accurate eigenvalue inversion—operations highly sensitive to gate noise, decoherence, and restricted circuit depth. Moreover, the need for fault-tolerant mechanisms to encode real-valued matrices and mitigate condition-number amplification makes the full implementation of HHL infeasible on current hardware. Thus, while HHL stands as a theoretically powerful algorithm with strong asymptotic promises, its practical use is postponed to the era of large-scale, error-corrected quantum computers.


\section*{Acknowledgments}

Portions of this work were developed with the assistance of ChatGPT, an AI language model by OpenAI, in accordance with the CC-BY 4.0 license.

\newpage
\appendix

\section{Two-athlete strategy}
\label{strategy}

\begin{figure}
\begin{subfigure}{.48\textwidth}
  \centering
  \includegraphics[width=1.0\linewidth]{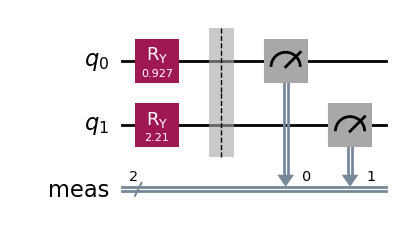}
  \caption{Quantum circuit}
  \label{uncorr-athletes:sfig1}
\end{subfigure}
\begin{subfigure}{.48\textwidth}
  \centering
  \includegraphics[width=1.0\linewidth]{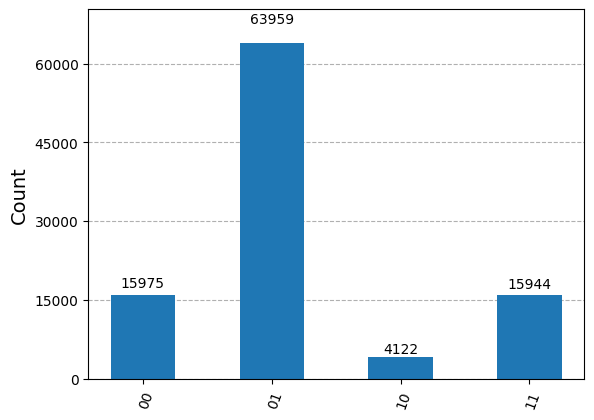}
  \caption{Simulated results}
  \label{uncorr-athletes:sfig2}
\end{subfigure}
\caption{Uncorrelated athletes/qubits $q_0$ and $q_1$. For each athlete, a proper gate represents the training, which realizes a single-qubit rotation about the $Y$-axis ($R_Y$ gate) and ensures the expected performance probability according to Eq.~(\ref{qubit2}). The dashed vertical line is used to demarcate logical gates from measurement.}
\label{uncorr-athletes}
\end{figure}

Let us imagine that two athletes must both participate in a preliminary qualifying tournament to advance to the final stage of a sports competition. Unfortunately, the first athlete has not had enough time to train properly and therefore has a 20\% chance of qualifying, while the second athlete has prepared adequately and thus has a 80\% chance of achieving the same result. Let us indicate by $\ket{1}$ the qualified state for the final stage and by $\ket{0}$ the unqualified state, after the measurement certified by the preliminary tournament. 
In this regard, the state of the two athletes, before the preliminary tournament, can be expressed by the superposition of unqualified state $\ket{0}$ and qualified state $\ket{1}$, namely
\begin{eqnarray}\label{athletes}
\ket{\psi_0} &=& \sqrt{0.8}\,\ket{0}+\sqrt{0.2}\,\ket{1},\\
\ket{\psi_1} &=& \sqrt{0.2}\,\ket{0}+\sqrt{0.8}\,\ket{1}.
\end{eqnarray}
If the two athletes compete in the qualifying tournament independently, the expected outcome will be
\begin{equation}\label{uncorrelated-athletes}
\ket{\psi_0}\otimes\ket{\psi_1} =
\sqrt{0.16}\,\ket{00} + 
\sqrt{0.64}\,\ket{01} + 
\sqrt{0.04}\,\ket{10} + 
\sqrt{0.16}\,\ket{11}.
\end{equation}
The previous formula means that there is a 64\% probability that the under-prepared athlete does not qualify while the well-trained athlete qualifies, which is the most probable outcome of the tournament.
The opposite outcome changing both predictions at the same time is quite unlikely (4\% probability). 
Mixed outcomes, where one event is aligned with the most likely expectation and the other one changing the expected outcome, have the same (intermediate) 16\% probability. This outcome can be obtained also by sampling properly a purposely-designed quantum circuit. Let us assign a qubit for each athlete, i.e. $q_0$ and $q_1$ respectively. For each athlete, let us design a proper gate representing the training, which realizes a single-qubit rotation about the $Y$-axis and ensures the expected performance probability according to Eq.~(\ref{qubit2}). The obtained quantum circuit and the corresponding simulated results are reported in Fig. (\ref{uncorr-athletes}).
It is possible to prove that the previous predictions are correct by recalling the general formula for combining the probabilities of uncorrelated events by tensor product, namely 
\begin{equation}\label{uncorrelated-generic}
\ket{\psi_0}\otimes\ket{\psi_1} =
\sqrt{p_0^{\ket{0}} p_1^{\ket{0}}}\,\ket{00} + 
\sqrt{p_0^{\ket{0}} p_1^{\ket{1}}}\,\ket{01} + 
\sqrt{p_0^{\ket{1}} p_1^{\ket{0}}}\,\ket{10} + 
\sqrt{p_0^{\ket{1}} p_1^{\ket{1}}}\,\ket{11}.
\end{equation}
Putting aside the sports for a while, the previous formula is consistent with the kinetic theory of gases and, in particular, with the assumption of molecular chaos in deriving the Boltzmann equation (also known as the \textit{Stosszahlansatz}), which states that before a collision occurs, the velocities of two colliding particles are uncorrelated.

\begin{figure}
\begin{subfigure}{.48\textwidth}
  \centering
  \includegraphics[width=1.0\linewidth]{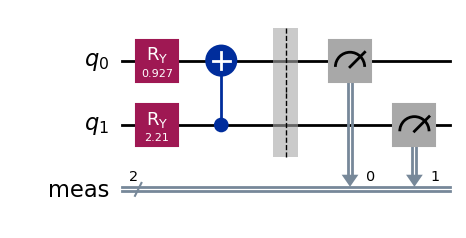}
  \caption{Circuit}
  \label{corr-athletes:sfig1}
\end{subfigure}
\begin{subfigure}{.48\textwidth}
  \centering
  \includegraphics[width=1.0\linewidth]{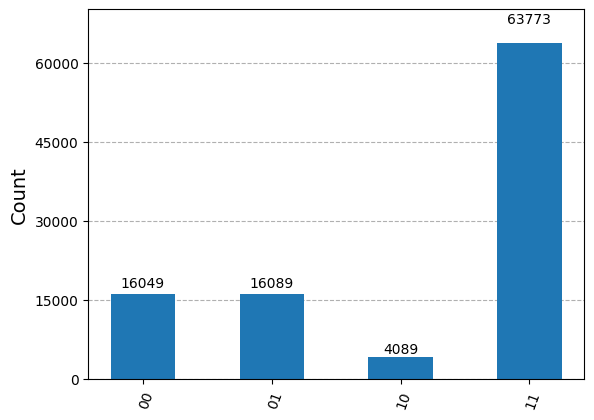}
  \caption{Results}
  \label{corr-athletes:sfig2}
\end{subfigure}
\caption{Correlated athletes/qubits due to entanglement. In this second case, the quantum circuit was updated by adding a controlled $NOT$ gate (also called controlled-$X$ gate or $CNOT$ gate): every time that $q_1$ is equal to one, the athlete is convincing enough to flip the second athlete’s performance outcome (from $0$ to $1$ but also vice versa).}
\label{corr-athletes}
\end{figure}

Coming back to the example, let us suppose now that the better-prepared athlete decides to help the less-prepared one by sharing advice on how to tackle the various challenges of the qualification tournament and perhaps provides some insights about their opponents. 
This time, the chances of success for the less-prepared athlete increase significantly. 
However, there is also a price to pay: in some cases, the better-prepared athlete may provide misleading advice, leading to failures that would not have occurred otherwise. 
Now the state representing the tournament outcome for the two athletes becomes correlated (i.e. entangled). Because of entanglement, when the well-trained athlete qualifies (i.e. the second qubit is equal to $\ket{\cdot 1}$) then the athlete is convincing enough to flip the other athlete’s performance outcome (from $0$ to $1$ but also vice versa). This means that the probabilities of the outcome $\ket{01}$ and $\ket{11}$ are swapped, namely
\begin{equation}\label{correlated-athletes}
\ket{\psi_0\psi_1} =
\sqrt{0.16}\,\ket{00} + 
\sqrt{0.16}\,\ket{01} + 
\sqrt{0.04}\,\ket{10} + 
\sqrt{0.64}\,\ket{11}.
\end{equation}
Now the probability that both athletes qualify is increased to 64\%, meaning that this strategy is anyway advantageous. In this second case, the quantum circuit must be updated by adding a controlled $NOT$ gate (also called controlled-$X$ gate or $CNOT$ gate): as already pointed out, the $CNOT$ gate implies that, whenever $q_1$ equals one, the athlete is persuasive enough to reverse the other’s performance outcome (switching between 0 and 1 but also vice versa). The quantum circuit and the corresponding simulated results in this second case are reported in Fig. (\ref{corr-athletes}).
The key point is that, in presence of correlation -- also called entanglement --, the outcome state is not separable, which means that it cannot be expressed as the tensor product of two independent states, namely $\ket{\psi_0\psi_1}\neq\ket{\psi_0^?}\otimes\ket{\psi_1^?}$, where $\ket{\psi_0^?}$ and $\ket{\psi_1^?}$ are hypothetical states which do not exist. In order to prove that such separated states do not exist, let us compare Eq.~(\ref{correlated-athletes}) with the tensor product definition given by Eq.~(\ref{uncorrelated-generic}), namely
\begin{eqnarray}\label{impossible}
p_0^{\ket{0}?} p_1^{\ket{0}?} &=& 0.16,\label{im00}\\
p_0^{\ket{0}?} p_1^{\ket{1}?} &=& 0.16,\label{im01}\\
p_0^{\ket{1}?} p_1^{\ket{0}?} &=& 0.04,\label{im10}\\
p_0^{\ket{1}?} p_1^{\ket{1}?} &=& 0.64.\label{im11}
\end{eqnarray}
Let us combine Eq.~(\ref{im00}) and Eq.~(\ref{im01}), which yields $p_1^{\ket{0}?} = p_1^{\ket{1}?}$, where $p_0^{\ket{0}?}$ was simplified. The last relation can be used in Eq.~(\ref{im10}) to yield
\begin{eqnarray}\label{impossible2}
p_0^{\ket{1}?} p_1^{\ket{1}?} &=& 0.04,\label{10}\\
p_0^{\ket{1}?} p_1^{\ket{1}?} &=& 0.64,\label{11}
\end{eqnarray}
which are two relations clearly incompatible with each other. Hence, the state given by Eq.~(\ref{correlated-athletes}) is not separable because the two athletes are entangled. This metaphor of the two-athlete strategy aligns well with the numerical example shown in Fig. \ref{fig:entanglement}, which can therefore also be interpreted as a visual representation of the current example.

\section{Prisoner's dilemma}
\label{dilemma}

In the classical prisoner's dilemma, two players (Alice and Bob) must independently decide whether to Cooperate (C) or Defect (D). If both cooperate, they receive a moderate penalty (e.g., 1 year in prison each). If one defects while the other cooperates, the defector goes free (0 years) while the cooperator gets the maximum penalty (3 years in prison). If both defect, they receive a higher penalty (typically 2 years in prison each). In classical game theory, defection is the dominant strategy, leading to a situation where both players receive a worse outcome than if they had cooperated.

Now, suppose Alice and Bob behave as an entangled Bell state \citep{nielsen_quantum_2010}. This state introduces non-classical correlations between their choices, namely
\begin{equation}
\ket{\psi} = \frac{1}{\sqrt{2}} \left(\ket{CC}+\ket{DD}\right).
\end{equation}
In this state, their decisions are no longer independent: if Alice is measured and goes for cooperation, Bob’s measurement in the same basis will necessarily yield cooperation as well, and the same holds for defection. By leveraging quantum operations, Alice and Bob can reach a new equilibrium where cooperation becomes as likely as defection, leading to a better outcome than in the classical case. Quantum entanglement thus enhances cooperation and offers a possible resolution to the prisoner’s dilemma beyond classical strategies.

\section{Hilbert space versus Bloch sphere}
\label{hilbert-bloch}

Here, we aim to intuitively explain the construction of the Bloch sphere representation of a single qubit, and its relationship to the Hilbert space representation.
As previously discussed, a single qubit state vector can be expressed using Eq.~\eqref{qubit}:
\begin{equation*}
    \ket{\psi_q} = \delta_q^{\ket{0}}\ket{0} + \delta_q^{\ket{1}}\ket{1},
\end{equation*}
where $\delta_q^{\ket{0}}, \delta_q^{\ket{1}} \in \mathbb{C}$ are complex numbers. 
These coefficients define a state in a two-dimensional Hilbert space over the complex numbers, which can be thought of as having four real parameters; thus, even in this simple case, it is difficult to visualize the state. 
Fortunately, we can overcome this limitation by using the polar (Euler) representation of complex numbers and applying the normalization condition, $|\delta_q^{\ket{0}}|^2 + |\delta_q^{\ket{1}}|^2 = 1$.
We first express the complex amplitudes as $\delta_q^{\ket{0}} = A e^{i\alpha}$ and $\delta_q^{\ket{1}} = B e^{i\beta}$. Substituting into Eq.~\eqref{qubit}, we obtain:
\begin{equation}\label{qubit_euler}
    \ket{\psi_q} = A e^{i\alpha} \ket{0} + B e^{i\beta} \ket{1}.
\end{equation}
The normalization condition constrains the amplitudes to lie on the unit circle in real two-dimensional space: $A^2 + B^2 = 1$.

\begin{figure}[htbp]
    \centering
    \includegraphics[scale=0.95]{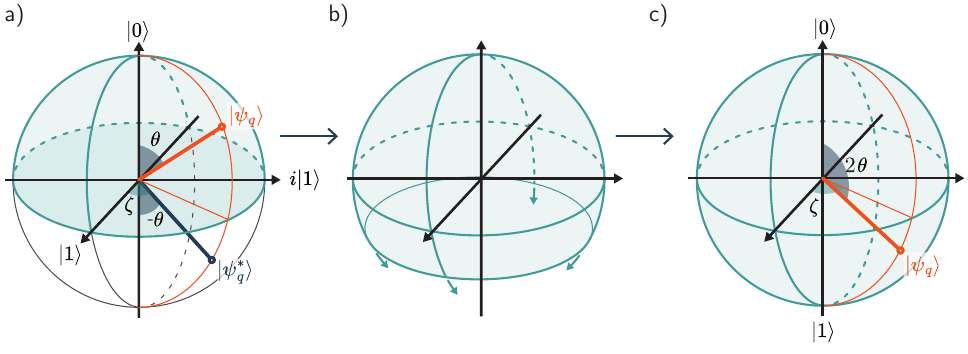}
    \caption{
        Intuitive construction of the Bloch sphere. (a) The qubit vector state $ \ket{\psi_q} $ (and the reflected state $ \ket{\psi_q^*} $) represented in Hilbert space. (b) The conceptual stretching of the upper hemisphere onto a sphere (and the equator of the upper hemisphere in a point). (c) The final representation of $ \ket{\psi_q} $ as a vector state on the Bloch sphere.
    }\label{fig:hilber_bloch}
    
\end{figure}

Therefore, we can parameterize the amplitudes using a single \textit{polar} angle $\theta$:
\begin{equation*}\label{qubit_euler2}
    \ket{\psi_q} = \cos\theta \, e^{i\alpha} \ket{0} + \sin\theta \, e^{i\beta} \ket{1}.
\end{equation*}
Since an overall (global) phase factor has no physical effect, we can factor out $e^{i\alpha}$ and ignore it, defining a relative phase $\zeta = \beta - \alpha$. We then obtain:
\begin{equation*}\label{qubit_euler3}
    \ket{\psi_q} = e^{i\alpha} \left( \cos\theta \ket{0} + \sin\theta \, e^{i\zeta} \ket{1} \right),
\end{equation*}
where the term $e^{i\alpha}$ can be safely omitted, since it represents a phase shift, and does not affect measurement outcomes \citep{nielsen_quantum_2010}, yielding the simplified and physically equivalent expression:
\begin{equation}\label{qubit_hilbert}
    \ket{\psi_q} = \cos\theta \ket{0} + \sin\theta \, e^{i\zeta} \ket{1}.
\end{equation}
Comparing Eq.~(\ref{qubit_hilbert}) with Eq.~(\ref{qubit2}) in the main text yields $\theta = \varphi_q/2$ and $\zeta = \zeta_q$, which will be clearer at the end of this appendix and is the main point of this derivation. Now, using Euler's formula $ e^{i\zeta} = \cos\zeta + i\sin\zeta$, we can rewrite the qubit state:
\begin{equation*}\label{qubit_hilbert2}
    \ket{\psi_q} = \cos\theta \ket{0} + \sin\theta \cos\zeta \ket{1} + i \sin\theta \sin\zeta \ket{1},
\end{equation*}
and comparing to the spherical coordinate vector $\vec{r}=(\sin\theta \cos\zeta ,\sin\theta \sin\zeta, \cos\theta  )^T$.
We recognize that it is possible to visualize the state $\ket{\psi_q}$ in a tridimensional Hilbert space with the bases $\ket{1}=(1,0,0)^T$, $i\ket{1}=(0,i,0)^T$ (the generic axis of the Hilbert space may be complex, but the inner product must be still non-commutative\footnote{In a two-dimensional complex space, the inner product is non-commutative because it is conjugate symmetric, meaning $\braket{\psi|\phi} = \braket{\phi|\psi}^*$ (complex conjugate), so swapping the vectors changes the result unless the inner product is real. For example, if $\ket{a} = \ket{1}$ and $\ket{b} = i\ket{1}$, then $\braket{a|b} = i \neq \braket{b|a} = -i$, therefore $\braket{a|b} = \braket{b|a}^*$.}), $\ket{0}=(0,0,1)^T$, as shown in Fig.~\ref{fig:hilber_bloch}(a).
Using the full sphere in Hilbert space has the problem that multiple states can result in the same measurement outcomes. 
For example, consider the state $\ket{\psi_q}$ and the  reflected state through the equatorial plane  $\ket{\psi_q^*} = \cos(-\theta) \ket{0} + \sin(-\theta) \, e^{i\varphi} \ket{1}$, depicted in Fig.~\ref{fig:hilber_bloch}(a), which yield the same measurement probabilities:
\begin{equation*}
    \begin{split}
    p_0 &= \braket{\psi_q|P_0|\psi_q} = \braket{\psi_q^*|P_0|\psi_q^*} = \cos^2\theta,\\
    p_1 &= \braket{\psi_q|P_1|\psi_q} = \braket{\psi_q^*|P_1|\psi_q^*} = \sin^2\theta,
    \end{split}
\end{equation*}
where the projectors operators of the measurement are defined as $P_0 = \ket{0}\bra{0}$ and $P_1 = \ket{1}\bra{1}$.
Thus, we see that $\ket{\psi_q}$ and $\ket{\psi_q^*}$ are physically indistinguishable by measurement. 
As a result, only the \textit{upper hemisphere} of the sphere in the Hilbert space is needed to uniquely describe a qubit state, which corresponds to restricting the angle $\theta$ to the interval $\theta \in \left[0, \pi/2\right]$.

Furthermore, all the states that lie along the equator ($\theta = \pi/2$ in the Hilbert space) represent the same measurement outcome (i.e., for $\theta = \pi/2$ we have $\ket{\psi_q'} = \cos\zeta\ket{1} + \sin\zeta\;i\ket{1}$, so the measurement probability is $p_1 = \braket{\psi_q'|P_1|\psi_q'}=\cos^2\zeta+\sin^2\zeta=1$). 
Therefore, we can conceptually imagine ``pulling'' this circumference downward until it collapses into a single point. 
This transformation simplifies the visualization and results in the familiar Bloch sphere representation, as illustrated in Fig.~\ref{fig:hilber_bloch}(b) and Fig.~\ref{fig:hilber_bloch}(c).
In this representation, the polar angle $ \theta $ from the Hilbert space mapping is effectively \textit{doubled} on the Bloch sphere.
To maintain consistency between the two representations, we can introduce the Bloch polar angle $\varphi = 2\theta \Leftrightarrow \theta = \varphi /2$, with $ \varphi \in [0, \pi] $, as used in Eq.~\eqref{qubit2}.

It is important to note that this is an illustrative non-rigorous explanation of the Bloch sphere construction. 
Mathematically, the Bloch sphere corresponds to the complex projective line of the two-dimensional Hilbert space, constructed using a stereographic projection of the qubit state onto a plane and topologically represented as the \emph{Riemann sphere}.

\section{Real data loading/encoding}
\label{appedix-real-data-loading}

A fundamental aspect of quantum computing is the ability to efficiently load real-world data into a quantum system and extract results back to classical computing. Naively one would like to have a procedure for correlating the parameters in $\vec{\theta}$ with the real amplitudes in $\ket{x(\vec{\theta}\,)}$ by means of some analytical formulas. This approach is usually called real data loading or better encoding, and some algorithms have been proposed in literature, e.g. a divide‑and‑conquer algorithm for quantum state preparation \citep{araujo_divide-and-conquer_2021}. Real data loading/encoding is a good way to understand how a quantum computer works and hence it will be discussed here.

Loading real-world data into a quantum system requires a quantum state preparation. Many algorithms to create arbitrary quantum states require quantum circuits with depth $O(N)$ to load an $N$-dimensional vector \citep{araujo_divide-and-conquer_2021}. In the context of quantum circuits, depth refers to the number of sequential (time-ordered) layers of quantum gates that must be applied to execute an algorithm. It measures how many steps a quantum circuit takes to process information. Some algorithms have been proposed in the literature based on a divide-and-conquer strategy to load a $N$-dimensional vector using a quantum circuit with poly-logarithmic depth \citep{araujo_divide-and-conquer_2021}. The problem is that these algorithms usually require a large number of parameters to compute and therefore are less suitable for variational problems as VQE.

\begin{figure}
  \includegraphics[width=1.0\linewidth]{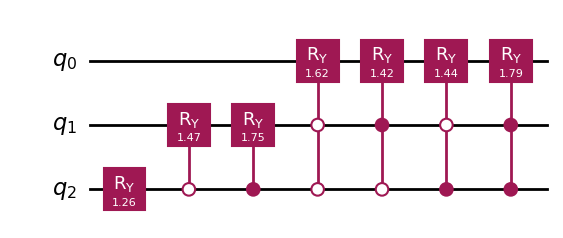}
\caption{Quantum circuit for loading real data in a quantum system state $\ket{x}$ based on a divide‑and‑conquer algorithm proposed in Ref. \citep{araujo_divide-and-conquer_2021}. Note that the proposed circuit uses only rotations about the $Y$-axis ($R_Y$ gates), which are designed by means of a proper binary tree data structure \citep{araujo_divide-and-conquer_2021}. For clarity, horizontal lines represent quantum wires which correspond to qubits in the circuit, red squares are the $R_Y$ gates (see Eq.~(\ref{RY})), red dots represent control points in controlled gates {and empty red circles represent anti-control points (i.e. activated by zero value). The little-endian convention is used \citep{nielsen_quantum_2010}, meaning that the topmost qubit represents the least significant bit (LSB), while the bottom qubit corresponds to the most significant bit (MSB), which, in this case, forms the trunk of the state tree in the divide-and-conquer strategy \citep{araujo_divide-and-conquer_2021}.}}
\label{divide‑and‑conquer-circuit}
\end{figure}

\begin{figure}
  \includegraphics[width=1.0\linewidth]{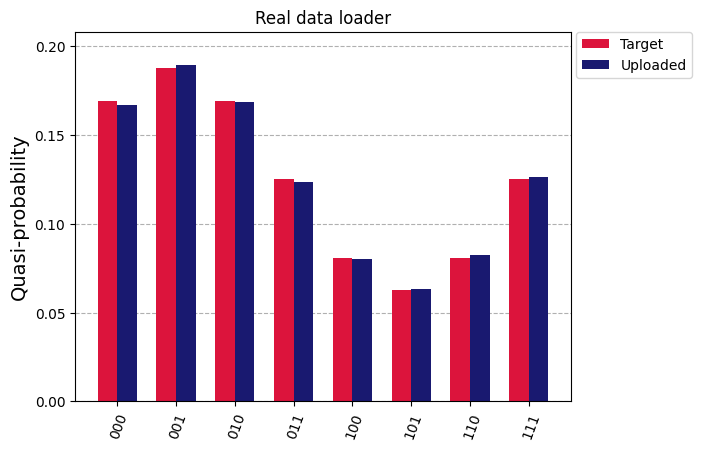}
\caption{Example of real data loading / encoding (simulated). The red bars represent the real data to be loaded. The blue bar represents the actual loaded data by a divide‑and‑conquer algorithm proposed in Ref. \citep{araujo_divide-and-conquer_2021}. These results are based on a quantum simulator provided in the Qiskit package.}
\label{divide‑and‑conquer-probabilities}
\end{figure}

The divide-and-conquer paradigm is used in efficient algorithms for sorting, computing the discrete Fourier
transform, and others \citep{araujo_divide-and-conquer_2021}. The main idea is to divide a problem into subproblems of the same class and combine the solutions of the subproblems, in a recursive way, to obtain the solution of the original problem \citep{araujo_divide-and-conquer_2021}. In particular, one of the standard methods for loading information in a quantum device is based on using controlled rotations \citep{araujo_divide-and-conquer_2021}. These controlled rotations can be designed by means of a proper binary tree data structure for the data to be loaded and consequently, by means of Eq.~(\ref{qubit2}), for the expected rotations. In the case of $n=3$ qubits, the resulting quantum circuit, which consists solely of rotations about the $Y$-axis ($R_Y$ gates), is shown in Fig. \ref{divide‑and‑conquer-circuit}. As an example of its application, let us suppose to load the following real data
\begin{equation}\label{datatobeloaded}
    L_l = \frac{1}{N}+\frac{1}{2N}\sin{\left(\frac{2\pi}{N}\,l+\frac{2\pi}{N}\right)}, \quad l = 0, 1, \dots, N-1,
\end{equation}
where the elements $L_l$ are designed such that $\sum_l L_l = 1$. The last condition makes possible to load the previous data as (quasi) probabilities of a quantum state, namely 
\begin{equation}\label{loadedasprobabilities}
    \ket{x_L} = \sum_{{j\in\{0,\,1\}^n}} \sqrt{L_j}\,\ket{j}.
\end{equation}
Please note that this is different from what is done in the main text (see section \ref{normalization}), where the elements of vector $\vec{T}$ are loaded as amplitudes of a quantum state $\ket{x_T}=\sum_j (T_j/\sqrt{\vec{T}\cdot\vec{T}})\,\ket{j}$, e.g. see Eq.~(\ref{x}). In quantum computing, real data can be loaded into quantum states either as probabilities or amplitudes, each with distinct implications. Probability encoding represents data as a quantum probability distribution, where measurement outcomes follow predefined likelihoods, making it useful for probabilistic modeling and quantum sampling. In contrast, amplitude encoding directly maps data values into quantum state amplitudes, enabling powerful applications in quantum machine learning and linear algebra but requiring more complex state reconstruction. While probability encoding is more intuitive and measurement-friendly, amplitude encoding offers greater expressive power for quantum algorithms. Coming back to the original data loading/encoding problem, in case of $N=8$, the values $L_l$ can be encoded into a three-qubit quantum state $\ket{L}$ using the proposed circuit by means of $N_\theta^{d\&c} = 7$ parameters. The quasi-probabilities obtained from simulating this circuit, based on $100,000$ measurements of system replicas, are presented in Fig. \ref{divide‑and‑conquer-probabilities}, demonstrating the circuit's effectiveness in accurately encoding the target real data.

In the divide-and-conquer approach, the good point is that the parameters $\vec{\theta}$ have a clear physical interpretation, thanks to the binary tree data structure, and they can be analytically computed to recover the target data to be loaded. The problem is that the number of parameters in this ansatz grows as $N_\theta^{d\&c} = 2^n-1$, which is as large as the number of real data to be loaded (recall that $\braket{x|x}=1$). For comparison, the efficient ansatz discussed in section \ref{step3qiskit}, called \say{\textit{EfficientSU2}} circuit in Qiskit \citep{javadi-abhari_quantum_2024}, requires instead a number of parameters which grows only linearly with the number of qubits, namely $N_\theta = 8\,n$. It is clear that, in case of a large number of qubits, the divide-and-conquer approach is less efficient than \say{\textit{EfficientSU2}} --  and therefore impractical -- for variational problems because $N_\theta^{d\&c} \gg N_\theta$. 

\begin{figure}
\begin{subfigure}{.48\textwidth}
  \centering
  \includegraphics[width=1.0\linewidth]{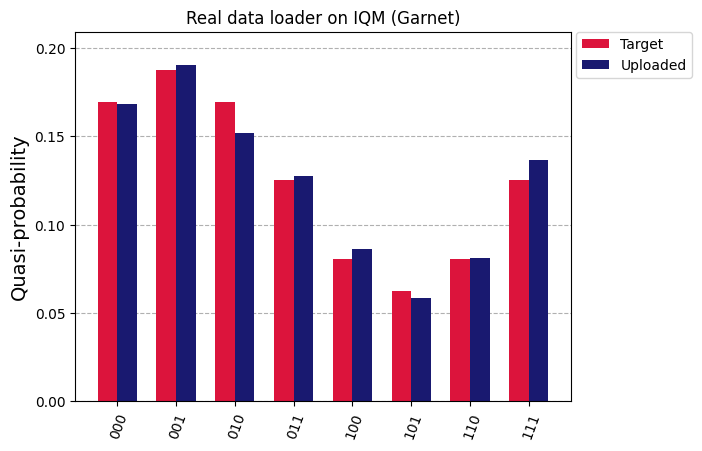}
  \caption{3 qubits (real hardware)}
  \label{real-probabilities-3qubits}
\end{subfigure}
\begin{subfigure}{.48\textwidth}
  \centering
  \includegraphics[width=1.0\linewidth]{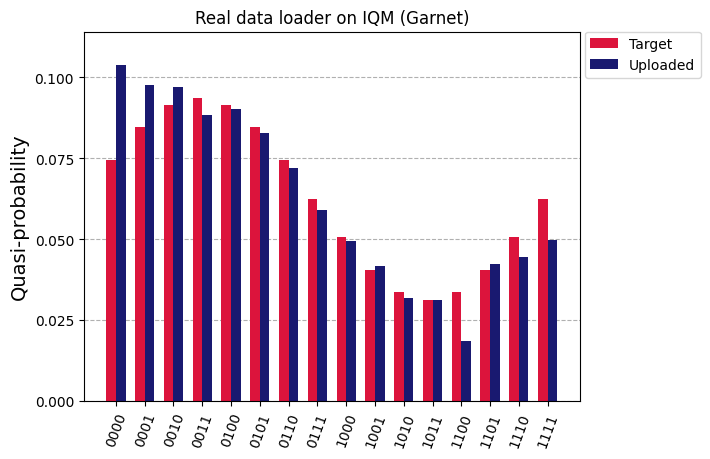}
  \caption{4 qubits (real hardware)}
  \label{real-probabilities-4qubits}
\end{subfigure}
\caption{Example of real data loading / encoding (real hardware). The red bars represent the real data to be loaded. The blue bar represents the actual loaded data by a divide‑and‑conquer algorithm proposed in Ref. \citep{araujo_divide-and-conquer_2021}. These results (based on $20,000$ shots) are obtained by \href{https://www.iqmacademy.com/qpu/garnet/}{IQM Garnet} machine developed by \href{https://www.meetiqm.com/}{IQM}, a Finnish-German quantum computer manufacturer.}
\label{divide‑and‑conquer-real-probabilities}
\end{figure}

Clearly, the problem of data loading and encoding is quite general and extends far beyond variational applications. In particular, all software development kits have efficient routines for performing this task. For example, Qiskit \citep{javadi-abhari_quantum_2024} has a state preparation routine based on the decomposition of arbitrary isometries into a sequence of single-qubit and controlled-not ($CNOT$) gates \citep{iten_quantum_2016}. This approach is tested here for loading the data given by Eq.~(\ref{datatobeloaded}) in case of $N=8$ (3 qubits) and $N=16$ (4 qubits) on a real hardware. In particular, let us use the \href{https://www.iqmacademy.com/qpu/garnet/}{IQM Garnet} machine developed by \href{https://www.meetiqm.com/}{IQM}, a Finnish-German quantum computer manufacturer. The experimental results (based on $20,000$ shots) of the state preparation algorithm proposed in Ref. \citep{iten_quantum_2016} on \href{https://www.iqmacademy.com/qpu/garnet/}{IQM Garnet} are reported in Fig. \ref{divide‑and‑conquer-real-probabilities}. These results should be considered indicative, as real hardware in \href{https://en.wikipedia.org/wiki/Noisy_intermediate-scale_quantum_era}{the NISQ era} is influenced by environmental conditions, causing actual outcomes to vary slightly from run to run.

\section{How discrete FT works}
\label{appedix-discrete-FT-works}

\begin{figure}
    \includegraphics[scale=0.95]{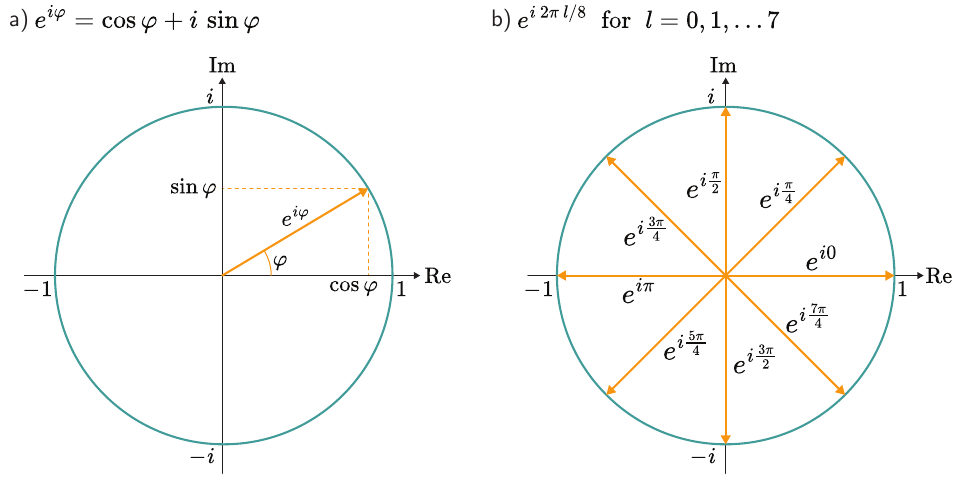}
\caption{Euler's formula for representing complex roots of unity in the polar form on complex plane.}
\label{fig-euler}
\end{figure}

\begin{figure}[htbp]
  \includegraphics[scale=0.95]{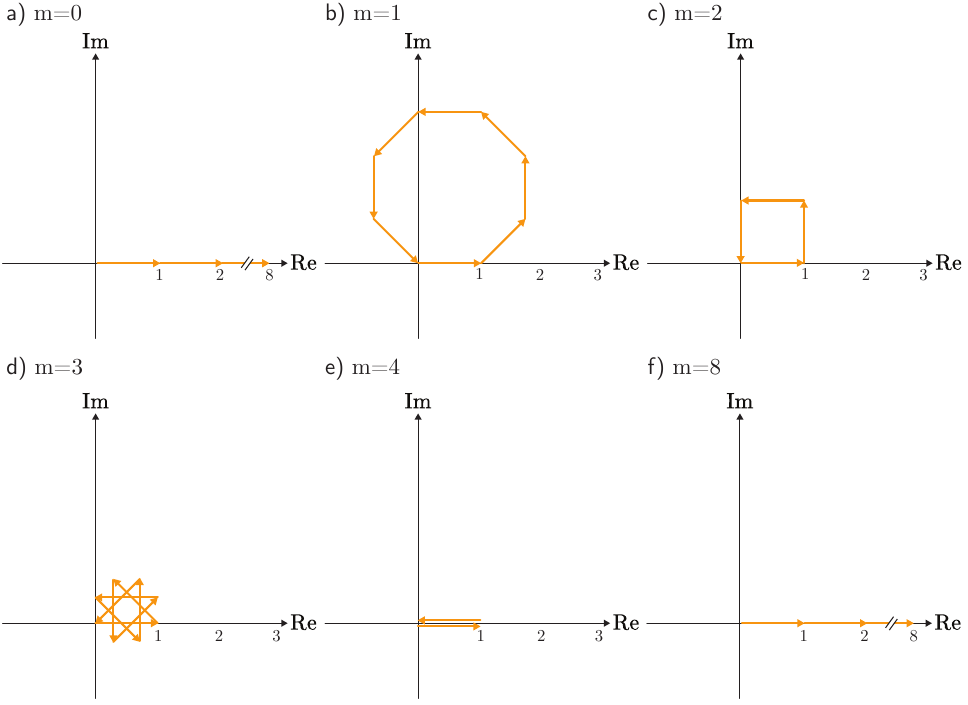}
\caption{Summation of geometric series of complex roots of unity, i.e. $\sum_{l = 0}^{N-1} e^{i\,2\pi\,lm/N}$. In case of $N=8$, the sub-figures report two examples where the summation is non-zero, i.e. $m=0$ and $m=8$, as well as three examples where the summation is zero, i.e. $m=1$, $m=2$ and $m=3$.}
\label{fig-geometric-series}
\end{figure}

In this Appendix, let us try to clarify the meaning of the discrete Fourier transform with complex output. First of all, let us introduce the Euler's formula which establishes the fundamental relationship between the trigonometric functions and the complex exponential function, namely
\begin{equation}\label{euler}
    e^{i \varphi}=\cos\varphi + i\,\sin\varphi,
\end{equation}
where $i=\sqrt{-1}$ is the imaginary unit and $\varphi$ is a generic angle between the line connecting the unitary complex number and the real axis on the complex plane in Fig. \ref{fig-euler}(a). Next, we need to understand the meaning of $\omega_N = e^{i\,2\pi/N}$ given by Eq.~(\ref{omega}). Dividing the full angle $2\pi$ into $N$ parts, $\omega_N$ is obtained by performing, in the complex plane, a rotation corresponding to a slice of $2\pi/N$. In polar form, $\omega_N$ identifies one of the possible complex roots of unity. 

The $l$-th power of $\omega_N$, i.e. $\omega_N^l$, corresponds to a rotation equal to $l$ slices, i.e. $2\pi\,l/N$. Also the power $\omega_N^l$ is a complex root of unity. This property allows to define a discrete geometrical series, namely $1, \omega_N^1, \omega_N^2, \dots, \omega_N^{N-1}$, which is represented as a series of vectors in Fig. \ref{fig-euler}(b). The linear combination of these vectors implies
\begin{equation}\label{sum-geometric-series-m-1}
    \sum_{l = 0}^{N-1} \omega_N^{l} = 0.
\end{equation}
The intuitive meaning is that the net effect of a sum over all discrete angular vectors (with norm one), which are angularly equally spaced is equal to zero, because they are obtained by adding progressively the slice $\omega_N$ until covering the full angle $2\pi$. It is possible to generalize the previous result by a generic parameter $m$ as
\begin{equation}\label{sum-geometric-series}
    \sum_{l = 0}^{N-1} \omega_N^{lm} = 
    \sum_{l = 0}^{N-1} e^{2\pi\,i\,l\,m/N} = 
    \begin{cases} 
    N, & m = 0 \\
    0, & m \neq 0
    \end{cases}.
\end{equation}
In order to understand the role of the parameter $m$ and hence the meaning of the previous result, let us consider the examples reported in Fig. \ref{fig-geometric-series}. In top of those cases where the net effect is zero, one has also to add the trivial cases where the exponent is equal to zero and therefore all terms in the summation become equal to unity. Some powers of $\omega_N$ appear also in the definition of operator $\hat{U}_{FT}$ given by Eq.~(\ref{FT-matrix}). The sum over all the components of the generic $m$-th row of the operator $\hat{U}_{FT}$ is given by Eq.~(\ref{sum-geometric-series}) multiplied by the normalization factor $1/\sqrt{N}$. Eq.~(\ref{sum-geometric-series}) is also an immediate consequence of Vieta's formulas. 

The property given by Eq.~(\ref{sum-geometric-series}) is particularly useful in the discrete FT to select the harmonic components of a generic field (in our case, a temperature field $\vec{T}$). Let us suppose that the temperature field $\vec{T}$ is defined by a proper orthonormal basis $\vec{e}_0, \vec{e}_1, \vec{e}_2, \dots \vec{e}_{N-1}$, as reported in Eq.~(\ref{T_field}), where $T_l$ is the nodal value for the $l$-th mesh node. Let us suppose to decompose each nodal value $T_l$ in harmonic components $\tilde{T}_{{m}}$ where ${{m}}$ goes from $0$ to $N-1$, as prescribed by the inverse transform given by Eq.~(\ref{antiFT}). These components altogether makes the transformed field, which is a column vector of complex numbers $\vec{\tilde{T}}$ defined with regards to the same orthonormal basis by Eq.~(\ref{tildeT_field}). As it will be clear by the following example, Eq.~(\ref{sum-geometric-series}) allows one to project the vector $\vec{T}$ on those components of the transformed vector with wavenumbers ${{m}}$ such that $m = f({{m}}) = 0$. As an example, let us consider a normalized (dimensionless) temperature profile, where the $l$-th generic nodal value is given by the following expression
\begin{equation}\label{T_old-example}
    T_l = 1 + \frac{1}{2}\sin{\left[\frac{2\pi}{N}\,(l+1)\right]}.
\end{equation}
Again from the Euler's formula, namely $e^{i\,\theta}=\cos{\theta}+i\,\sin{(\theta)}$, it is possible to express the sine function in the previous example as $\sin{(\theta)} = (e^{i\,\theta}-e^{-i\,\theta})/(2\,i)$, which yields equivalently
\begin{equation}\label{T_old-example2}
    T_l = 1 + \frac{1}{4\,i}\left(
    -\omega_N^{-1-l}
    +\omega_N^{1+l}\right).
\end{equation}
Let us apply the discrete FT given by Eq.~(\ref{FT}), namely
\begin{equation}\label{tildeT_old-example}
    \tilde{T}_{{m}} = \frac{1}{\sqrt{N}}\,\sum_{l = 0}^{N-1} \left[\omega_N^{{{m}}l}  
    - \frac{\omega_N^{-1}}{4\,i}\omega_N^{l({{m}}-1)}
    + \frac{\omega_N}{4\,i}\omega_N^{l({{m}}+1)}\right].
\end{equation}
Because the previous result is invariant under cyclic shifts, as evident from Fig. \ref{fig-euler}(b), substituting the equivalence $\omega_N^{{{m}}(l+1)} = \omega_N^{{{m}}(l+1-N)}$ {(which is valid because $\omega_N^N = 1$)} in the previous expression yields
\begin{equation}\label{tildeT_old-example2}
    \tilde{T}_{{m}} = \frac{1}{\sqrt{N}}\,\sum_{l = 0}^{N-1} \left[\omega_N^{{{m}}l}  
    - \frac{\omega_N^{-1}}{4\,i}\omega_N^{l({{m}}-1)}
    + \frac{\omega_N}{4\,i}\omega_N^{l({{m}}+1-N)}\right].
\end{equation}
Using the property for geometric series given by Eq.~(\ref{sum-geometric-series}) and combining the results by the orthonormal basis used in Eq.~(\ref{tildeT_field}) yields
\begin{equation}\label{tildeT_old-example3}
    \vec{\tilde{T}} = \sqrt{N}\,\left(
    \vec{e}_0 
    -\frac{\omega_N^{-1}}{4\,i}\,\vec{e}_1
    +\frac{\omega_N}{4\,i}\,\vec{e}_{N-1}
    \right).
\end{equation}
Substituting the definition of $\omega_N$ yields
\begin{equation}\label{tildeT_old-example-final}
    \vec{\tilde{T}} = 
    \sqrt{N}\,\vec{e}_0 
    +\left(\frac{\sqrt{2N}}{8}+i\,\frac{\sqrt{2N}}{8}\right)\,\vec{e}_1
    +\left(\frac{\sqrt{2N}}{8}-i\,\frac{\sqrt{2N}}{8}\right)\,\vec{e}_{N-1}.
\end{equation}
From the application point of view, let us introduce the shift operator $S$, which is the circulant matrix defined as 
\begin{equation}\label{fourier-shift}
    \hat{S} : \vec{e}_l \rightarrow \vec{e}_{l+N/2\bmod N},
\end{equation}
where $\bmod$ is the modulo operation, which returns the remainder of a division. By means of the shifted operator, it is possible to define the result in the standard (shifted) form, namely
\begin{equation}\label{tildeT_old-example-shifted}
    \vec{\tilde{T}}_s = \hat{S}\,\vec{\tilde{T}} =
    \left(\frac{\sqrt{2N}}{8}-i\,\frac{\sqrt{2N}}{8}\right)\,\vec{e}_{N/2-1}
    +\sqrt{N}\,\vec{e}_{N/2} 
    +\left(\frac{\sqrt{2N}}{8}+i\,\frac{\sqrt{2N}}{8}\right)\,\vec{e}_{N/2+1}.
\end{equation}
It is also useful to compute the wavenumber spectrum, which describes how the variance of the temperature
field is distributed over different harmonic components, by means of Eq.~(\ref{spectrum}). The (shifted) wavenumber spectrum $\vec{p}_s^{\;c}$ is defined as 
\begin{equation}\label{wavenumber-example-shifted}
    \vec{p}_s^{\;c} = \frac{1}{N}\,\vec{\tilde{T}}_s \odot \vec{\tilde{T}}_s^*  =
    \frac{1}{16}\,\vec{e}_{N/2-1}
    + \vec{e}_{N/2} 
    +\frac{1}{16}\,\vec{e}_{N/2+1}.
\end{equation}
where $\odot$ represents the Hadamard (element-wise) product and the superscript $^*$ means the complex conjugate. 

\begin{figure}
\begin{subfigure}{.48\textwidth}
  \centering
  \includegraphics[width=1.0\linewidth]{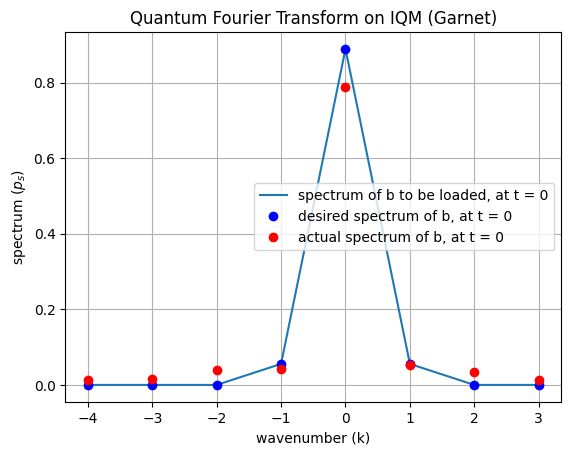}
  \caption{3 qubits (real hardware)}
  \label{fig-QFT-3qubits}
\end{subfigure}
\begin{subfigure}{.48\textwidth}
  \centering
  \includegraphics[width=1.0\linewidth]{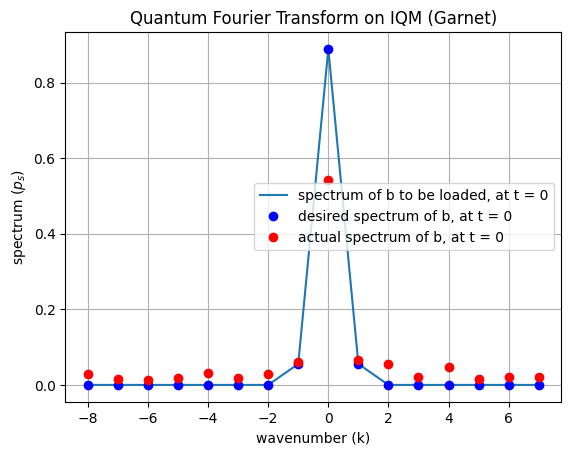}
  \caption{4 qubits (real hardware)}
  \label{fig-QFT-4qubits}
\end{subfigure}
\caption{Wavenumber spectrum, given by Eq.~(\ref{spectrum-qft}), computed by QFT routine provided in Qiskit \citep{javadi-abhari_quantum_2024}. These results (based on $20,000$ shots) are obtained by \href{https://www.iqmacademy.com/qpu/garnet/}{IQM Garnet} machine developed by \href{https://www.meetiqm.com/}{IQM}, a Finnish-German quantum computer manufacturer.}
\label{fig-qft}
\end{figure}

It may be interesting to compute the (shifted) wavenumber spectrum $\vec{p}_s^{\;c}$ by the quantum spectrum $\vec{p}_s$. Taking into account Eq.~(\ref{b}) and Eq.~(\ref{theta}) yields $\ket{b} = (1/\theta)\,\vec{T}$, which means that state $\ket{b}$ is obtained by normalizing $\vec{T}$ by the scaling factor $\theta$. Because the FT is a linear transformation, then $\ket{\tilde{b}} = (1/\theta)\,\vec{\tilde{T}}$ and consequently $\vec{\tilde{T}} = \theta\,\ket{\tilde{b}}$. Using the latter relation into Eq.~(\ref{wavenumber-example-shifted}) yields
\begin{equation}\label{wavenumber-example-shifted2}
    \vec{p}_s^{\;c} = \frac{1}{N}\,\vec{\tilde{T}}_s \odot \vec{\tilde{T}}_s^* = \frac{\theta^2}{N}\,\vec{\tilde{b}}_s \odot \vec{\tilde{b}}_s^* =
    \frac{\theta^2}{N}\,\vec{p}_s,
\end{equation}
where shifted $\vec{\tilde{b}}_s = \hat{S}\,\vec{\tilde{b}}$, shifted $\vec{p}_s = \hat{S}\,\vec{p}$ and $\vec{p}$ is the quantum spectrum given by Eq.~(\ref{spectrum-qft}). For our example temperature profile, given by Eq.~(\ref{T_old-example}), $\theta^2/N = 9/8$ holds, which makes possible to compute $\vec{p}_s^{\;c}$ by $\vec{p}_s$. In particular, for the example considered in this Appendix, $\vec{p}_s$ is computed in case of $N=8$ (3 qubits) and $N=16$ (4 qubits) on a real hardware. In particular, let us use the \href{https://www.iqmacademy.com/qpu/garnet/}{IQM Garnet} machine developed by \href{https://www.meetiqm.com/}{IQM}, a Finnish-German quantum computer manufacturer. The experimental results (based on $20,000$ shots) of the QFT routine provided in Qiskit by IBM \citep{javadi-abhari_quantum_2024} on \href{https://www.iqmacademy.com/qpu/garnet/}{IQM Garnet} are reported in Fig. \ref{fig-qft}. These results should be considered indicative, as real hardware in \href{https://en.wikipedia.org/wiki/Noisy_intermediate-scale_quantum_era}{the NISQ era} is influenced by environmental conditions, causing actual outcomes to vary slightly from run to run.

\section{Example codes for VQE}
\label{example-codes}

In this Appendix, we provide some example codes for implementing the Variational Quantum Eigensolver (VQE) algorithm. Let us start with the Qiskit language \citep{javadi-abhari_quantum_2024} by IBM as a popular open-source software development kit. In Qiskit, the VQE can be implemented as follows. 

\vspace{0.5cm}
\begin{lstlisting}[language=Python, frame=none]
import numpy as np
from qiskit.circuit.library import EfficientSU2
from qiskit.quantum_info import SparsePauliOp
from qiskit.primitives import StatevectorEstimator as Estimator
from scipy.optimize import minimize

num_qubits = 3 # number of qubits
N = pow(2,num_qubits) # number of mesh nodes
for i in range(N):
    T_old[i] = 1 + (1/2)*np.sin(2*np.pi*(i+1)/N)
TT_old = np.sum(T_old**2)
b0 = np.sqrt(TT_old)
b = T_old/b0 # initial profile

# (1) ANSATZ
raw_ansatz = EfficientSU2(num_qubits)
# Initial (arbitrary) set of parameter
theta0 = np.ones(raw_ansatz.num_parameters)

# (2) OBSERVABLE = HAMILTONIAN = "ENERGY"
# Conduction matrix
r = 0.5 # = delta_t*alpha/(delta_x**2) = Fo, Fourier number 
d = np.ones(N)*(1+2*r)
od = np.ones(N-1)*(-r)
C = np.diag(d, 0) + np.diag(od, -1) + np.diag(od, 1)
C[0,N-1] = -r
C[N-1,0] = -r
O = np.identity(N)-np.outer(b,b)
O = np.matmul(O,C)
C_dag = np.transpose(C)
O = np.matmul(C_dag,O)
observable = SparsePauliOp.from_operator(O)

# (3) ESTIMATOR, quantum simulator
estimator = Estimator()

# LOSS FUNCTION
def cost_func_vqe(params, ansatz, hamiltonian, estimator):
    """Return estimate of energy from estimator

    Parameters:
        params (ndarray): Array of ansatz parameters
        ansatz (QuantumCircuit): Parameterized ansatz circuit
        hamiltonian (SparsePauliOp): Operator representation of Hamiltonian
        estimator (Estimator): Estimator primitive instance

    Returns:
        float: Energy estimate
    """
    pub = (ansatz, hamiltonian, params)
    cost = estimator.run([pub]).result()[0].data.evs

    return cost

# MINIMIZATION STEP
result = minimize(cost_func_vqe, theta0, args=(raw_ansatz.decompose(), observable, estimator), 
                  method="COBYLA", # minimization method
                  tol = 1e-3, # which affects iterations/time
                  options={'maxiter': 1000, 'disp': True})                  
\end{lstlisting}
\vspace{0.5cm}

While the physics side of quantum computing makes significant progress, the support for high-level quantum programming abstractions is still in its infancy compared to modern classical languages and frameworks \citep{seidel_qrisp_2022}. An interesting example is provided by Qrisp, which is a high-level programming language developed by Fraunhofer for creating and compiling quantum algorithms \citep{seidel_qrisp_2022}. Its structured programming model enables scalable development and maintenance \citep{seidel_qrisp_2022}. In Qrisp, the VQE can be implemented as follows.

\vspace{0.5cm}
\begin{lstlisting}[language=Python, frame=none]
import numpy as np
from qrisp import *
from qrisp.operators import QubitOperator
from qrisp.vqe.vqe_problem import *

num_qubits = 3 # number of qubits
N = pow(2,num_qubits) # number of mesh nodes

# (1) ANSATZ
def ansatz(qv,theta):
    for i in range(num_qubits):
        ry(theta[i],qv[i])
    for i in range(num_qubits-1):
        cx(qv[i],qv[i+1])
    cx(qv[num_qubits-1],qv[0])

# (2) OBSERVABLE = HAMILTONIAN = "ENERGY"
H = QubitOperator.from_matrix(O).to_pauli()

# (3) ESTIMATOR, quantum simulator
# Default, if 'backend' is not specified

# VQE PROBLEM
vqe = VQEProblem(hamiltonian = H,
                 ansatz_function = ansatz,
                 num_params = 3,
                 callback = True)

# MINIMIZATION STEP
qarg = QuantumVariable(num_qubits)
energy = vqe.run(qarg,
            depth = 4,
            max_iter = 1000,
            mes_{{m}}wargs={'precision':0.1,'diagonalisation_method':'commuting'})
                 
\end{lstlisting}
\vspace{0.5cm}

\section{Phase kickback}
\label{kickback}

\begin{figure}[htbp]
    \centering
    \includegraphics[scale=0.95]{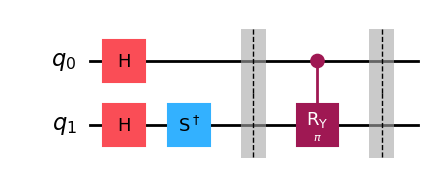}
    \caption{Example for explaining the phase kickback phenomenon. For clarity, horizontal lines represent quantum wires which correspond to qubits in the circuit, orange squares are the Hadamard gates $H$, blue square is the Hermitian adjoint of the phase gate $S^\dagger$, red square is the $R_Y$ gate (see Eq.~(81)) and red dot represents the control point in the controlled gate.
    }\label{fig:kickback}    
\end{figure}

Phase kickback is a fundamental quantum phenomenon in which a phase acquired by a target quantum state during a controlled operation is effectively transferred back onto the control qubit \citep{nielsen_quantum_2010}. When a controlled-unitary acts on an eigenstate of the unitary gate $U$, the eigenvalue’s phase factor—normally applied to the target—appears instead as a relative phase on the control qubit, leaving the target unchanged. This surprising “kickback” of phase enables many key quantum algorithms, most notably Quantum Phase Estimation (QPE), where the eigenphase of a unitary is written onto the clock qubits, allowing extraction of otherwise inaccessible phase information using interference and measurement.

In order to understand phase kickback, let us consider the eigenstates of the Pauli-Y matrix (see Eqs. (\ref{Pauli-operators}))
\begin{equation}\label{Y-eigenvectors}
    \ket{y\pm} =
    \frac{\ket{0}\pm i\,\ket{1}}{\sqrt{2}},
\end{equation}
which ensure by definition $Y\ket{y\pm} = \pm\ket{y\pm}$. Applying rotation $R_Y (\theta_Y)$ to the previous states yields
\begin{equation}\label{RY-eigenvectors}
    R_Y (\theta_Y)\ket{y\pm} =
    e^{\mp\,i\,\theta_Y/2} \ket{y\pm}.
\end{equation}
This looks like an eigenvalue equation, but the eigenvalue is only a global phase, which has no physical effect because it cannot be measured. In particular, let us consider
\begin{equation}\label{Y-eigenvector-m}
    \ket{y-} =
    \frac{\ket{0}- i\,\ket{1}}{\sqrt{2}},
\end{equation}
which can be prepared by applying a Hadamard gate $H$ followed by the Hermitian adjoint of the phase gate $S^\dagger$, as shown for qubit $q_1$ in Fig. (\ref{fig:kickback}). The system quantum state at the first barrier of the circuit show in Fig. (\ref{fig:kickback}) is
\begin{eqnarray}\label{backward-before}
    \ket{\Psi_{\mathrm{before}}} &=&
    \ket{y-}
    \otimes
    \frac{\ket{0}+\ket{1}}{\sqrt{2}}=
    \frac{\ket{0}-i\,\ket{1}}{\sqrt{2}}
    \otimes
    \frac{\ket{0}+\ket{1}}{\sqrt{2}} = 
    \nonumber\\
    &=&\frac{1}{2}\,\left(
    \ket{00}+\ket{01}-i\,\ket{10}-i\,\ket{11}
    \right).
\end{eqnarray}
This state can also be represented in column vector representation with regards to the computational basis $\{\ket{00}, \ket{01}, \ket{10}, \ket{11}\}$, namely
\begin{equation}\label{backward-before-vec}
    \ket{\Psi_{\mathrm{before}}} 
= \frac{1}{2}\,
\begin{pmatrix}
    1 \\
    1 \\
    -i \\
    -i
\end{pmatrix}
= \frac{1}{2}\,
\begin{pmatrix}
    1 \\
    -i
\end{pmatrix}
\otimes
\begin{pmatrix}
    1 \\
    1
\end{pmatrix}.
\end{equation}
In this example, let us now suppose to consider the rotation $R_Y(\pi)$ as the unitary gate $U_\pi$, namely
\begin{equation}\label{U_pi}
    U_\pi := 
    R_Y(\pi) = 
    \begin{pmatrix}
    0 & -1 \\
    1 & 0
\end{pmatrix}.
\end{equation}
Let us now couple the two qubits of the example in Fig. (\ref{fig:kickback}). When composing physical systems, the sequential labeling of their components (e.g., $\ket{\psi_0}, \ket{\psi_1}, \dots, \ket{\psi_{n-1}}$) may differ from the mathematical notation used to represent the bit strings, i.e., $\beta_{n-1}\dots\beta_1\beta_0$. In this case, in order to avoid confusion with the HHL algorithm in section \ref{HHL}, let us align the two notations, namely $\ket{q_1 q_0}$. The controlled version of $U_\pi$ can then be constructed as
\begin{equation}\label{CU_pi}
    CU_\pi := 
    I \otimes \ket{0}\bra{0} + 
    R_Y (\pi) \otimes \ket{1}\bra{1},
\end{equation}
where $q_0$ is the control qubit and $q_1$ is the target qubit. Equivalently, in the vector representation, the controlled-unitary looks as 
\begin{equation}\label{controlled-CU_pi}
CU_\pi = 
\begin{pmatrix}
    1 & 0 \\
    0 & 1 
\end{pmatrix}
\otimes
\begin{pmatrix}
    1 & 0 \\
    0 & 0 
\end{pmatrix}
+\begin{pmatrix}
    0 & -1 \\
    1 & 0
\end{pmatrix}
\otimes
\begin{pmatrix}
    0 & 0 \\
    0 & 1 
\end{pmatrix}
=
\begin{pmatrix}
    1 & 0 & 0 & 0\\
    0 & 0 & 0 & -1\\
    0 & 0 & 1 & 0\\
    0 & 1 & 0 & 0
\end{pmatrix}.
\end{equation}
Consequently, the system quantum state at the second barrier of the circuit shown in Fig. (\ref{fig:kickback}) becomes
\begin{equation}\label{kickback-after-vec}
    \ket{\Psi_{\mathrm{after}}} = 
    CU_\pi\,\ket{\Psi_{\mathrm{before}}}=
        \frac{1}{2}\,
\begin{pmatrix}
    1 \\
    i \\
    -i \\
    1
\end{pmatrix}.
\end{equation}
In order to rationalize the previous result, let consider that, when the control is active, as in the second state (corresponding to $\ket{01}$) and in the fourth state (corresponding to $\ket{11}$), then the unitary $U_\pi$ is applied to the target, leading to 
\begin{equation}\label{RY-eigenvector-m}
    U_\pi\,\ket{y-} = R_Y (\pi)\ket{y-} =
    e^{i\,\pi/2} \ket{y-}=
    i\,\ket{y-},
\end{equation}
which is equivalent to multiply by $i$ the corresponding state before the controlled-unitary (i.e. $1\,i = i$ for the second state and $-i\,i=-i^2 = 1$ for the fourth state). Surprisingly, the system quantum state given by the vector reported in Eq.~(\ref{kickback-after-vec}) is separable
\begin{equation}\label{kickback-after-vec-sep}
    \ket{\Psi_{\mathrm{after}}} = 
    \frac{1}{2}\,
\begin{pmatrix}
    1 \\
    i \\
    -i \\
    1
\end{pmatrix}
=
    \frac{1}{2}\,
\begin{pmatrix}
    1 \\
    -i
\end{pmatrix}
\otimes
\begin{pmatrix}
    1 \\
    i
\end{pmatrix}=
\ket{y-}\otimes
\frac{1}{\sqrt{2}}\,
\begin{pmatrix}
    1 \\
    i
\end{pmatrix},
\end{equation}
where the multiplying factor $i$ moves to the control qubit, which is somehow counterintuitive and is the main point here. When a controlled-unitary acts on an eigenstate of the unitary gate, the eigenvalue's phase factor appears as a relative phase on the control qubit, leaving the target unchanged. With other words, one can say that the phase accumulates on the control qubit. This effect, called “kickback” effect, can be made even more evident by recalling that
\begin{eqnarray}\label{backward-after2}
    \ket{\Psi_{\mathrm{after}}} &=&
    \frac{1}{2}\,\left(
    \ket{00}+i\,\ket{01}-i\,\ket{10}+\ket{11}
    \right)\nonumber\\
    &=&
    \frac{\ket{0}-i\,\ket{1}}{\sqrt{2}}
    \otimes
    \frac{\ket{0}}{\sqrt{2}}
    + i\,\frac{\ket{0}-i\,\ket{1}}{\sqrt{2}}
    \otimes
    \frac{\ket{1}}{\sqrt{2}}\nonumber\\
    &=&
    \frac{\ket{0}-i\,\ket{1}}{\sqrt{2}}
    \otimes
    \frac{\ket{0} + i\,\ket{1}}{\sqrt{2}} =
    \ket{y-}
    \otimes
    \frac{\ket{0} + i\,\ket{1}}{\sqrt{2}}.
\end{eqnarray}
Again, in the previous formula, the kickback effect modifies the control state but leaves the target state unchanged.

\newpage

\begin{thebibliography}{21}
\providecommand{\natexlab}[1]{#1}
\providecommand{\url}[1]{\texttt{#1}}
\expandafter\ifx\csname urlstyle\endcsname\relax
  \providecommand{\doi}[1]{doi: #1}\else
  \providecommand{\doi}{doi: \begingroup \urlstyle{rm}\Url}\fi

\bibitem[Wang et~al.(2018)Wang, Zhang, Liu, Cheng, Zhuang, and Chronopoulos]{wang_performance_2018}
Yong-Xian Wang, Li-Lun Zhang, Wei Liu, Xing-Hua Cheng, Yu~Zhuang, and Anthony~T. Chronopoulos.
\newblock Performance optimizations for scalable {CFD} applications on hybrid {CPU}+{MIC} heterogeneous computing system with millions of cores.
\newblock \emph{Computers \& Fluids}, 173:\penalty0 226--236, September 2018.
\newblock ISSN 0045-7930.
\newblock \doi{10.1016/j.compfluid.2018.03.005}.
\newblock URL \url{https://www.sciencedirect.com/science/article/pii/S0045793018301038}.

\bibitem[Nielsen and Chuang(2010)]{nielsen_quantum_2010}
Michael~A. Nielsen and Isaac~L. Chuang.
\newblock \emph{Quantum {Computation} and {Quantum} {Information}: 10th {Anniversary} {Edition}}.
\newblock Cambridge University Press, USA, 2010.

\bibitem[Virtanen et~al.(2020)Virtanen, Gommers, Oliphant, Haberland, Reddy, Cournapeau, Burovski, Peterson, Weckesser, Bright, van~der Walt, Brett, Wilson, Millman, Mayorov, Nelson, Jones, Kern, Larson, Carey, Polat, Feng, Moore, VanderPlas, Laxalde, Perktold, Cimrman, Henriksen, Quintero, Harris, Archibald, Ribeiro, Pedregosa, and van Mulbregt]{virtanen_scipy_2020}
Pauli Virtanen, Ralf Gommers, Travis~E. Oliphant, Matt Haberland, Tyler Reddy, David Cournapeau, Evgeni Burovski, Pearu Peterson, Warren Weckesser, Jonathan Bright, Stéfan~J. van~der Walt, Matthew Brett, Joshua Wilson, K.~Jarrod Millman, Nikolay Mayorov, Andrew R.~J. Nelson, Eric Jones, Robert Kern, Eric Larson, C.~J. Carey, İlhan Polat, Yu~Feng, Eric~W. Moore, Jake VanderPlas, Denis Laxalde, Josef Perktold, Robert Cimrman, Ian Henriksen, E.~A. Quintero, Charles~R. Harris, Anne~M. Archibald, Antônio~H. Ribeiro, Fabian Pedregosa, and Paul van Mulbregt.
\newblock {SciPy} 1.0: fundamental algorithms for scientific computing in {Python}.
\newblock \emph{Nat Methods}, 17\penalty0 (3):\penalty0 261--272, March 2020.
\newblock ISSN 1548-7105.
\newblock \doi{10.1038/s41592-019-0686-2}.
\newblock URL \url{https://www.nature.com/articles/s41592-019-0686-2}.
\newblock Publisher: Nature Publishing Group.

\bibitem[Javadi-Abhari et~al.(2024)Javadi-Abhari, Treinish, Krsulich, Wood, Lishman, Gacon, Martiel, Nation, Bishop, Cross, Johnson, and Gambetta]{javadi-abhari_quantum_2024}
Ali Javadi-Abhari, Matthew Treinish, Kevin Krsulich, Christopher~J. Wood, Jake Lishman, Julien Gacon, Simon Martiel, Paul~D. Nation, Lev~S. Bishop, Andrew~W. Cross, Blake~R. Johnson, and Jay~M. Gambetta.
\newblock Quantum computing with {Qiskit}, June 2024.
\newblock URL \url{http://arxiv.org/abs/2405.08810}.
\newblock arXiv:2405.08810 [quant-ph].

\bibitem[Xu et~al.(2021)Xu, Sun, Endo, Li, Benjamin, and Yuan]{xu_variational_2021}
Xiaosi Xu, Jinzhao Sun, Suguru Endo, Ying Li, Simon~C. Benjamin, and Xiao Yuan.
\newblock Variational algorithms for linear algebra.
\newblock \emph{Science Bulletin}, 66\penalty0 (21):\penalty0 2181--2188, November 2021.
\newblock ISSN 2095-9273.
\newblock \doi{10.1016/j.scib.2021.06.023}.
\newblock URL \url{https://www.sciencedirect.com/science/article/pii/S2095927321004631}.

\bibitem[Guseynov et~al.(2023)Guseynov, Zhukov, Pogosov, and Lebedev]{guseynov_depth_2023}
N.~M. Guseynov, A.~A. Zhukov, W.~V. Pogosov, and A.~V. Lebedev.
\newblock Depth analysis of variational quantum algorithms for the heat equation.
\newblock \emph{Phys. Rev. A}, 107\penalty0 (5):\penalty0 052422, May 2023.
\newblock \doi{10.1103/PhysRevA.107.052422}.
\newblock URL \url{https://link.aps.org/doi/10.1103/PhysRevA.107.052422}.
\newblock Publisher: American Physical Society.

\bibitem[Ingelmann et~al.(2024)Ingelmann, Bharadwaj, Pfeffer, Sreenivasan, and Schumacher]{ingelmann_two_2024}
Julia Ingelmann, Sachin~S. Bharadwaj, Philipp Pfeffer, Katepalli~R. Sreenivasan, and Jörg Schumacher.
\newblock Two quantum algorithms for solving the one-dimensional advection–diffusion equation.
\newblock \emph{Computers \& Fluids}, 281:\penalty0 106369, August 2024.
\newblock ISSN 0045-7930.
\newblock \doi{10.1016/j.compfluid.2024.106369}.
\newblock URL \url{https://www.sciencedirect.com/science/article/pii/S0045793024002019}.

\bibitem[Huang et~al.(2021)Huang, Bharti, and Rebentrost]{huang_near-term_2021}
Hsin-Yuan Huang, Kishor Bharti, and Patrick Rebentrost.
\newblock Near-term quantum algorithms for linear systems of equations with regression loss functions.
\newblock \emph{New J. Phys.}, 23\penalty0 (11):\penalty0 113021, November 2021.
\newblock ISSN 1367-2630.
\newblock \doi{10.1088/1367-2630/ac325f}.
\newblock URL \url{https://doi.org/10.1088/1367-2630/ac325f}.
\newblock Publisher: IOP Publishing.

\bibitem[Hantzko et~al.(2024)Hantzko, Binkowski, and Gupta]{hantzko_tensorized_2024}
Lukas Hantzko, Lennart Binkowski, and Sabhyata Gupta.
\newblock Tensorized {Pauli} decomposition algorithm.
\newblock \emph{Phys. Scr.}, 99\penalty0 (8):\penalty0 085128, July 2024.
\newblock ISSN 1402-4896.
\newblock \doi{10.1088/1402-4896/ad6499}.
\newblock URL \url{https://doi.org/10.1088/1402-4896/ad6499}.
\newblock Publisher: IOP Publishing.

\bibitem[Araujo et~al.(2021)Araujo, Park, Petruccione, and da~Silva]{araujo_divide-and-conquer_2021}
Israel~F. Araujo, Daniel~K. Park, Francesco Petruccione, and Adenilton~J. da~Silva.
\newblock A divide-and-conquer algorithm for quantum state preparation.
\newblock \emph{Sci Rep}, 11\penalty0 (1):\penalty0 6329, March 2021.
\newblock ISSN 2045-2322.
\newblock \doi{10.1038/s41598-021-85474-1}.
\newblock URL \url{https://www.nature.com/articles/s41598-021-85474-1}.
\newblock Publisher: Nature Publishing Group.

\bibitem[Seidel et~al.(2022)Seidel, Bock, Tcholtchev, and Hauswirth]{seidel_qrisp_2022}
Raphael Seidel, Sebastian Bock, Nikolay~Vassilev Tcholtchev, and Manfred Hauswirth.
\newblock Qrisp: a framework for compilable high-level programming of gate-based quantum computers.
\newblock In \emph{International Workshop on Programming Languages for Quantum Computing 2022}, 2022.
\newblock \doi{10.24406/publica-1631}.
\newblock URL \url{https://publica.fraunhofer.de/handle/publica/445649}.

\bibitem[Larocca et~al.(2025)Larocca, Thanasilp, Wang, Sharma, Biamonte, Coles, Cincio, McClean, Holmes, and Cerezo]{larocca_barren_2025}
Martín Larocca, Supanut Thanasilp, Samson Wang, Kunal Sharma, Jacob Biamonte, Patrick~J. Coles, Lukasz Cincio, Jarrod~R. McClean, Zoë Holmes, and M.~Cerezo.
\newblock Barren plateaus in variational quantum computing.
\newblock \emph{Nat Rev Phys}, 7\penalty0 (4):\penalty0 174--189, April 2025.
\newblock ISSN 2522-5820.
\newblock \doi{10.1038/s42254-025-00813-9}.
\newblock URL \url{https://www.nature.com/articles/s42254-025-00813-9}.
\newblock Publisher: Nature Publishing Group.

\bibitem[Bravo-Prieto et~al.(2023)Bravo-Prieto, LaRose, Cerezo, Subaşı, Cincio, and Coles]{Bravo-Prieto2023}
Carlos Bravo-Prieto, Ryan LaRose, M.~Cerezo, Yiğit Subaşı, Lukasz Cincio, and Patrick~J. Coles.
\newblock Variational quantum linear solver.
\newblock \emph{Quantum}, 7, 2023.
\newblock \doi{10.22331/q-2023-11-22-1188}.
\newblock URL \url{https://quantum-journal.org/papers/q-2023-11-22-1188/}.

\bibitem[Cerezo et~al.(2021)Cerezo, Sone, Volkoff, Cincio, and Coles]{Cerezo2021}
M.~Cerezo, Akira Sone, Tyler Volkoff, Lukasz Cincio, and Patrick~J. Coles.
\newblock Cost function dependent barren plateaus in shallow parametrized quantum circuits.
\newblock \emph{Nature Communications}, 12\penalty0 (1), 2021.
\newblock \doi{10.1038/s41467-021-21728-w}.
\newblock URL \url{https://www.nature.com/articles/s41467-021-21728-w}.

\bibitem[Harrow et~al.(2009)Harrow, Hassidim, and Lloyd]{PhysRevLett.103.150502}
Aram~W. Harrow, Avinatan Hassidim, and Seth Lloyd.
\newblock Quantum algorithm for linear systems of equations.
\newblock \emph{Phys. Rev. Lett.}, 103:\penalty0 150502, 10 2009.
\newblock \doi{10.1103/PhysRevLett.103.150502}.
\newblock URL \url{https://link.aps.org/doi/10.1103/PhysRevLett.103.150502}.

\bibitem[Turro et~al.(2025)Turro, Lignarolo, and Dragoni]{leonardo2025}
Francesco Turro, Alessandra Lignarolo, and Daniele Dragoni.
\newblock Practical application of the quantum carleman lattice boltzmann method in industrial cfd simulations, 2025.
\newblock URL \url{https://arxiv.org/abs/2504.13033}.

\bibitem[Zaman et~al.(2023)Zaman, Morrell, and Wong]{hhl_tutorial}
Anika Zaman, Hector~Jose Morrell, and Hiu~Yung Wong.
\newblock A step-by-step hhl algorithm walkthrough to enhance understanding of critical quantum computing concepts.
\newblock \emph{IEEE Access}, 11:\penalty0 77117--77131, 2023.
\newblock \doi{10.1109/ACCESS.2023.3297658}.
\newblock URL \url{https://ieeexplore.ieee.org/document/10189828}.

\bibitem[Agliardi and Prati(2025)]{Agliardi2025}
Gabriele Agliardi and Enrico Prati.
\newblock Quantum data encoding as a distinct abstraction layer in the design of quantum circuits.
\newblock \emph{Quantum Science and Technology}, 10\penalty0 (2), 2025.
\newblock \doi{10.1088/2058-9565/ada6f8}.
\newblock URL \url{https://iopscience.iop.org/article/10.1088/2058-9565/ada6f8}.

\bibitem[Wang et~al.(2020)Wang, Wang, Li, Fan, Wei, and Gu]{Wang2020}
Shengbin Wang, Zhimin Wang, Wendong Li, Lixin Fan, Zhiqiang Wei, and Yongjian Gu.
\newblock Quantum fast poisson solver: the algorithm and complete and modular circuit design.
\newblock \emph{Quantum Information Processing}, 19\penalty0 (6), 2020.
\newblock \doi{10.1007/s11128-020-02669-7}.
\newblock URL \url{https://link.springer.com/article/10.1007/s11128-020-02669-7}.

\bibitem[Zimborás et~al.(2025)Zimborás, Koczor, Holmes, Borrelli, Gilyén, Huang, Cai, Acín, Aolita, Banchi, Brandão, Cavalcanti, Cubitt, Filippov, García-Pérez, Goold, Kálmán, Kyoseva, Rossi, Sokolov, Tavernelli, and Maniscalco]{zimboras_myths_2025}
Zoltán Zimborás, Bálint Koczor, Zoë Holmes, Elsi-Mari Borrelli, András Gilyén, Hsin-Yuan Huang, Zhenyu Cai, Antonio Acín, Leandro Aolita, Leonardo Banchi, Fernando G. S.~L. Brandão, Daniel Cavalcanti, Toby Cubitt, Sergey~N. Filippov, Guillermo García-Pérez, John Goold, Orsolya Kálmán, Elica Kyoseva, Matteo A.~C. Rossi, Boris Sokolov, Ivano Tavernelli, and Sabrina Maniscalco.
\newblock Myths around quantum computation before full fault tolerance: {What} no-go theorems rule out and what they don't, January 2025.
\newblock URL \url{http://arxiv.org/abs/2501.05694}.
\newblock arXiv:2501.05694 [quant-ph].

\bibitem[Iten et~al.(2016)Iten, Colbeck, Kukuljan, Home, and Christandl]{iten_quantum_2016}
Raban Iten, Roger Colbeck, Ivan Kukuljan, Jonathan Home, and Matthias Christandl.
\newblock Quantum circuits for isometries.
\newblock \emph{Phys. Rev. A}, 93\penalty0 (3):\penalty0 032318, March 2016.
\newblock \doi{10.1103/PhysRevA.93.032318}.
\newblock URL \url{https://link.aps.org/doi/10.1103/PhysRevA.93.032318}.
\newblock Publisher: American Physical Society.
\end{thebibliography}

\end{document}